\documentclass[aip,amsmath,amssymv,reprint]{revtex4-2}
\usepackage{graphicx}
\usepackage{float}
\usepackage{subfig}
\graphicspath{{./figures/}}

\usepackage{dcolumn}
\usepackage{bm}

\usepackage[utf8]{inputenc}
\usepackage[T1]{fontenc}
\usepackage{mathptmx}
\usepackage{etoolbox}

\usepackage[version=4]{mhchem}
\usepackage{braket}
\usepackage{multirow} 

\usepackage{subfiles}
\usepackage{todonotes}
\usepackage{amsmath, amssymb}
\usepackage{ulem}
\usepackage{orcidlink}
\usepackage{lipsum}

\newcommand{\Endo}[2]{\ce{#1}@\ce{#2}}
\newcommand{\wavenumber}{ cm$^{-1}$}

\makeatletter
\def\@email#1#2{%
 \endgroup
 \patchcmd{\titleblock@produce}
  {\frontmatter@RRAPformat}
  {\frontmatter@RRAPformat{\produce@RRAP{*#1\href{mailto:#2}{#2}}}\frontmatter@RRAPformat}
  {}{}
}%
\makeatother

\begin{document}
    \title[\Endo{Ne}{C70}]{Targeting spectroscopic accuracy for dispersion bound systems from \textit{ab initio} techniques: translational eigenstates of \Endo{Ne}{C70}} 

    \author{K. Panchagnula*\orcidlink{0009-0004-3952-073X}}
    \email{ksp31@cam.ac.uk}
        \affiliation{Yusuf Hamied Department of Chemistry, University of Cambridge, Cambridge, United Kingdom}
    \author{D. Graf \orcidlink{0000-0002-7640-4162}}
        \affiliation{Yusuf Hamied Department of Chemistry, University of Cambridge, Cambridge, United Kingdom}
        \affiliation{Department of Chemistry, University of Munich (LMU), Munich, Germany}
    \author{E.R. Johnson \orcidlink{0000-0002-5651-468X}}
        \affiliation{Yusuf Hamied Department of Chemistry, University of Cambridge, Cambridge, United Kingdom}
        \affiliation{Department of Chemistry, Dalhousie University, 6243 Alumni Crescent, Halifax, Nova Scotia, Canada }
    \author{A.J.W. Thom \orcidlink{0000-0002-2417-7869}}
        \affiliation{Yusuf Hamied Department of Chemistry, University of Cambridge, Cambridge, United Kingdom}

    \date{\today}

    \begin{abstract}

        We investigate the endofullerene system \Endo{Ne}{C70}, by constructing a three-dimensional Potential Energy Surface (PES) describing the translational motion of the \ce{Ne} atom. This is constructed from electronic structure calculations from a plethora of methods including: MP2, SCS-MP2, SOS-MP2, RPA@PBE, C(HF)-RPA, which were previously used for \Endo{He}{C60} in J. Chem. Phys. 160, 104303 (2024), alongside B86bPBE-25X-XDM and B86bPBE-50X-XDM. The reduction in symmetry moving from \ce{C60} to \ce{C70} introduces a double well potential along the anisotropic direction, which forms a test of the sensitivity and effectiveness of the electronic structure methods. Due to the large cost of these calculations, the PES is interpolated using Gaussian Process Regression due to its effectiveness with sparse training data. The nuclear Hamiltonian is diagonalised using a symmetrised double minimum basis set outlined in J. Chem. Phys. 159, 164308 (2023), with translational energies having error bars $\pm 1$ and $\pm 2$ \wavenumber. We quantify the shape of the ground state wavefunction by considering its prolateness and kurtosis, and compare the eigenfunctions between electronic structure methods from their Hellinger distance. We find no consistency between electronic structure methods as they find a range of barrier heights and minima positions of the double well, and different translational eigenspectra which also differ from the Lennard--Jones (LJ) PES given in J. Chem. Phys. 101, 2126–2140 (1994). We find that generating effective LJ parameters for each electronic structure method cannot reproduce the full PES, nor recreate the eigenstates and this suggests that the LJ form of the PES, while simple, may not be best suited to describe these systems. Even though MP2 and RPA@PBE performed best for \Endo{He}{C60}, due to the lack of concordance between all electronic structure methods we require more experimental data in order to properly validate the choice.

    \end{abstract}

    \maketitle
    
    \section{Introduction \label{sec: Intro}}    
    	Endofullerenes are a collection of species, where atom(s) or small molecule(s) are trapped inside fullerene cages. This entrapment provides a bounding potential for the encapsulated species, quantising its translational motion.\cite{bacicPerspectiveAccurateTreatment2018} The development of a technique known as ``molecular surgery''\cite{murataSynthesisReactionFullerene2008, bloodworthSynthesisEndohedralFullerenes2022} has allowed for the controlled synthesis and characterisation of these species. 

    Experimentally and theoretically, the vast majority of studies have focused on using \ce{C60} as the encapsulating fullerene\cite{wangFormulationsClosedshellInteractions2010, pyykkoLondontypeFormulaDispersion2007} for species such as \Endo{He}{C60},\cite{bacanuExperimentalDeterminationInteraction2021c, jafariTerahertzSpectroscopyHelium2022} \Endo{H2}{C60}\cite{xuCoupledTranslationrotationEigenstates2009, xuH2HDD22008, xuQuantumDynamicsCoupled2008, xuInelasticNeutronScattering2013, xuLightMoleculesNanocavities2020, felkerTranslationrotationStates602016} and \Endo{H2O}{C60}.\cite{xuFullerene60Inelastic2022, felkerFlexibleWaterMolecule2020, felkerCommunicationQuantumSixdimensional2016, carrillo-bohorquezEncapsulationWaterMolecule2021, rashedInteractionsWaterMolecule2019} Its attractiveness can be attributed to its much larger abundance and high $I_h$ symmetry, but this is not the only possible choice. The next most easily synthesisable is \ce{C70}, which can be conceptualised as elongating a single axis of \ce{C60}, and thereby reducing the symmetry to $D_{5h}$. 

    There have been fewer investigations where \ce{C70} has been used as the encapsulating cage,\cite{sebastianelliHydrogenMoleculesFullerene2010, foroutan-nejadDipolarMoleculesC702016, caliskanStructuralElectronicAdsorption2021} and of those, the Potential Energy Surface (PES) is usually approximated as a pairwise Lennard--Jones (LJ) summation.\cite{xuH2HDD22008, xuQuantumDynamicsCoupled2008,xuCoupledTranslationrotationEigenstates2009, xuInelasticNeutronScattering2013, felkerCommunicationQuantumSixdimensional2016, xuLightMoleculesNanocavities2020, felkerFlexibleWaterMolecule2020, jafariNeArKr2023a, panchagnulaExploringParameterSpace2023b, mandziukQuantumThreeDimensional1994} The difficulty in accurately describing the dominant dispersion interactions in these endofullerene systems, alongside the high dimensionality of the PES, puts considerable strain on highly accurate electronic structure (ES) calculations. In order to keep the dimensionality of the system as low as possible, we here investigate the translational eigenstates of the noble gas endofullerene \Endo{Ne}{C70}. To a good approximation,\cite{panchagnulaTranslationalEigenstatesHe2024a} we can keep the \ce{C70} cage fixed and only consider the translational motion of the single Ne atom. 

    For \Endo{He}{C60}, the spectroscopic observations motivated the validation of choice of electronic structure methods.\cite{panchagnulaTranslationalEigenstatesHe2024a} The lack of experimental data for \Endo{Ne}{C70} dictates that a variety of ES methods need to be tested in order to gauge the feasibility of the method, alongside gaining confidence in the effectiveness in any particular method. 

    Due to the large cost of the electronic structure methods, it is impractical to run thousands of calculations in order to generate the PES. Instead, an interpolation scheme can be used. Due to their efficiency with sparse input data and conceptual simplicity, Gaussian Processes (GPs) have risen in popularity for PES evaluation.\cite{behlerPerspectiveMachineLearning2016, musilPhysicsInspiredStructuralRepresentations2021, dralStructurebasedSamplingSelfcorrecting2017} The integral part of the GP is the choice of covariance function, or kernel, whose hyperparameters are optimised during its training. These are what introduce the flexibility into the GP and allow it to efficiently model many different systems.\cite{rasmussenGaussianProcessesMachine2005} 

    After learning the PES, the translational eigenstates can be found by diagonalising the nuclear Hamiltonian. The ES methods can be compared against each other by either considering the translational energies, or the wavefunctions.\cite{panchagnulaTranslationalEigenstatesHe2024a} By ensuring the energy zero for each ES method is the minimum of the PES, the obtained energy values are comparable, allowing each ES method to be scrutinised on an equal footing. The wavefunctions, or nuclear orbitals, can be compared by considering the distance between them.

    In this paper, we take methods from previous studies and apply the techniques to the little studied \Endo{Ne}{C70} system. We use the electronic structure and PES generation methodology used for \Endo{He}{C60} as described in Ref \citenum{panchagnulaTranslationalEigenstatesHe2024a}, and the nuclear Hamiltonian diagonalisation technique presribed for \Endo{X}{C70} in Ref \citenum{panchagnulaExploringParameterSpace2023b}. The key components of these methodologies are reproduced in Section \ref{sec: Methodology}. The PES and eigenstates are presented in Section \ref{sec: Results}, alongside a discussion of the quality of the electronic structure methods as compared to using a simple Lennard--Jones PES. Concluding remarks and future prospects are outlined in Section \ref{sec: Conc}.

    \section{Theory \label{sec: Methodology}}
    	\subsection{Electronic Structure Calculations\label{sec: TheoryES}}

        Building upon prior investigations of \Endo{He}{C60},\cite{panchagnulaTranslationalEigenstatesHe2024a} we will employ an expanded array of electronic structure methods to generate the PES for \Endo{Ne}{C70}. In the case of \Endo{He}{C60}, the availability of experimental data\cite{bacanuExperimentalDeterminationInteraction2021c, jafariTerahertzSpectroscopyHelium2022} facilitated a robust comparison between the outcomes from various computational approaches and experimental results. While awaiting experimental data for \Endo{Ne}{C70}, we have increased the repertoire of ES methods to enhance the reliability estimation of our computational findings. This approach is necessitated by the challenges encountered by various electronic structure methods in accurately capturing the interactions within these systems, as highlighted in a recent review.\cite{cioslowskiElectronicStructureCalculations2023a}

        Accurately modeling the pivotal dispersion effects in this system from first principles presents significant challenges. Moreover, due to the system's size and the large number of points required to construct the PES, the ES method must also be highly efficient, which substantially narrows the choice of applicable ES methods. Based on our experiences with \Endo{He}{C60}, we will employ $\omega$-RI-CDD-MP2\cite{glasbrennerEfficientReducedScalingSecondOrder2020}--- a highly efficient variant of second-order Møller–Plesset perturbation theory (MP2), utilizing resolution of identity (RI) and Cholesky decomposed densities (CDD) --- along with its scaled opposite spin (SOS)\cite{jungScaledOppositespinSecond2004} and spin component scaled (SCS)\cite{grimmeImprovedSecondorderMoller2003} variants. Additionally, we will use $\omega$-CDGD-RPA\cite{grafAccurateEfficientParallel2018} --- an efficient random phase approximation (RPA) variant based on the Cholesky decomposed ground state density (CDGD) --- and its Hartree--Fock (HF) corrected version,\cite{grafCorrectedDensityFunctional2023} C(HF)-RPA. In the following, we will provide a succinct rationale for the method choices we have made.

        Wavefunction (WF) methods are known for their high accuracy but are often prohibitive to use due to their substantial computational demands. These demands arise from the steep scaling of computational costs with system size and the need for large basis sets to achieve reliable results.\cite{ginerCuringBasissetConvergence2018} Among WF methods, MP2 has been established as both accurate and computationally efficient\cite{pulayLocalizabilityDynamicElectron1983, haserMollerPlessetMP2Perturbation1993, maslenNoniterativeLocalSecond1998,ayalaLinearScalingSecondorder1999,schutzLoworderScalingLocal1999, saeboLowscalingMethodSecond2001, wernerFastLinearScaling2003, jungFastCorrelatedElectronic2006,jungFastEvaluationScaled2007, doserTighterMultipolebasedIntegral2008, doserLinearscalingAtomicOrbitalbased2009, zienauCholeskydecomposedDensitiesLaplacebased2009, kristensenMP2EnergyDensity2012, maurerCholeskydecomposedDensityMP22014, pinskiSparseMapsSystematic2015,nagyIntegralDirectLinearScalingSecondOrder2016, baudinEfficientLinearscalingSecondorder2016, phamHybridDistributedShared2019, barcaQMP2OSMollerPlesset2020, glasbrennerEfficientReducedScalingSecondOrder2020, forsterQuadraticPairAtomic2020} --- particularly in its most efficient formulations --- and is thus our preferred WF method. However, in recent work it was reported that MP2 can significantly overestimate the strengths of non-covalent interactions, with errors increasing with system size and discrepancies reaching up to 100\,\% of the binding energies.\cite{nguyenDivergenceManyBodyPerturbation2020a} The semi-empirical variants SOS-MP2 and SCS-MP2 were shown to reduce this erratic behavior\cite{nguyenDivergenceManyBodyPerturbation2020a} and hence we include them as well.   

        The RPA is a method on the border between wave function theory (WFT) and density functional theory (DFT) and combines several appealing features: it is entirely ab initio; it offers a robust intrinsic description of dispersion effects; and, due to recent advancements, it has achieved high computational efficiency. Furthermore, RPA has been demonstrated to counteract the erratic behavior observed in MP2, proving its reliability across various system sizes.\cite{nguyenDivergenceManyBodyPerturbation2020a} Given these advantages, incorporating RPA into our computational toolkit is a reasonable choice.

        RPA calculations are typically conducted in a post-Kohn--Sham manner. As a rung five functional, this involves using orbitals and orbital energies derived from a preceding density functional approximation (DFA) calculation. The preferred DFA for this purpose is the functional developed by Perdew, Burke, and Ernzerhof (PBE)\cite{perdewGeneralizedGradientApproximation1996, perdewGeneralizedGradientApproximation1997} and is also the choice for the current work; we will denote the approach as RPA@PBE. However, recent studies into the inaccuracies of DFAs have revealed that particularly DFAs of lower rank like PBE can generate substantial errors in the densities.\cite{namExplainingFixingDFT2021, namMeasuringDensityDrivenErrors2020, simQuantifyingDensityErrors2018, simImprovingResultsImproving2022, vuckovicDensityFunctionalAnalysis2019, songDensitySensitivityEmpirical2021, kimUnderstandingReducingErrors2013, martin-fernandezUseNormalizedMetrics2021, kimHalogenChalcogenBinding2019, kimIonsSolutionDensity2014, wassermanImportanceBeingInconsistent2017} It follows, then, that the associated orbitals and orbital energies might also be compromised, potentially affecting the accuracy of subsequent RPA calculations. Addressing this issue, two of us have recently developed an approach termed C(HF)-RPA\cite{grafCorrectedDensityFunctional2023}, the results of which will also be presented in this study.

        To account for the basis set incompleteness error which can be prominent in dispersion bound systems,\cite{eshuisBasisSetConvergence2012} the electronic structure calculations are extrapolated to the complete basis set (CBS) limit. The SCF\cite{fellerUseSystematicSequences1993} and correlation\cite{eshuisBasisSetConvergence2012, fellerEffectivenessCCSDComplete2011, halkierBasissetConvergenceCorrelated1998, helgakerBasissetConvergenceCorrelated1997} energies are extrapolated separately, as
            \begin{align}
                E_{\text{SCF}}^X &= E_{\text{SCF}}^{\infty} + a\exp(-bX) \label{eq: CBS SCF}\\
                E_{\text{corr}}^X &= E_{\text{corr}}^{\infty} + (X+d)^{-3} \label{eq: CBS correlation}
            \end{align}
        using the correlation consistent basis sets cc-pVXZ.\cite{balabanovSystematicallyConvergentBasis2005, dunningGaussianBasisSets1989, koputInitioPotentialEnergy2002, prascherGaussianBasisSets2011, wilsonGaussianBasisSets1999, woonGaussianBasisSets1993, woonGaussianBasisSets1994} For the SCF energies, we use $X\in\{\text{T,Q,5}\}$, whereas for the correlation energies we use $X\in\{\text{T,Q}\}$ with $d=0$ and $-1.17$ for the MP2 type and RPA type methods, respectively.

        In addition to the methods used in our investigation of \Endo{He}{C60}, this study also includes the B86bPBE-25-XDM and B86bPBE-50-XDM functionals. The B86bPBE-25-XDM and B86bPBE-50-XDM functionals represent advanced hybrid DFAs that incorporate 25\% and 50\% exact exchange, respectively, and dispersion corrections using the exchange-hole dipole moment (XDM) model. These functionals have shown high accuracy in predicting lattice energies and intermolecular interactions, significantly outperforming many other DFAs.\cite{priceRequirementsAccurateDispersioncorrected2021} The excellent performance is attributed to the accurate description of dispersion effects and the proper handling of electronic many-body interactions, making these functionals particularly suitable for complex molecular systems. Calculations using the MP2 and RPA type methods were carried out using the FermiONs++ package, and the B86bPBE calculations were carried out using the Fritz Haber Institute \textit{ab initio} materials simulations (FHI-aims) package with \textit{tight} settings.\cite{blumInitioMolecularSimulations2009, ihrigAccurateLocalizedResolution2015, priceXDMcorrectedHybridDFT2023}

    \subsection{Potential Energy Surface Generation\label{sec: TheoryPES}}
            \begin{figure}
                \centering
                \includegraphics[scale=0.33]{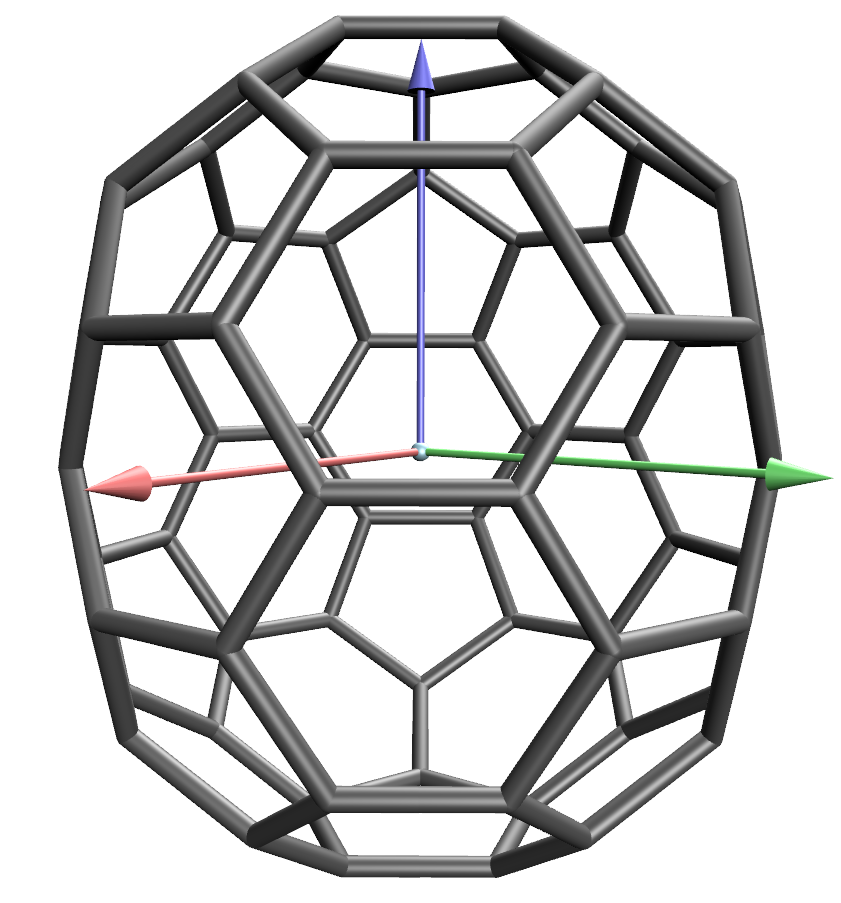}
                \caption{Orientation of Cartesian $x,y,z$ axes as the red, green and blue arrows within the fixed \ce{C70} cage.}
                \label{fig: C70}
            \end{figure}

        With the \ce{C70} geometry fixed and symmetrised,\cite{bakerStructurePropertiesC701991, huynhQSym2QuantumSymbolic2024}, it is oriented as in a previous study and shown in Fig \ref{fig: C70} with the $z$-axis and $x$-axis coinciding with the $C_5$ and $C_2'$ rotation axes respectively dictated by the $D_{5h}$ symmetry of the system.\cite{panchagnulaExploringParameterSpace2023b}  Whilst the shape of the cage and cavity could motivate the use of cylindrical polar coordinates, instead for ease of intuition and simplicity, we choose to work in a Cartesian representation. 

        As mentioned in Sec \ref{sec: Intro}, the core component of the GP is its kernel. While there are a host of suitable choices, following previous work we choose to build it from a Mat\'{e}rn covariance function, with $\nu$ fixed at 2.5 to ensure twice differentiability, with an anisotropic distance function. The full form of the kernel is 
            \begin{equation}
                K_{ij} = \sigma^2M_{2.5}(d(\mathbf{X}_i, \mathbf{X}_j)) + \nu^2\delta_{ij}, \label{eq: GP kernel}
            \end{equation}
        where $\sigma^2$ and $\nu^2$ are the amplitude and noise hyperparameters which are optimised alongside the three Cartesian length scales. Once trained, the GP can be queried to evaluate the value of the PES and an associated covariance as\cite{rasmussenGaussianProcessesMachine2005}
            \begin{align}
                E(\mathbf{X}) &= \mathbf{K}_{pt}\mathbf{K}_{tt}^{-1}\mathbf{E}_{t},\label{eq: GP mean}\\
                \boldsymbol{\Sigma}(\mathbf{X}) &= \mathbf{K}_{pp}-\mathbf{K}_{pt}\mathbf{K}_{tt}^{-1}\mathbf{K}_{pt},\label{eq: GP covariance}
            \end{align}
        where $\mathbf{K}$ and $\mathbf{X}$ refer to the kernel matrix and the endohedral \ce{Ne} position respectively, with the subscripts $p$ and $t$ referring to the prediction and training points respectively. As the training coordinates are chosen randomly (as outlined within the supplementary information), the spatial symmetry of the system is not guaranteed to be enforced when sampling the trained surface. A PES that obeys the symmetry correctly will transform purely as the totally symmetric irreducible representation (irrep) within the point group of the system. The amount of symmetry breaking (which must be removed before calculating properties from the surface) can be quantified by calculating the projection of the PES onto each irrep of the point group. We first define the inner product
            \begin{equation}
                \braket{\hat{g}V(\mathbf{X})|V(\mathbf{X})}=\int_{\mathbb{R}^3}V(\hat{g}^{-1}\mathbf{X})V(\mathbf{X})w(\mathbf{r})d^3\mathbf{X}, \label{eq: PES symmetry}
            \end{equation}
        where $\hat{g}$ is a symmetry operation in the point group $\mathcal{G}$ (in this specific case $\mathcal{G}$ is $D_{5h}$) and $w(\mathbf{r})$ is a weight function which can be chosen to ensure convergence of the integral, and any other specified constraints. By calculating this for all symmetry operations $\hat{g} \in \mathcal{G}$, then renormalising by the value of $\braket{V(\mathbf{r})|V(\mathbf{r})}$ and applying the reduction formula, this calculates the projection of the PES onto each irrep. This allows for assessment as to whether the PES obeys the symmetry of the system. 

    \subsection{Nuclear Hamiltonian Diagonalisation\label{sec: TheoryEigenstates}}

        The Hamiltonian for the \Endo{Ne}{C70} system in atomic units can be decomposed into
            \begin{align}
                \hat{H}&=-\frac{1}{2M}\nabla^2+V(\mathbf{r})\nonumber\\
                &=\left(-\frac{1}{2M}\frac{\partial^2}{\partial x^2}+\frac{1}{2}k_xx^2\right)+\left(-\frac{1}{2M}\frac{\partial^2}{\partial y^2}+\frac{1}{2}k_yy^2\right)\nonumber\\
                &+\left(-\frac{1}{2M}\frac{\partial^2}{\partial z^2}\right) + \left(V(x,y,z)-\frac{1}{2}k_xx^2-\frac{1}{2}k_yy^2\right)\nonumber\\
                &=\hat{h}^0_x+\hat{h}^0_y+\hat{k}_z+\Delta V, \label{eq: Hamiltonian}
            \end{align}
        where $M$ is the effective two-particle reduced mass of the \ce{Ne} and \ce{C70}. This partitioning is motivated by assuming a harmonic well in the $x$ and $y$ directions, with a double well in the $z$ direction and a correction $\Delta V$ to recover the true three-dimensional PES. The effective force constants $k_x$ and $k_y$ can be chosen as required, either within a discrete variable representation\cite{lightGeneralizedDiscreteVariable1985, lightDiscreteVariableRepresentationsTheir2000} (DVR) or a potential optimised discrete variable representation\cite{echavePotentialOptimizedDiscrete1992} (PODVR) framework. The former selects a region within the fullerene which encapsulates the region that the endohedral \ce{Ne} atom can explore which is usually constrained by a potential energy cutoff, whereas the latter selects this region by taking into account the shape of the PES. 

        The translational basis set is constructed as a direct product of one-dimensional basis functions
            \begin{align}
                \ket{n_i^{+/-/0}}&=\frac{1}{\sqrt{2^nn!\sqrt{\pi}}}H_n(q_i^{+/-/0})\exp\left(-\frac{1}{2}(q_i^{+/-/0})^2\right), \label{eq: 1D basis functions}\\
                \ket{n_{x/y}}&:=\ket{n_{x/y}^0}, \label{eq: x/y basis}\\
                \ket{n_z}&:=\frac{1}{\sqrt{2}}\left(\Ket{\left\lfloor\frac{n_z}{2}\right\rfloor^-}+(-1)^{\lfloor\frac{n_z}{2}\rfloor+(n_z \text{mod 2})}\Ket{\left\lfloor\frac{n_z}{2}\right\rfloor^+}\right), \label{eq: z basis}
            \end{align}
        where the subscript $i$ refers to a Cartesian direction, the superscripts $+/-/0$ refer to the position of the minima located along the coordinate $q_i=\sqrt{\alpha}(r_i-r_i^{+/i/0})$, which is a centred PODVR scaled Cartesian coordinate, with $r_i$ referring to the $x$, $y$ or $z$ direction.\cite{panchagnulaExploringParameterSpace2023b} In the $x$ and $y$ directions, the functions are one-dimensional harmonic oscillators, with $x_i^0=0$. Owing to the double well along the $z$ direction,\cite{mandziukQuantumThreeDimensional1994} the symmetrised double minimum functions are used instead, as in Ref \citenum{panchagnulaExploringParameterSpace2023b}. This basis set is non-orthogonal but this can be treated by canonically orthogonalising the basis set. If there are any linearly dependencies, these can be removed by projecting into the linearly independent subspace.

        Alongside the translational energy levels, this procedure also gives the eigenfunctions, which can be thought of as nuclear orbitals. The ground state wavefunction can be classified using its prolateness defined in Eq \eqref{eq: prolateness} which compares the spread of the wavefunction in the $z$ and $x$ directions which is influenced by the location of the minima of the double well; and the kurtosis as defined in Eq \eqref{eq: kurtosis} which describes how squashed or double-peaked the wavefuntion is along the $z$ direction, which is influenced by the precise shape of the double well.\cite{panchagnulaExploringParameterSpace2023b}        \begin{align}
                \varsigma^2&=\frac{\sigma_z^2}{\sigma_x^2}=\frac{\braket{0|z^2|0}}{\braket{0|x^2|0}}, \label{eq: prolateness}\\
                \kappa_z&=\frac{1}{\sigma_z^4}\braket{0|z^4|0}. \label{eq: kurtosis}
            \end{align}
            
        Comparing the wavefunctions between different ES methods can also be achieved by considering their Hellinger distance 
            \begin{align}
                H(\Phi,\Psi)&=\sqrt{1-|\braket{\Phi|\Psi}|}\nonumber\\
                &=\sqrt{1-|\sum_{\mathbf{m},\mathbf{n}}c_{\mathbf{m}}c_{\mathbf{n}}\braket{m_x|n_x}\braket{m_y|n_y}\braket{m_z|n_z}|},\label{eq: Hellinger}
            \end{align}
        where $\mathbf{m}$ and $\mathbf{n}$ refer to the tuples of quantum numbers $(m_x,m_y,m_z)$ and $(n_x,n_y,n_z)$ respectively. Owing to the differing PODVR scale factors of the dimensionless $q_i$ coordinates between ES methods, the orthonormality of $\braket{m_x|n_x}$ and $\braket{m_y|n_y}$ is not guaranteed. Along the $z$ direction, the use of the symmetrised double minimum basis already removes the orthonormality constraint as it is a non-orthogonal basis set, even without considering the scaling or the centres of expansion.

    \section{Results \label{sec: Results}}
    	\subsection{Potential Energy Surface\label{sec: ResultsPES}}

        \begin{figure*}
            \centering
            \includegraphics[scale=1]{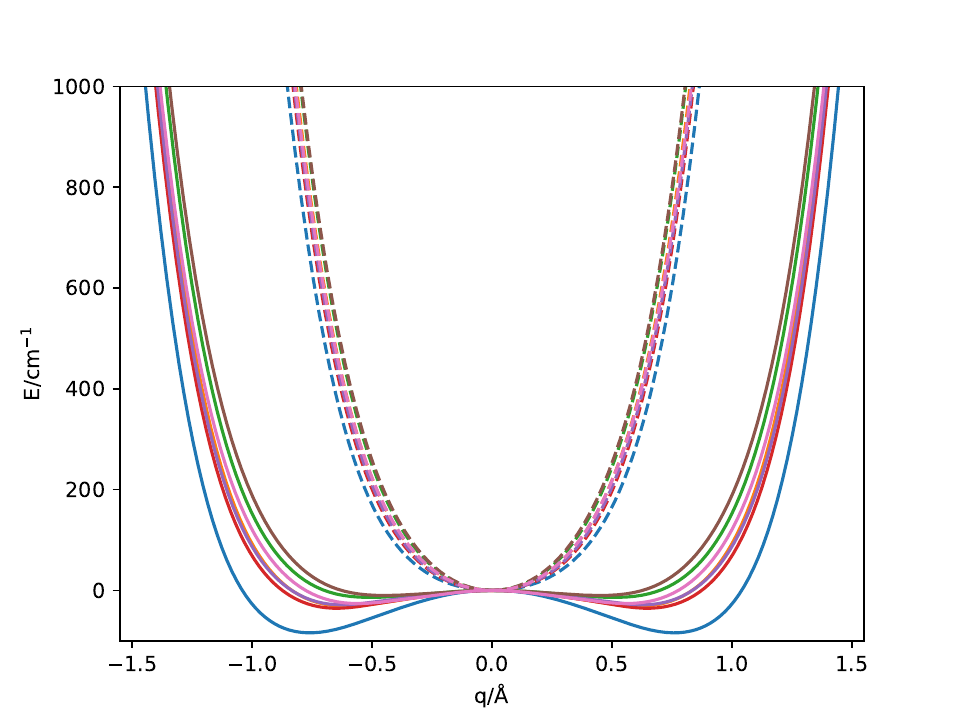}
            \caption{One dimensional slices in the $x$ (dashed) and $z$ (solid) directions between $\pm1$\AA\, and $\pm1.5$\AA\, respectively for all electronic structure methods used. MP2 in blue, SCS-MP2 in orange, SOS-MP2 in green, RPA@PBE in red, C(HF)-RPA in purple, B86bPBE-25X-XDM in brown, and B86bPBE-25X-XDM in pink.}
            \label{fig: contours}
        \end{figure*}
    
        \begin{table}
            \centering
            \begin{tabular}{ccccccccc}
                 \hline Method&$A_1'$&$A_2'$&$E_1'$&$E_2'$&$A_1''$&$A_2''$&$E_1''$&$E_2''$  \\
                 \hline MP2&0.99998&0&0&1&0&0&0&0\\
                 SCS-MP2&0.99991&0&1&6&0&0&0&0\\
                 SOS-MP2&0.99984&0&2&11&0&1&0&0\\
                 RPA@PBE&0.99988&0&1&10&0&0&1&0\\
                 C(HF)-RPA&0.99986&0&1&11&0&0&1&0\\
                 B86bPBE-25X-XDM&0.99988&0&8&2&0&0&2&0\\
                 B86bPBE-50X-XDM&0.99978&0&16&1&0&0&3&0\\
                 \hline
            \end{tabular}
            \caption{Projections of the unsymmetrised PESs for each ES method on all irreps of $D_{5h}$, calculated using Eq \eqref{eq: PES symmetry}, with an anisotropic Gaussian weight function. The magnitude of non $A_1'$ components, which are scaled by a factor of $10^5$, quantifies the amount of ``symmetry breaking'' in the PES for each ES.}
            \label{table: PES symmetry}
        \end{table}

        As the $D_{5h}$ symmetry of the true PES was not enforced by the GP, its projection onto each irrep as defined in Eq \eqref{eq: PES symmetry} was calculated \textsc{QSym\textsuperscript{2}} package.\cite{huynhQSym2QuantumSymbolic2024} Taking inspiration from the translational basis set based on harmonic oscillators, an anisotropic Gaussian was used as the weight function, with further details given in the supplementary information. It is important to note that the ``symmetry breaking'' in the GP learned surfaces is a byproduct of what is contained within the ES calculations. That is to say, the symmetry breaking is present within the training data, due to the tolerances on the calculations which is propagated through the learning of the PES. For the unsymmetrised PESes, the amount of ``symmetry breaking'' for each ES method is given in Table \ref{table: PES symmetry} for all irreps in the point group $D_{5h}$. Even though all ES methods have a coefficient larger than 0.999 on the $A_1$' irrep, the contribution of the other irreps gives more insight into the amount of ``symmetry breaking''. Generally, the MP2 type methods (MP2, SCS-MP2, SOS-MP2) have a smaller contribution of the non totally symmetric irreps as compared to the RPA type methods (RPA@PBE, C(HF)-RPA). For the B86bPBE methods, these are sensitive to the amount of Hartree--Fock exchange used, with 25\% being comparable with the MP2 type methods, and 50\% more so with the RPA type methods. This is a propagation of the range of energies that each ES method calculates for a set of symmetrically equivalent points. For the MP2 type methods, while larger basis sets which include more diffuse functions have a larger spread, these are still sub-wavenumber. For the RPA methods, these do not show the same dependence on basis set size and have a spread of approximately 1.5\wavenumber. This is much larger than the $\mu$\wavenumber spread of energies for the LJ type PES, which given the enforced $D_{5h}$ of the cage, shows no ``symmetry breaking''. Before calculating properties, the PES needs to have its symmetry restored; this is achieved by averaging each point over its symmetrically equivalent copies. The precise details of this procedure are outlined in the supplementary information. Henceforth, references to the PES will refer to the versions appropriately symmetrised under $D_{5h}$.

       \begin{table}
           \centering
           \begin{tabular}{ccc}
                \hline Method&Barrier Height/\wavenumber&$z_{\mathrm{min}}$/\AA  \\
                \hline MP2&84.21&0.760\\
                SCS-MP2&31.21&0.610\\
                SOS-MP2&13.89&0.506\\
                RPA@PBE&35.46&0.648\\
                C(HF)-RPA&28.57&0.626\\
                B86bPBE-25X-XDM&10.05&0.450\\
                B86bPBE-50X-XDM&26.07&0.558\\
                LJ&7.61&0.41\\
                \hline
           \end{tabular}
           \caption{Barrier heights and positions of the minima of the double well along the anisotropic $z$ direction for each ES method. The LJ values are calculated from the parameters used in Ref \citenum{mandziukQuantumThreeDimensional1994}.}
           \label{table: PES features}
       \end{table}

       One-dimensional slices of the PES along the $x$ (dashed) and $z$ (solid) directions for all ES methods used are shown in Fig \ref{fig: contours}. Along the $x$ direction, all seven methods seem very similar, with very little discernible difference, with the only noticeable feature being the MP2 (blue) being shallower and lying lower in energy than all the other methods. The ES methods do not appear to cross each other, indicating their curvature in the radially symmetric $xy$ plane is very similar, which would lead to very similar translational frequencies along this mode. In the $z$ direction, with the appearance of the double well, the differences between all the ES methods become more apparent, in the depths and positions of the double well. The obvious outlier is the MP2, which has the most prominent double well, with the minima furthest apart at $\pm$0.760~\AA\, and a barrier height of 84.21\wavenumber. The remaining six ES methods are more consistent with each other but they still have barrier heights ranging between 10.05\wavenumber and 35.46\wavenumber, with the minima placed between $\pm$0.450~\AA\, and 0.648~\AA. These features are summarised in Table \ref{table: PES features} which highlights the lack of concordance between the full set of ES methods, and the disagreement with the LJ PES. An important feature to note is the discrepancy between the B86bPBE methods, illustrating the dependence on this method on the exact amount of Hartree--Fock exchange used. Even though all ES methods disagree on the precise parameters that describe the double well, they do corroborate the existence of the double well. While for \Endo{He}{C60} the MP2 and RPA@PBE described the endohedral interaction well with the latter being preferred for its computational efficiency, \cite{panchagnulaTranslationalEigenstatesHe2024a} the spread of values in \Endo{Ne}{C70} makes it unclear which ES method is accurately describing the \ce{Ne-C} endohedral interaction. The values of barrier heights motivates the grouping of ES methods into three distinct sets: \{MP2\}, \{SCS-MP2, RPA@PBE, C(HF)-RPA, B86bPBE-50X-XDM\}, and \{SOS-MP2, B86bPBE-25X-XDM\} referring to large, medium and low barrier heights.

       Previous studies have made extensive use of LJ parameters for this endohedral \ce{Ne-C} interaction,\cite{panchagnulaExploringParameterSpace2023b, mandziukQuantumThreeDimensional1994} and it can be asked what LJ parameters can be used to represent the PES of each ES method, and whether this simplified functional form is a good representation. These effective LJ parameters can be found by matching any two features of the \Endo{Ne}{C70} system, whether from the PES or from properties derived from it such as the translational eigenstates. The simplest set of properties to choose are the features that define the double well given in Table \ref{table: PES features}. Fitting to these, we find that the energy scale, $\varepsilon$ ranges between 42.27\wavenumber and 67.41\wavenumber, and the length scale, $\sigma$ ranges between 2.897~\AA\, and 3.003~\AA. Whether these effective LJ parameters form a good representation of the system can be quantified by calculating the translational eigenspectrum, and comparing it to the true ES methods and is considered in Sec \ref{sec: ResultsEigenstates}. Details of the effective LJ parameters and the translational eigenspectra are provided in the supplementary information.

    \subsection{\Endo{Ne}{C70} Eigenstates \label{sec: ResultsEigenstates}}
        \begin{figure*}
            \centering
            \includegraphics[scale=1.2]{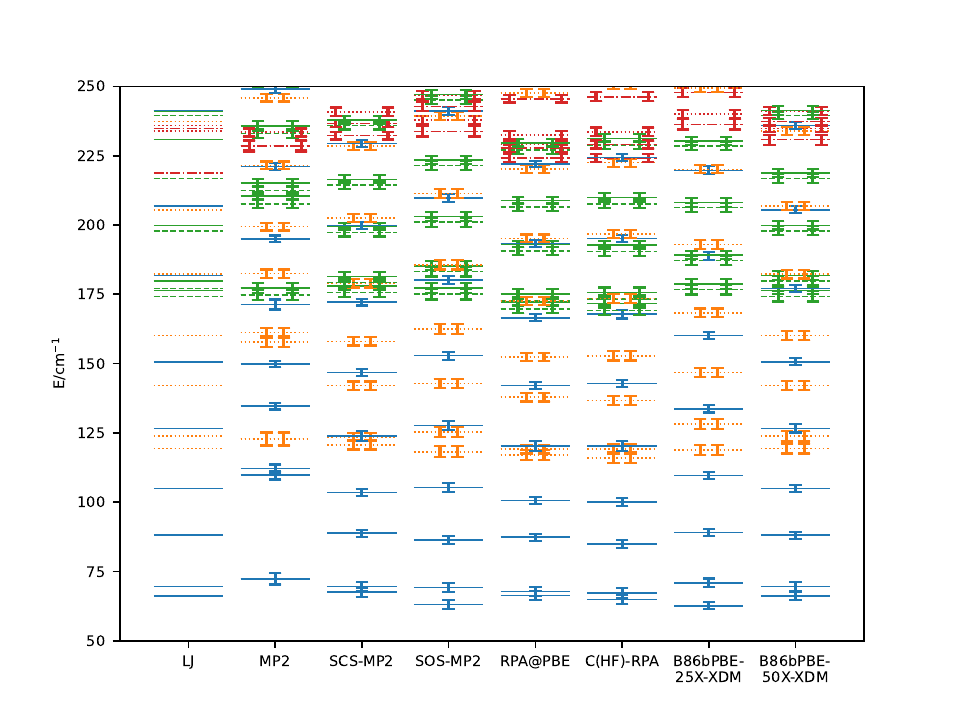}
            \caption{Lowest 50 translational energies of \Endo{Ne}{C70} for MP2, SCS-MP2, SOS-MP2, RPA@PBE, C(HF)-RPA, B86bPBE-25X-XDM and B86bPBE-50X-XDM PESs, alongside the LJ counterparts. The energy zero is set to the minimum of each PES, ensuring this is an absolute energy scale. The colours and linestyle correspond to the principal and angular momentum quantum numbers $(n,|l|)$ respectively. $n\in[0,1,2,3]$ is shown in blue, orange, green and red lines and error bars; $|l|\in[0,1,2,3]$ is shown by solid, dotted, dashed, dash-dot lines.}
            \label{fig: Energy Levels}
        \end{figure*}

        \begin{figure*}
            \centering
            \setcounter{subfigure}{0}
            \subfloat[MP2]{\includegraphics[scale=0.42]{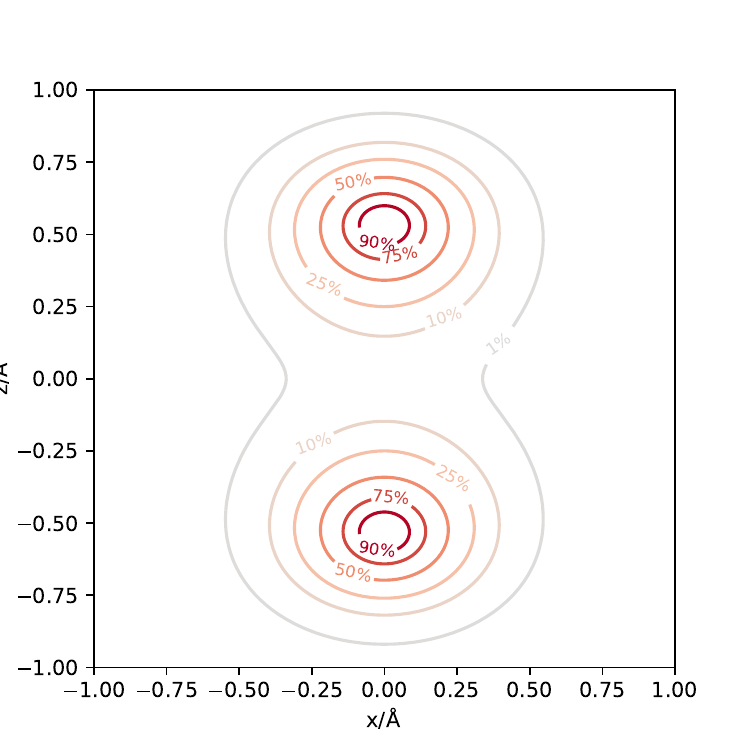}}\,
            \subfloat[RPA@PBE]{\includegraphics[scale=0.42]{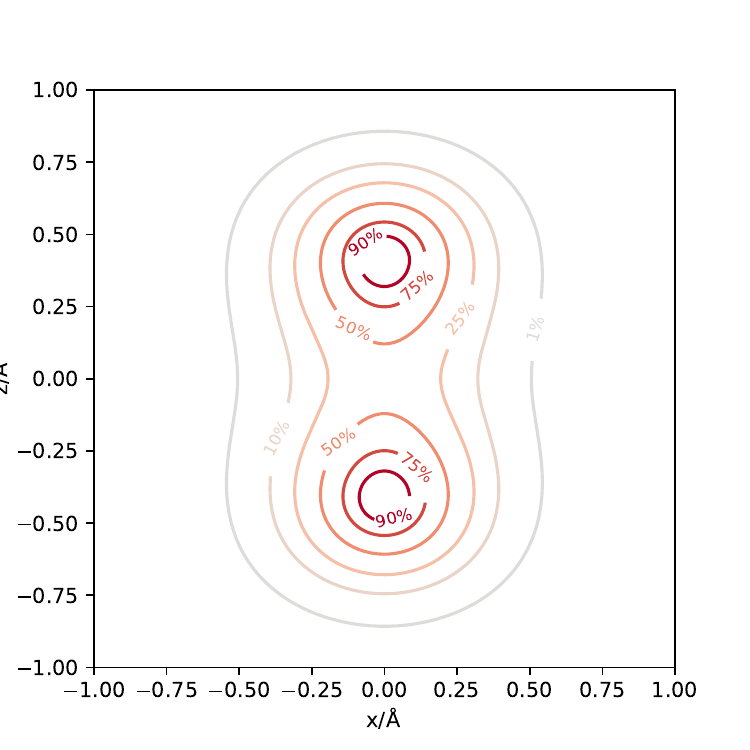}}\,
            \subfloat[B86bPBE-25X-XDM]{\includegraphics[scale=0.42]{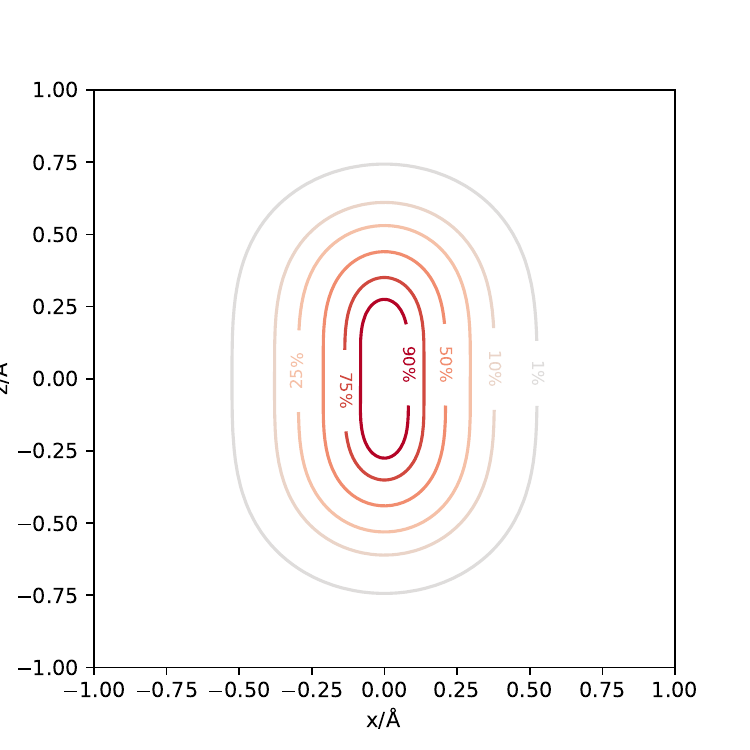}}
            \caption{Two dimensional slices of the ground state wavefunction in the $xz$ plane for (a) MP2, (b) RPA@PBE, and (c) B86bPBE-25X-XDM ES methods. Contours are taken at 1\%, 10\%, 25\%, 50\%, 75\%, 90\% and 99\% of the maximum amplitude of the wavefunction in this plane.}
            \label{fig: wavefunctions}
        \end{figure*}

        A basis set consisting of 14 functions in each of the $x$ and $y$ directions alongside 28 in the $z$ direction was used to diagonalise the Hamiltonian, ensuring convergence of the lowest 50 eigenstates to within 0.002\wavenumber. Resetting the energy zero of each ES method to be the minimum value of the PES, the 50 lowest energies of the translational eigenstates up to 250\wavenumber\ for each ES method are shown in Fig \ref{fig: Energy Levels}, alongside the ones for the LJ PES. This linear shift of energy zero makes the energy scale absolute, and allows for direct comparison of energies between all ES methods. The colours and linestyles correspond to the quantum numbers for each state. Although a Cartesian basis set was used suggesting the use of $(n_x,n_y,n_z)$ quantum numbers, the cylindrical symmetry of the system can be leveraged by using the quantum numbers of an isotropic two-dimensional harmonic oscillator $(n,l)$, referring to the principal and angular momentum quantum numbers respectively, in place of $(n_x,n_y)$.\cite{panchagnulaExploringParameterSpace2023b, mandziukQuantumThreeDimensional1994} Discussion referring to quantum numbers uses the $(n,l,n_z)$ notation. These quantum numbers have the restriction that $|l|\leq n$, and $(n,l)$ have the same parity. These can be assigned by analysing the nodal structure of the eigenfunctions. The energy levels are coloured such that $n\in[0,1,2,3]$ are shown by blue, orange, green, red lines and error bars; $|l|\in[0,1,2,3]$ are shown by solid, dotted, dashed, dash-dot lines. For the same line colour and style traversing the energy scale corresponds to an increase in the $n_z$ quantum number. The states with $l=0$ are singly degenerate, with the rest doubly degenerate due to the symmetry of $\pm l$. 
        
        The error bars in the ES method eigenstates arise from the covariance in the PES, are calculated alongside the mean PES prediction from the GP given by Eqns \eqref{eq: GP covariance} and \eqref{eq: GP mean} respectively. The translational energies are calculated from the GP mean prediction. The covariance matrix is used to generate 200 samples of the Hamiltonian matrix, by sampling the potential from the posterior distribution (goverened by the covariance matrix) of the GP. The error bars are then calculated by diagonalising each sample of the Hamiltonian matrix and calculating the standard deviation in each eigenvalue. More precise details of this procedure can be found in the suppplementary information. All the error bars on the lowest 50 eigenstates are under $\pm$2\wavenumber, indicating a tightly constrained confidence interval on the GP mean of each PES. 

        Comparing the eigenspectra between ES methods, the grouping as suggested by barrier height is pertinent. This is due to considering the values of the fundamental frequencies, the energy gap between the (0,0,0) and (0,0,1) eigenstates, with the MP2 having the smallest, of 0.04\wavenumber. The SCS-MP2, RPA@PBE, C(HF)-RPA and B86bPBE-50X-XDM predict this value to be in the range [1,3]\wavenumber, whereas the SOS-MP2 and B86bPBE-25X-XDM put this value in the interval [6,9]\wavenumber. These are smaller than the LJ equivalent, which is approximately 10\wavenumber. The large apparent fundamental frequency (energy gap between the ground and first excited state) in the MP2 is the $(0,0,0)\rightarrow(0,0,2)$ transition. The true fundamental frequency is very small as the zero-point energy (ZPE) lies below the barrier height and this is the only method where this is the case. The very small $n_z:0\rightarrow1$ transition appears again in the first doubly degenerate state at around 125\wavenumber, with two doubly degenerate sets present separated by under 0.1\wavenumber. Going upwards in energy, the $n_z\rightarrow n_z+1$ gaps are increasing, indicating negative anharmonicity. This is to be expected given the shape of the one-dimensional slices in Fig \ref{fig: contours} as the bounding potential grows as a monotonic polynomial oscillator and there is no ``dissociation'' type event. This trend is only not seen for the MP2 $n_z:2\rightarrow3$ transitions, as the ground state lies below the barrier height and the previous transition is the \ce{Ne} breaking free of the double well.\cite{panchagnulaExploringParameterSpace2023b}

        The ordering of states by quantum number is not preserved between ES methods, nor with the LJ. Due to the prominent double well, the first difference can be seen in the MP2 where the $(0,0,5)$ state lies below the $(1,\pm1,2)$ states unlike the other ES methods. This can be justified by its ZPE lying below the barrier height, leading to a very small frequency of transition in the $n_z$ coordinate. The large gap to the $(0,0,2)$ state is indicative of the \ce{Ne} breaking free of the confining double well into polynomial oscillator, resetting the transition frequency to its smaller value before the negative anharmonicity takes effect and the gaps increase. The most apparent mixing of states occurs as the $n=3$ states start appearing\cite{panchagnulaExploringParameterSpace2023b}. This is the region where the interplay between $\Delta n$, $\Delta l$ and $\Delta n_z$ frequencies become more cluttered leading to a jumbling in the order of states. 

        The quality of effective LJ parameters for each PES described in Section \ref{sec: ResultsPES} can be quantified by calculating the eigenspectrum and comparing to the true translational energy levels in Fig \ref{fig: Energy Levels}. An analogous figure of the translational eigenspectra using the effective LJ PESes is given in the supplementary information. As these effective LJ parameters were found by matching features of the double well, the transition frequencies corresponding to increases in the $n_z$ quantum number show excellent agreement with the ES transition frequencies. However, this is the end of the similarities, as the numerical values of energy levels and the frequencies of the radially symmetric $xy$ modes also differ. For the MP2, these differences for the lowest 10 eigenstates are approximately 3\wavenumber. On the other end of the scale, the RPA@PBE has a difference of roughly 8\wavenumber. The general trend is that the effective LJ PESs better match the MP2 and B86bPBE type methods compared to the RPA and its derivatives. The difference between the ES and their effective LJ type PESes suggests that only using $R^{-6}$ and $R^{-12}$ terms to describe the attractive and repulsive effects of the endohedral interaction are too simplistic, and more detailed formulae would be required.

        From an experimental perspective, these LJ type PESs are popular due to their conceptual simplicity and ease of use. From the set of noble gas --- \ce{C} set of LJ parameters, the \ce{Ne} had the best match to spectroscopic data. While these LJ parameters may be suited to matching spectroscopic observations, they do not necessarily represent the true PES. \cite{jafariNeArKr2023a, xuCoupledTranslationrotationEigenstates2009} Based on these theoretical calculations, it would seem prudent to dissuade the extensive use of any LJ parameterised PESs to describe these systems. Another reason to avoid these LJ parameters is that while the intuitive choice was to match the double well characteristics, other choices are also possible. These include (but not necessarily limited to): matching the energies of two specific eigenstates, matching two frequencies or the prolateness and kurtosis of the ground state wavefunction.

        As well as comparing the energies of each ES method, the wavefunctions, or nuclear orbitals generated from the diagonalisation can also be examined. Considering the partitioning of the ES methods by barrier height: \{MP2\}, \{SCS-MP2, RPA@PBE, C(HF)-RPA, B86bPBE-50X-XDM\}, and \{SOS-MP2, B86bPBE-25X-XDM\}, the ground state wavefunction of one method from each set is shown in Fig \ref{fig: wavefunctions}. There is a smooth transition between the types of wavefunction between all three ES methods. The MP2 has the furthest apart minima and largest barrier height, leading to the wavefunction with the most concentrated density around the minima. The RPA@PBE has the minima closer together, and there is a moving of the wavefunction density away from these towards the centre of the cage. The B86bPBE-25X-XDM has the closest together minima, and the two maxima in the wavefunction seen for the previous methods have coalesced into a single peak at the origin implying the \ce{Ne} atom is completely delocalised over both minima.

       \begin{table}
           \centering
           \begin{tabular}{ccc}
                \hline Method&$\varsigma$&$\kappa_z$  \\
                \hline MP2&5.335&1.162\\
                SCS-MP2&3.795&1.534\\
                SOS-MP2&2.824&2.051\\
                RPA@PBE&3.997&1.452\\
                C(HF)-RPA&3.640&1.598\\
                B86bPBE-25X-XDM&2.518&2.236\\
                B86bPBE-50X-XDM&3.311&1.722\\
                LJ&2.302&2.371\\
                \hline
           \end{tabular}
           \caption{Prolateness and kurtosis in the anisotropic $z$ direction as defined in Eqns \eqref{eq: prolateness} and \eqref{eq: kurtosis} for the ground state wavefunction for all ES methods and LJ.}
           \label{table: Ground State Stats}
       \end{table}

       The shape of the ground state function can also be quantified by two main statistics, its prolateness and kurtosis as defined in Eqns \eqref{eq: prolateness} and \eqref{eq: kurtosis} respectively. These describe the ellipsoidal and double-peakedness of the wavefunction respectively. As seen in Table \ref{table: Ground State Stats} the grouping of ES methods by barrier height partitions them into the same sets as if done by either statistic. Considering the prolateness, which is a measure of how stretched the wavefunction is along the $z$ direction as compared to the $x$ direction, the MP2 is the most stretched, as its minima are the furthest apart. The SCS-MP2, RPA@PBE, C(HF)-RPA and B86bPBE-50X-XDM have $\varsigma$ lie in the interval [3.31,4.00], indicating a moderate stretch and the remaining SOS-MP2 and B86bPBE-25X-XDM have this value lie in the interval [2.51,2.83]. The kurtosis range can also be partitioned, with the most double peaked, the MP2 having a kurtosis of 1.16, the four intermediate ES methods having this value between 1.45 and 1.60. The two methods with the lowest barrier heights, SOS-MP2 and B86bPBE-25X-XDM have kurtoses of over 2 which, noticeably, are still lower than the LJ value. However, these are all below 3 which is the value for a true Gaussian, further confirming the anharmonicity in the potential. These values indicate that the wavefunctions are more delocalised than a pure Gaussian, with a shallower central peak and broader shoulders of the distribution.

       \begin{table*}
           \centering
            \begin{tabular}{c|cccccccc}
                 $\bra{\downarrow}$, $\ket{\rightarrow}$& MP2 & SCS-MP2 & SOS-MP2 & RPA@PBE & C(HF)-RPA&B86bPBE-25X-XDM&B86bPBE-50X-XDM&LJ \\
                 \hline MP2&-&0.487&0.674&0.398&0.484&0.735&0.574&0.767\\
                 SCS-MP2&0.487&-&0.257&0.078&0.033&0.337&0.124&0.382\\
                 SOS-MP2&0.674&0.257&-&0.327&0.231&0.083&0.137&0.131\\
                 RPA@PBE&0.398&0.078&0.327&-&0.099&0.406&0.199&0.449\\
                 C(HF)-RPA&0.484&0.033&0.231&0.099&-&0.311&0.101&0.355\\
                 B86bPBE-25X-XDM&0.735&0.337&0.083&0.406&0.311&-&0.219&0.050\\
                 B86bPBE-50X-XDM&0.574&0.124&0.137&0.199&0.101&0.219&-&0.266\\
                 LJ&0.767&0.382&0.131&0.449&0.355&0.050&0.266&-
            \end{tabular}
           \caption{Hellinger distance of the ground state wavefunction defined in Eq \eqref{eq: Hellinger} between all ES methods, and LJ. Diagonal elements are necessarily zero, and this is symmetric with respect to the bra and ket.}
           \label{table: Ground State Hellinger}
       \end{table*}

       The Hellinger distances between the ground state wavefunctions, as defined by Eq \eqref{eq: Hellinger} are shown in Table \ref{table: Ground State Hellinger}. The groupings of ES methods previously used for the barrier height, energy levels and ground state statistics can also be applied here. As with the other properties, the MP2 is substantially different to the other ES methods, with its closest ground state wavefunction being the RPA@PBE but even that is still 0.398 apart, with it being furthest away from the B86bPBE-25X-XDM at 0.735. Of the middling barrier height group the SCS-MP2, RPA@PBE and C(HF)-RPA are all under 0.1 away from each other, about 0.3 away from the smallest barrier height set of methods but these lie far away from the MP2.  All these distances, barring the SCS-MP2 to C(HF)-RPA distance are much larger than what was previously seen for \Endo{He}{C60}.\cite{panchagnulaTranslationalEigenstatesHe2024a} This may be attributed to the presence of the double well and use of the symmetrised double minimum basis, as the differing positions of minima will more heavily influence the $\braket{m_z|n_z}$ integral.

       Considering the variations between the plethora of ES methods in all the calculated quantities it is not immediately obvious which is the most promising technique. The discrepancies can arise from a few sources of error including basis set superposition error (BSSE, commonly treated by counterpoise correction), CBS extrapolation error, or error with the method itself. When considering encapulsation energies, the BSSE can be up to 44\%\cite{cioslowskiElectronicStructureCalculations2023a} but that figure is misleading in the context of calculating the translational eigenstates as if this correction is flat (or close to) throughout the PES, it can be accounted for when setting the energy zero. Extra details of this are provided in the supplementary information. The CBS extrapolation error is more important here for the MP2 and RPA type methods, as while the SCF extrapolation was identical to what was done for \Endo{He}{C60},\cite{panchagnulaTranslationalEigenstatesHe2024a} the correlation extrapolation was only possible from cc-pVTZ and cc-pVQZ calculations whereas cc-pV5Z was attainable for \Endo{He}{C60}. While important, this is likely to be less significant that the error of the ES method itself and the approximations made within its framework which have been earlier outlined\cite{panchagnulaTranslationalEigenstatesHe2024a} and in Sec \ref{sec: TheoryES}. For the B86bPBE methods, these have a parameter built into them which is the amount of Hartree--Fock exchange. The choices of 25\% and 50\%, while common, may not be optimal and this parameter may vary depending on a variety of features including the geometry of the system.\cite{bryentonDelocalizationErrorGreatest2023, gouldPoisoningDensityFunctional2022, khanAdaptiveHybridDensity2024} All this put together suggests alternative ES methods need to also be applied in order to achieve spectroscopic agreement.

    \section{Conclusion \label{sec: Conc}}
    	In this paper, we have investigated the endofullerene \Endo{Ne}{C70} using a multitude of electronic structure (ES) techniques including: MP2, SCS-MP2, SOS-MP2, RPA@PBE and C(HF)-RPA which were all used for \Endo{He}{C60},\cite{panchagnulaTranslationalEigenstatesHe2024a} with B86bPBE-25X-XDM and B86bPBE-50X-XDM as state-of-the-art empirically corrected DFT added to the set. The translational eigenstates were calculated in order to test the spectroscopic accuracy of the methods. The elongation of \ce{C60} along a single axis to form \ce{C70} unveiled a double well potential along this unique anisotropic direction for all methods. Trying to achieve an accurate description of this double well by correctly placing the minima and gauging the barrier height forms a good test of the sensitivity and effectiveness of the ES method at describing the dispersion interaction, making \Endo{Ne}{C70} a good system to probe other ES methods.

    Due to the high cost of the ES calculations, the PES is interpolated using Gaussian Process (GP) Regression because of its effectiveness with sparse training data. The PESes are symmetrised within the $D_{5h}$ symmetry of the system, before further properties were calculated. We find the MP2 has the most prominent double well, with a barrier height of 84.21\wavenumber and minima at $\pm$0.76~\AA. The remaining six ES methods, while not in complete agreement, can be collated together with barrier heights ranging between 10\wavenumber\ and 35\wavenumber, and minima between $\pm$0.506~\AA\, and $\pm$0.648~\AA. These are all vastly different from the LJ PES \cite{mandziukQuantumThreeDimensional1994} shown in Table \ref{table: PES features}.

    The translational eigenstates for each ES method were calculated by diagonalising the nuclear Hamiltonian using the symmetrised double minimum basis set due to the double well potential. The eigenspectrum for the methods is shown in Fig \ref{fig: Energy Levels}. Combined with the double well features, these together motivate the grouping of the ES methods into three groups: \{MP2\}, \{SCS-MP2, RPA@PBE, C(HF)-RPA, B86bPBE-50X-XDM\}, and \{SOS-MP2, B86bPBE-25X-XDM\}. This is further compounded by considering the shape of their ground state wavefunctions, shown by the statistics in Table \ref{table: Ground State Stats}, and the distances between them given in Table \ref{table: Ground State Hellinger}. Effective LJ parameters were found by matching the double well features: barrier height and minima positions. The eigenspectra generated by diagonalising these effective LJ PESs differed to their ES counterparts by varying amounts. For MP2, these were sub-wavenumber apart, whereas for RPA@PBE, they differed by over 7\wavenumber. This difference suggests that while a LJ type surface may be able to reproduce spectroscopically observed data, it is not necessarily an appropriate description for the true PES. This is further compounded by the fact that there is no reason to pick the LJ parameters to match the double well features, as other properties of the PES or eigenspectra can be picked as equally valid choices. As a consequence, we would warn against the ubiquitous use of the LJ type PESes, as despite their simplicity and ability to match experimental data, they do not necessarily accurately describe the endohedral species and cage interaction.

    While multiple ES methods were used, it is not an exhaustive set. The lack of concordance between the methods elucidates the necessity of more accurate calculations, especially with a different cage than \ce{C60}.\cite{cioslowskiElectronicStructureCalculations2023a}. 
    For \Endo{Ne}{C70}, MP2 is an outlier; however, it should not be discounted as it was the best performing for \Endo{He}{C60}, along with RPA@PBE.\cite{panchagnulaTranslationalEigenstatesHe2024a} The drastic change between the performance of the ES methods between these two systems motivates further investigation into whether the larger endohedral atom (\ce{Ne}) or the larger cage (\ce{C70}) is causing the bulk of the difficulties. More accurate calculations on this \Endo{Ne}{C70} system could be achieved using differing ES methods and techniques (e.g. coupled cluster), or an improvement on the methodology used. For example, these systems can be sensitive to the CBS extrapolation scheme used with alternative formulations possible\cite{johnsonTheoreticalStudyDispersionbound2004}. Alternatively, the CBS extrapolation of the correlation energy can be made more accurate with energies from cc-pV5Z calculations which were computationally tractable for \Endo{He}{C60} but not as much for \Endo{Ne}{C70}. Regarding the B86bPBE calculations, rather than using heuristics to determine the amount of Hartree--Fock exhange to include, this value could be optimised.\cite{bryentonDelocalizationErrorGreatest2023, gouldPoisoningDensityFunctional2022, khanAdaptiveHybridDensity2024} However, in order to eventually be confident in the choice of ES method, this requires experimental and spectroscopic data for the system, analogous to what is available for \Endo{He}{C60}.\cite{bacanuExperimentalDeterminationInteraction2021c, jafariTerahertzSpectroscopyHelium2022} 

    \section*{Supplementary Information}
        See the supplementary information for more details on generating the Potential Energy Surfaces, including the symmetrisation procedure. Extra figures and results for the other electronic structure methods are also presented.
    
    \begin{acknowledgements}
        D.~G.~acknowledges funding by the Deutsche Forschungsgemeinschaft (DFG, German Research Foundation) -- 498448112. D.~G.~thanks J.~Kussmann (LMU Munich) for providing a development version of the FermiONs++ software package. K.~P.~ would like to thank Dr.~Bang C.~Huynh for their comments and providing a development version of \textsc{QSym\textsuperscript{2}}. E.~R.~J.\ is grateful for financial support from the Natural Sciences and Engineering Research Council (NSERC) of Canada and from the Royal Society through a Wolfson Visiting Fellowship. E.~R.~J.\ also thanks the Atlantic Computational Excellence Network (ACENET) for computational resources.
    \end{acknowledgements}

    \section*{Data Availability Statement}
    The data that support the findings of this study are openly available in Apollo - University of Cambridge Repository at https://doi.org/10.17863/CAM.109314\cite{KripaPanchagnulaDataset2024}, reference number \citenum{KripaPanchagnulaDataset2024}.

    \bibliography{NeC70}

\begin{thebibliography}{96}%
\makeatletter
\providecommand \@ifxundefined [1]{%
 \@ifx{#1\undefined}
}%
\providecommand \@ifnum [1]{%
 \ifnum #1\expandafter \@firstoftwo
 \else \expandafter \@secondoftwo
 \fi
}%
\providecommand \@ifx [1]{%
 \ifx #1\expandafter \@firstoftwo
 \else \expandafter \@secondoftwo
 \fi
}%
\providecommand \natexlab [1]{#1}%
\providecommand \enquote  [1]{``#1''}%
\providecommand \bibnamefont  [1]{#1}%
\providecommand \bibfnamefont [1]{#1}%
\providecommand \citenamefont [1]{#1}%
\providecommand \href@noop [0]{\@secondoftwo}%
\providecommand \href [0]{\begingroup \@sanitize@url \@href}%
\providecommand \@href[1]{\@@startlink{#1}\@@href}%
\providecommand \@@href[1]{\endgroup#1\@@endlink}%
\providecommand \@sanitize@url [0]{\catcode `\\12\catcode `\$12\catcode
  `\&12\catcode `\#12\catcode `\^12\catcode `\_12\catcode `\%12\relax}%
\providecommand \@@startlink[1]{}%
\providecommand \@@endlink[0]{}%
\providecommand \url  [0]{\begingroup\@sanitize@url \@url }%
\providecommand \@url [1]{\endgroup\@href {#1}{\urlprefix }}%
\providecommand \urlprefix  [0]{URL }%
\providecommand \Eprint [0]{\href }%
\providecommand \doibase [0]{https://doi.org/}%
\providecommand \selectlanguage [0]{\@gobble}%
\providecommand \bibinfo  [0]{\@secondoftwo}%
\providecommand \bibfield  [0]{\@secondoftwo}%
\providecommand \translation [1]{[#1]}%
\providecommand \BibitemOpen [0]{}%
\providecommand \bibitemStop [0]{}%
\providecommand \bibitemNoStop [0]{.\EOS\space}%
\providecommand \EOS [0]{\spacefactor3000\relax}%
\providecommand \BibitemShut  [1]{\csname bibitem#1\endcsname}%
\let\auto@bib@innerbib\@empty
\bibitem [{\citenamefont {Ba{\v
  c}i{\'c}}(2018)}]{bacicPerspectiveAccurateTreatment2018}%
  \BibitemOpen
  \bibfield  {author} {\bibinfo {author} {\bibfnamefont {Z.}~\bibnamefont
  {Ba{\v c}i{\'c}}},\ }\bibfield  {title} {\enquote {\bibinfo {title}
  {Perspective: {{Accurate}} treatment of the quantum dynamics of light
  molecules inside fullerene cages: {{Translation-rotation}} states,
  spectroscopy, and symmetry breaking},}\ }\href
  {https://doi.org/10.1063/1.5049358} {\bibfield  {journal} {\bibinfo
  {journal} {The Journal of Chemical Physics}\ }\textbf {\bibinfo {volume}
  {149}},\ \bibinfo {pages} {100901} (\bibinfo {year} {2018})}\BibitemShut
  {NoStop}%
\bibitem [{\citenamefont {Murata}\ \emph {et~al.}(2008)\citenamefont {Murata},
  \citenamefont {Maeda}, \citenamefont {Morinaka}, \citenamefont {Murata},\
  and\ \citenamefont {Komatsu}}]{murataSynthesisReactionFullerene2008}%
  \BibitemOpen
  \bibfield  {author} {\bibinfo {author} {\bibfnamefont {M.}~\bibnamefont
  {Murata}}, \bibinfo {author} {\bibfnamefont {S.}~\bibnamefont {Maeda}},
  \bibinfo {author} {\bibfnamefont {Y.}~\bibnamefont {Morinaka}}, \bibinfo
  {author} {\bibfnamefont {Y.}~\bibnamefont {Murata}},\ and\ \bibinfo {author}
  {\bibfnamefont {K.}~\bibnamefont {Komatsu}},\ }\bibfield  {title} {\enquote
  {\bibinfo {title} {Synthesis and {{Reaction}} of {{Fullerene C70
  Encapsulating Two Molecules}} of {{H2}}},}\ }\href
  {https://doi.org/10.1021/ja8076846} {\bibfield  {journal} {\bibinfo
  {journal} {Journal of the American Chemical Society}\ }\textbf {\bibinfo
  {volume} {130}},\ \bibinfo {pages} {15800--15801} (\bibinfo {year}
  {2008})}\BibitemShut {NoStop}%
\bibitem [{\citenamefont {Bloodworth}\ and\ \citenamefont
  {Whitby}(2022)}]{bloodworthSynthesisEndohedralFullerenes2022}%
  \BibitemOpen
  \bibfield  {author} {\bibinfo {author} {\bibfnamefont {S.}~\bibnamefont
  {Bloodworth}}\ and\ \bibinfo {author} {\bibfnamefont {R.~J.}\ \bibnamefont
  {Whitby}},\ }\bibfield  {title} {\enquote {\bibinfo {title} {Synthesis of
  endohedral fullerenes by molecular surgery},}\ }\href
  {https://doi.org/10.1038/s42004-022-00738-9} {\bibfield  {journal} {\bibinfo
  {journal} {Communications Chemistry}\ }\textbf {\bibinfo {volume} {5}},\
  \bibinfo {pages} {1--14} (\bibinfo {year} {2022})}\BibitemShut {NoStop}%
\bibitem [{\citenamefont {Wang}, \citenamefont {Straka},\ and\ \citenamefont
  {Pyykk{\"o}}(2010)}]{wangFormulationsClosedshellInteractions2010}%
  \BibitemOpen
  \bibfield  {author} {\bibinfo {author} {\bibfnamefont {C.}~\bibnamefont
  {Wang}}, \bibinfo {author} {\bibfnamefont {M.}~\bibnamefont {Straka}},\ and\
  \bibinfo {author} {\bibfnamefont {P.}~\bibnamefont {Pyykk{\"o}}},\ }\bibfield
   {title} {\enquote {\bibinfo {title} {Formulations of the closed-shell
  interactions in endohedral systems},}\ }\href
  {https://doi.org/10.1039/B922808J} {\bibfield  {journal} {\bibinfo  {journal}
  {Physical Chemistry Chemical Physics}\ }\textbf {\bibinfo {volume} {12}},\
  \bibinfo {pages} {6187--6203} (\bibinfo {year} {2010})}\BibitemShut {NoStop}%
\bibitem [{\citenamefont {Pyykk{\"o}}\ \emph {et~al.}(2007)\citenamefont
  {Pyykk{\"o}}, \citenamefont {Wang}, \citenamefont {Straka},\ and\
  \citenamefont {Vaara}}]{pyykkoLondontypeFormulaDispersion2007}%
  \BibitemOpen
  \bibfield  {author} {\bibinfo {author} {\bibfnamefont {P.}~\bibnamefont
  {Pyykk{\"o}}}, \bibinfo {author} {\bibfnamefont {C.}~\bibnamefont {Wang}},
  \bibinfo {author} {\bibfnamefont {M.}~\bibnamefont {Straka}},\ and\ \bibinfo
  {author} {\bibfnamefont {J.}~\bibnamefont {Vaara}},\ }\bibfield  {title}
  {\enquote {\bibinfo {title} {A {{London-type}} formula for the dispersion
  interactions of endohedral {{A}}@{{B}} systems},}\ }\href
  {https://doi.org/10.1039/B704695B} {\bibfield  {journal} {\bibinfo  {journal}
  {Physical Chemistry Chemical Physics}\ }\textbf {\bibinfo {volume} {9}},\
  \bibinfo {pages} {2954--2958} (\bibinfo {year} {2007})}\BibitemShut {NoStop}%
\bibitem [{\citenamefont {Bacanu}\ \emph {et~al.}(2021)\citenamefont {Bacanu},
  \citenamefont {Jafari}, \citenamefont {Aouane}, \citenamefont {Rantaharju},
  \citenamefont {Walkey}, \citenamefont {Hoffman}, \citenamefont {Shugai},
  \citenamefont {Nagel}, \citenamefont {{Jim{\'e}nez-Ruiz}}, \citenamefont
  {Horsewill}, \citenamefont {Rols}, \citenamefont {R{\~o}{\~o}m},
  \citenamefont {Whitby},\ and\ \citenamefont
  {Levitt}}]{bacanuExperimentalDeterminationInteraction2021c}%
  \BibitemOpen
  \bibfield  {author} {\bibinfo {author} {\bibfnamefont {G.~R.}\ \bibnamefont
  {Bacanu}}, \bibinfo {author} {\bibfnamefont {T.}~\bibnamefont {Jafari}},
  \bibinfo {author} {\bibfnamefont {M.}~\bibnamefont {Aouane}}, \bibinfo
  {author} {\bibfnamefont {J.}~\bibnamefont {Rantaharju}}, \bibinfo {author}
  {\bibfnamefont {M.}~\bibnamefont {Walkey}}, \bibinfo {author} {\bibfnamefont
  {G.}~\bibnamefont {Hoffman}}, \bibinfo {author} {\bibfnamefont
  {A.}~\bibnamefont {Shugai}}, \bibinfo {author} {\bibfnamefont
  {U.}~\bibnamefont {Nagel}}, \bibinfo {author} {\bibfnamefont
  {M.}~\bibnamefont {{Jim{\'e}nez-Ruiz}}}, \bibinfo {author} {\bibfnamefont
  {A.~J.}\ \bibnamefont {Horsewill}}, \bibinfo {author} {\bibfnamefont
  {S.}~\bibnamefont {Rols}}, \bibinfo {author} {\bibfnamefont {T.}~\bibnamefont
  {R{\~o}{\~o}m}}, \bibinfo {author} {\bibfnamefont {R.~J.}\ \bibnamefont
  {Whitby}},\ and\ \bibinfo {author} {\bibfnamefont {M.~H.}\ \bibnamefont
  {Levitt}},\ }\bibfield  {title} {\enquote {\bibinfo {title} {Experimental
  determination of the interaction potential between a helium atom and the
  interior surface of a {{C60}} fullerene molecule},}\ }\href
  {https://doi.org/10.1063/5.0066817} {\bibfield  {journal} {\bibinfo
  {journal} {The Journal of Chemical Physics}\ }\textbf {\bibinfo {volume}
  {155}},\ \bibinfo {pages} {144302} (\bibinfo {year} {2021})}\BibitemShut
  {NoStop}%
\bibitem [{\citenamefont {Jafari}\ \emph {et~al.}(2022)\citenamefont {Jafari},
  \citenamefont {Bacanu}, \citenamefont {Shugai}, \citenamefont {Nagel},
  \citenamefont {Walkey}, \citenamefont {Hoffman}, \citenamefont {Levitt},
  \citenamefont {Whitby},\ and\ \citenamefont
  {R{\~o}{\~o}m}}]{jafariTerahertzSpectroscopyHelium2022}%
  \BibitemOpen
  \bibfield  {author} {\bibinfo {author} {\bibfnamefont {T.}~\bibnamefont
  {Jafari}}, \bibinfo {author} {\bibfnamefont {G.~R.}\ \bibnamefont {Bacanu}},
  \bibinfo {author} {\bibfnamefont {A.}~\bibnamefont {Shugai}}, \bibinfo
  {author} {\bibfnamefont {U.}~\bibnamefont {Nagel}}, \bibinfo {author}
  {\bibfnamefont {M.}~\bibnamefont {Walkey}}, \bibinfo {author} {\bibfnamefont
  {G.}~\bibnamefont {Hoffman}}, \bibinfo {author} {\bibfnamefont {M.~H.}\
  \bibnamefont {Levitt}}, \bibinfo {author} {\bibfnamefont {R.~J.}\
  \bibnamefont {Whitby}},\ and\ \bibinfo {author} {\bibfnamefont
  {T.}~\bibnamefont {R{\~o}{\~o}m}},\ }\bibfield  {title} {\enquote {\bibinfo
  {title} {Terahertz spectroscopy of the helium endofullerene
  {{He}}@{{C60}}},}\ }\href {https://doi.org/10.1039/D2CP00515H} {\bibfield
  {journal} {\bibinfo  {journal} {Physical Chemistry Chemical Physics}\
  }\textbf {\bibinfo {volume} {24}},\ \bibinfo {pages} {9943--9952} (\bibinfo
  {year} {2022})}\BibitemShut {NoStop}%
\bibitem [{\citenamefont {Xu}\ \emph {et~al.}(2009)\citenamefont {Xu},
  \citenamefont {Sebastianelli}, \citenamefont {Gibbons}, \citenamefont {Ba{\v
  c}i{\'c}}, \citenamefont {Lawler},\ and\ \citenamefont
  {Turro}}]{xuCoupledTranslationrotationEigenstates2009}%
  \BibitemOpen
  \bibfield  {author} {\bibinfo {author} {\bibfnamefont {M.}~\bibnamefont
  {Xu}}, \bibinfo {author} {\bibfnamefont {F.}~\bibnamefont {Sebastianelli}},
  \bibinfo {author} {\bibfnamefont {B.~R.}\ \bibnamefont {Gibbons}}, \bibinfo
  {author} {\bibfnamefont {Z.}~\bibnamefont {Ba{\v c}i{\'c}}}, \bibinfo
  {author} {\bibfnamefont {R.}~\bibnamefont {Lawler}},\ and\ \bibinfo {author}
  {\bibfnamefont {N.~J.}\ \bibnamefont {Turro}},\ }\bibfield  {title} {\enquote
  {\bibinfo {title} {Coupled translation-rotation eigenstates of {{H2}} in
  {{C60}} and {{C70}} on the spectroscopically optimized interaction potential:
  {{Effects}} of cage anisotropy on the energy level structure and
  assignments},}\ }\href {https://doi.org/10.1063/1.3152574} {\bibfield
  {journal} {\bibinfo  {journal} {The Journal of Chemical Physics}\ }\textbf
  {\bibinfo {volume} {130}},\ \bibinfo {pages} {224306} (\bibinfo {year}
  {2009})}\BibitemShut {NoStop}%
\bibitem [{\citenamefont {Xu}\ \emph {et~al.}(2008{\natexlab{a}})\citenamefont
  {Xu}, \citenamefont {Sebastianelli}, \citenamefont {Ba{\v c}i{\'c}},
  \citenamefont {Lawler},\ and\ \citenamefont {Turro}}]{xuH2HDD22008}%
  \BibitemOpen
  \bibfield  {author} {\bibinfo {author} {\bibfnamefont {M.}~\bibnamefont
  {Xu}}, \bibinfo {author} {\bibfnamefont {F.}~\bibnamefont {Sebastianelli}},
  \bibinfo {author} {\bibfnamefont {Z.}~\bibnamefont {Ba{\v c}i{\'c}}},
  \bibinfo {author} {\bibfnamefont {R.}~\bibnamefont {Lawler}},\ and\ \bibinfo
  {author} {\bibfnamefont {N.~J.}\ \bibnamefont {Turro}},\ }\bibfield  {title}
  {\enquote {\bibinfo {title} {H2, {{HD}}, and {{D2}} inside {{C60}}:
  {{Coupled}} translation-rotation eigenstates of the endohedral molecules from
  quantum five-dimensional calculations},}\ }\href
  {https://doi.org/10.1063/1.2967858} {\bibfield  {journal} {\bibinfo
  {journal} {The Journal of Chemical Physics}\ }\textbf {\bibinfo {volume}
  {129}},\ \bibinfo {pages} {064313} (\bibinfo {year}
  {2008}{\natexlab{a}})}\BibitemShut {NoStop}%
\bibitem [{\citenamefont {Xu}\ \emph {et~al.}(2008{\natexlab{b}})\citenamefont
  {Xu}, \citenamefont {Sebastianelli}, \citenamefont {Ba{\v c}i{\'c}},
  \citenamefont {Lawler},\ and\ \citenamefont
  {Turro}}]{xuQuantumDynamicsCoupled2008}%
  \BibitemOpen
  \bibfield  {author} {\bibinfo {author} {\bibfnamefont {M.}~\bibnamefont
  {Xu}}, \bibinfo {author} {\bibfnamefont {F.}~\bibnamefont {Sebastianelli}},
  \bibinfo {author} {\bibfnamefont {Z.}~\bibnamefont {Ba{\v c}i{\'c}}},
  \bibinfo {author} {\bibfnamefont {R.}~\bibnamefont {Lawler}},\ and\ \bibinfo
  {author} {\bibfnamefont {N.~J.}\ \bibnamefont {Turro}},\ }\bibfield  {title}
  {\enquote {\bibinfo {title} {Quantum dynamics of coupled translational and
  rotational motions of {{H2}} inside {{C60}}},}\ }\href
  {https://doi.org/10.1063/1.2828556} {\bibfield  {journal} {\bibinfo
  {journal} {The Journal of Chemical Physics}\ }\textbf {\bibinfo {volume}
  {128}},\ \bibinfo {pages} {011101} (\bibinfo {year}
  {2008}{\natexlab{b}})}\BibitemShut {NoStop}%
\bibitem [{\citenamefont {Xu}\ \emph {et~al.}(2013)\citenamefont {Xu},
  \citenamefont {Ye}, \citenamefont {Powers}, \citenamefont {Lawler},
  \citenamefont {Turro},\ and\ \citenamefont {Ba{\v
  c}i{\'c}}}]{xuInelasticNeutronScattering2013}%
  \BibitemOpen
  \bibfield  {author} {\bibinfo {author} {\bibfnamefont {M.}~\bibnamefont
  {Xu}}, \bibinfo {author} {\bibfnamefont {S.}~\bibnamefont {Ye}}, \bibinfo
  {author} {\bibfnamefont {A.}~\bibnamefont {Powers}}, \bibinfo {author}
  {\bibfnamefont {R.}~\bibnamefont {Lawler}}, \bibinfo {author} {\bibfnamefont
  {N.~J.}\ \bibnamefont {Turro}},\ and\ \bibinfo {author} {\bibfnamefont
  {Z.}~\bibnamefont {Ba{\v c}i{\'c}}},\ }\bibfield  {title} {\enquote {\bibinfo
  {title} {Inelastic neutron scattering spectrum of {{H2}}@{{C60}} and its
  temperature dependence decoded using rigorous quantum calculations and a new
  selection rule},}\ }\href {https://doi.org/10.1063/1.4817534} {\bibfield
  {journal} {\bibinfo  {journal} {The Journal of Chemical Physics}\ }\textbf
  {\bibinfo {volume} {139}},\ \bibinfo {pages} {064309} (\bibinfo {year}
  {2013})}\BibitemShut {NoStop}%
\bibitem [{\citenamefont {Xu}, \citenamefont {Felker},\ and\ \citenamefont
  {Ba{\v c}i{\'c}}(2020)}]{xuLightMoleculesNanocavities2020}%
  \BibitemOpen
  \bibfield  {author} {\bibinfo {author} {\bibfnamefont {M.}~\bibnamefont
  {Xu}}, \bibinfo {author} {\bibfnamefont {P.~M.}\ \bibnamefont {Felker}},\
  and\ \bibinfo {author} {\bibfnamefont {Z.}~\bibnamefont {Ba{\v c}i{\'c}}},\
  }\bibfield  {title} {\enquote {\bibinfo {title} {Light molecules inside the
  nanocavities of fullerenes and clathrate hydrates: Inelastic neutron
  scattering spectra and the unexpected selection rule from rigorous quantum
  simulations},}\ }\href {https://doi.org/10.1080/0144235X.2020.1794097}
  {\bibfield  {journal} {\bibinfo  {journal} {International Reviews in Physical
  Chemistry}\ }\textbf {\bibinfo {volume} {39}},\ \bibinfo {pages} {425--463}
  (\bibinfo {year} {2020})}\BibitemShut {NoStop}%
\bibitem [{\citenamefont {Felker}\ and\ \citenamefont {Ba{\v
  c}i{\'c}}(2016{\natexlab{a}})}]{felkerTranslationrotationStates602016}%
  \BibitemOpen
  \bibfield  {author} {\bibinfo {author} {\bibfnamefont {P.~M.}\ \bibnamefont
  {Felker}}\ and\ \bibinfo {author} {\bibfnamefont {Z.}~\bibnamefont {Ba{\v
  c}i{\'c}}},\ }\bibfield  {title} {\enquote {\bibinfo {title}
  {Translation-rotation states of {{H}} {\textsubscript{2}} in {{C}}
  {\textsubscript{60}} : {{New}} insights from a perturbation-theory
  treatment},}\ }\href {https://doi.org/10.1063/1.4961650} {\bibfield
  {journal} {\bibinfo  {journal} {The Journal of Chemical Physics}\ }\textbf
  {\bibinfo {volume} {145}},\ \bibinfo {pages} {084310} (\bibinfo {year}
  {2016}{\natexlab{a}})}\BibitemShut {NoStop}%
\bibitem [{\citenamefont {Xu}, \citenamefont {Felker},\ and\ \citenamefont
  {Ba{\v c}i{\'c}}(2022)}]{xuFullerene60Inelastic2022}%
  \BibitemOpen
  \bibfield  {author} {\bibinfo {author} {\bibfnamefont {M.}~\bibnamefont
  {Xu}}, \bibinfo {author} {\bibfnamefont {P.~M.}\ \bibnamefont {Felker}},\
  and\ \bibinfo {author} {\bibfnamefont {Z.}~\bibnamefont {Ba{\v c}i{\'c}}},\
  }\bibfield  {title} {\enquote {\bibinfo {title} {H {\textsubscript{2}} {{O}}
  inside the fullerene {{C}} {\textsubscript{60}} : {{Inelastic}} neutron
  scattering spectrum from rigorous quantum calculations},}\ }\href
  {https://doi.org/10.1063/5.0086842} {\bibfield  {journal} {\bibinfo
  {journal} {The Journal of Chemical Physics}\ }\textbf {\bibinfo {volume}
  {156}},\ \bibinfo {pages} {124101} (\bibinfo {year} {2022})}\BibitemShut
  {NoStop}%
\bibitem [{\citenamefont {Felker}\ and\ \citenamefont {Ba{\v
  c}i{\'c}}(2020)}]{felkerFlexibleWaterMolecule2020}%
  \BibitemOpen
  \bibfield  {author} {\bibinfo {author} {\bibfnamefont {P.~M.}\ \bibnamefont
  {Felker}}\ and\ \bibinfo {author} {\bibfnamefont {Z.}~\bibnamefont {Ba{\v
  c}i{\'c}}},\ }\bibfield  {title} {\enquote {\bibinfo {title} {Flexible water
  molecule in {{C}} {\textsubscript{60}} : {{Intramolecular}} vibrational
  frequencies and translation-rotation eigenstates from fully coupled
  nine-dimensional quantum calculations with small basis sets},}\ }\href
  {https://doi.org/10.1063/1.5138992} {\bibfield  {journal} {\bibinfo
  {journal} {The Journal of Chemical Physics}\ }\textbf {\bibinfo {volume}
  {152}},\ \bibinfo {pages} {014108} (\bibinfo {year} {2020})}\BibitemShut
  {NoStop}%
\bibitem [{\citenamefont {Felker}\ and\ \citenamefont {Ba{\v
  c}i{\'c}}(2016{\natexlab{b}})}]{felkerCommunicationQuantumSixdimensional2016}%
  \BibitemOpen
  \bibfield  {author} {\bibinfo {author} {\bibfnamefont {P.~M.}\ \bibnamefont
  {Felker}}\ and\ \bibinfo {author} {\bibfnamefont {Z.}~\bibnamefont {Ba{\v
  c}i{\'c}}},\ }\bibfield  {title} {\enquote {\bibinfo {title} {Communication:
  {{Quantum}} six-dimensional calculations of the coupled translation-rotation
  eigenstates of {{H}} {\textsubscript{2}} {{O}}@{{C}} {\textsubscript{60}}},}\
  }\href {https://doi.org/10.1063/1.4953180} {\bibfield  {journal} {\bibinfo
  {journal} {The Journal of Chemical Physics}\ }\textbf {\bibinfo {volume}
  {144}},\ \bibinfo {pages} {201101} (\bibinfo {year}
  {2016}{\natexlab{b}})}\BibitemShut {NoStop}%
\bibitem [{\citenamefont {{Carrillo-Boh{\'o}rquez}}, \citenamefont
  {Vald{\'e}s},\ and\ \citenamefont
  {Prosmiti}(2021)}]{carrillo-bohorquezEncapsulationWaterMolecule2021}%
  \BibitemOpen
  \bibfield  {author} {\bibinfo {author} {\bibfnamefont {O.}~\bibnamefont
  {{Carrillo-Boh{\'o}rquez}}}, \bibinfo {author} {\bibfnamefont
  {{\'A}.}~\bibnamefont {Vald{\'e}s}},\ and\ \bibinfo {author} {\bibfnamefont
  {R.}~\bibnamefont {Prosmiti}},\ }\bibfield  {title} {\enquote {\bibinfo
  {title} {Encapsulation of a {{Water Molecule}} inside {{C}}
  {\textsubscript{60}} {{Fullerene}}: {{The Impact}} of {{Confinement}} on
  {{Quantum Features}}},}\ }\href {https://doi.org/10.1021/acs.jctc.1c00662}
  {\bibfield  {journal} {\bibinfo  {journal} {Journal of Chemical Theory and
  Computation}\ }\textbf {\bibinfo {volume} {17}},\ \bibinfo {pages}
  {5839--5848} (\bibinfo {year} {2021})}\BibitemShut {NoStop}%
\bibitem [{\citenamefont {Rashed}\ and\ \citenamefont
  {Dunn}(2019)}]{rashedInteractionsWaterMolecule2019}%
  \BibitemOpen
  \bibfield  {author} {\bibinfo {author} {\bibfnamefont {E.}~\bibnamefont
  {Rashed}}\ and\ \bibinfo {author} {\bibfnamefont {J.~L.}\ \bibnamefont
  {Dunn}},\ }\bibfield  {title} {\enquote {\bibinfo {title} {Interactions
  between a water molecule and {{C60}} in the endohedral fullerene
  {{H2O}}@{{C60}}},}\ }\href {https://doi.org/10.1039/C8CP04390F} {\bibfield
  {journal} {\bibinfo  {journal} {Physical Chemistry Chemical Physics}\
  }\textbf {\bibinfo {volume} {21}},\ \bibinfo {pages} {3347--3359} (\bibinfo
  {year} {2019})}\BibitemShut {NoStop}%
\bibitem [{\citenamefont {Sebastianelli}\ \emph {et~al.}(2010)\citenamefont
  {Sebastianelli}, \citenamefont {Xu}, \citenamefont {Ba{\v c}i{\'c}},
  \citenamefont {Lawler},\ and\ \citenamefont
  {Turro}}]{sebastianelliHydrogenMoleculesFullerene2010}%
  \BibitemOpen
  \bibfield  {author} {\bibinfo {author} {\bibfnamefont {F.}~\bibnamefont
  {Sebastianelli}}, \bibinfo {author} {\bibfnamefont {M.}~\bibnamefont {Xu}},
  \bibinfo {author} {\bibfnamefont {Z.}~\bibnamefont {Ba{\v c}i{\'c}}},
  \bibinfo {author} {\bibfnamefont {R.}~\bibnamefont {Lawler}},\ and\ \bibinfo
  {author} {\bibfnamefont {N.~J.}\ \bibnamefont {Turro}},\ }\bibfield  {title}
  {\enquote {\bibinfo {title} {Hydrogen {{Molecules}} inside {{Fullerene C}}
  {\textsubscript{70}} : {{Quantum Dynamics}}, {{Energetics}}, {{Maximum
  Occupancy}}, {{And Comparison}} with {{C}} {\textsubscript{60}}},}\ }\href
  {https://doi.org/10.1021/ja103062g} {\bibfield  {journal} {\bibinfo
  {journal} {Journal of the American Chemical Society}\ }\textbf {\bibinfo
  {volume} {132}},\ \bibinfo {pages} {9826--9832} (\bibinfo {year}
  {2010})}\BibitemShut {NoStop}%
\bibitem [{\citenamefont {{Foroutan-Nejad}}, \citenamefont {Andrushchenko},\
  and\ \citenamefont {Straka}(2016)}]{foroutan-nejadDipolarMoleculesC702016}%
  \BibitemOpen
  \bibfield  {author} {\bibinfo {author} {\bibfnamefont {C.}~\bibnamefont
  {{Foroutan-Nejad}}}, \bibinfo {author} {\bibfnamefont {V.}~\bibnamefont
  {Andrushchenko}},\ and\ \bibinfo {author} {\bibfnamefont {M.}~\bibnamefont
  {Straka}},\ }\bibfield  {title} {\enquote {\bibinfo {title} {Dipolar
  molecules inside {{C70}}: An electric field-driven room-temperature
  single-molecule switch},}\ }\href {https://doi.org/10.1039/C6CP06986J}
  {\bibfield  {journal} {\bibinfo  {journal} {Physical Chemistry Chemical
  Physics}\ }\textbf {\bibinfo {volume} {18}},\ \bibinfo {pages} {32673--32677}
  (\bibinfo {year} {2016})}\BibitemShut {NoStop}%
\bibitem [{\citenamefont
  {Caliskan}(2021)}]{caliskanStructuralElectronicAdsorption2021}%
  \BibitemOpen
  \bibfield  {author} {\bibinfo {author} {\bibfnamefont {S.}~\bibnamefont
  {Caliskan}},\ }\bibfield  {title} {\enquote {\bibinfo {title} {Structural,
  {{Electronic}} and {{Adsorption Characteristics}} of {{Transition Metal}}
  doped {{TM}}@{{C70 Endohedral Fullerenes}}},}\ }\href
  {https://doi.org/10.1007/s10876-020-01762-2} {\bibfield  {journal} {\bibinfo
  {journal} {Journal of Cluster Science}\ }\textbf {\bibinfo {volume} {32}},\
  \bibinfo {pages} {77--84} (\bibinfo {year} {2021})}\BibitemShut {NoStop}%
\bibitem [{\citenamefont {Jafari}\ \emph {et~al.}(2023)\citenamefont {Jafari},
  \citenamefont {Shugai}, \citenamefont {Nagel}, \citenamefont {Bacanu},
  \citenamefont {Aouane}, \citenamefont {{Jim{\'e}nez-Ruiz}}, \citenamefont
  {Rols}, \citenamefont {Bloodworth}, \citenamefont {Walkey}, \citenamefont
  {Hoffman}, \citenamefont {Whitby}, \citenamefont {Levitt},\ and\
  \citenamefont {R{\~o}{\~o}m}}]{jafariNeArKr2023a}%
  \BibitemOpen
  \bibfield  {author} {\bibinfo {author} {\bibfnamefont {T.}~\bibnamefont
  {Jafari}}, \bibinfo {author} {\bibfnamefont {A.}~\bibnamefont {Shugai}},
  \bibinfo {author} {\bibfnamefont {U.}~\bibnamefont {Nagel}}, \bibinfo
  {author} {\bibfnamefont {G.~R.}\ \bibnamefont {Bacanu}}, \bibinfo {author}
  {\bibfnamefont {M.}~\bibnamefont {Aouane}}, \bibinfo {author} {\bibfnamefont
  {M.}~\bibnamefont {{Jim{\'e}nez-Ruiz}}}, \bibinfo {author} {\bibfnamefont
  {S.}~\bibnamefont {Rols}}, \bibinfo {author} {\bibfnamefont {S.}~\bibnamefont
  {Bloodworth}}, \bibinfo {author} {\bibfnamefont {M.}~\bibnamefont {Walkey}},
  \bibinfo {author} {\bibfnamefont {G.}~\bibnamefont {Hoffman}}, \bibinfo
  {author} {\bibfnamefont {R.~J.}\ \bibnamefont {Whitby}}, \bibinfo {author}
  {\bibfnamefont {M.~H.}\ \bibnamefont {Levitt}},\ and\ \bibinfo {author}
  {\bibfnamefont {T.}~\bibnamefont {R{\~o}{\~o}m}},\ }\bibfield  {title}
  {\enquote {\bibinfo {title} {Ne, {{Ar}}, and {{Kr}} oscillators in the
  molecular cavity of fullerene {{C60}}},}\ }\href
  {https://doi.org/10.1063/5.0152628} {\bibfield  {journal} {\bibinfo
  {journal} {The Journal of Chemical Physics}\ }\textbf {\bibinfo {volume}
  {158}},\ \bibinfo {pages} {234305} (\bibinfo {year} {2023})}\BibitemShut
  {NoStop}%
\bibitem [{\citenamefont {Panchagnula}\ and\ \citenamefont
  {Thom}(2023)}]{panchagnulaExploringParameterSpace2023b}%
  \BibitemOpen
  \bibfield  {author} {\bibinfo {author} {\bibfnamefont {K.}~\bibnamefont
  {Panchagnula}}\ and\ \bibinfo {author} {\bibfnamefont {A.~J.~W.}\
  \bibnamefont {Thom}},\ }\bibfield  {title} {\enquote {\bibinfo {title}
  {Exploring the parameter space of an endohedral atom in a cylindrical
  cavity},}\ }\href {https://doi.org/10.1063/5.0170010} {\bibfield  {journal}
  {\bibinfo  {journal} {The Journal of Chemical Physics}\ }\textbf {\bibinfo
  {volume} {159}},\ \bibinfo {pages} {164308} (\bibinfo {year}
  {2023})}\BibitemShut {NoStop}%
\bibitem [{\citenamefont {Mandziuk}\ and\ \citenamefont {Ba{\v
  c}i{\'c}}(1994)}]{mandziukQuantumThreeDimensional1994}%
  \BibitemOpen
  \bibfield  {author} {\bibinfo {author} {\bibfnamefont {M.}~\bibnamefont
  {Mandziuk}}\ and\ \bibinfo {author} {\bibfnamefont {Z.}~\bibnamefont {Ba{\v
  c}i{\'c}}},\ }\bibfield  {title} {\enquote {\bibinfo {title} {Quantum
  three-dimensional calculation of endohedral vibrational levels of atoms
  inside strongly nonspherical fullerenes: {{Ne}}@{{C70}}},}\ }\href
  {https://doi.org/10.1063/1.467719} {\bibfield  {journal} {\bibinfo  {journal}
  {The Journal of Chemical Physics}\ }\textbf {\bibinfo {volume} {101}},\
  \bibinfo {pages} {2126--2140} (\bibinfo {year} {1994})}\BibitemShut {NoStop}%
\bibitem [{\citenamefont {Panchagnula}\ \emph {et~al.}(2024)\citenamefont
  {Panchagnula}, \citenamefont {Graf}, \citenamefont {Albertani},\ and\
  \citenamefont {Thom}}]{panchagnulaTranslationalEigenstatesHe2024a}%
  \BibitemOpen
  \bibfield  {author} {\bibinfo {author} {\bibfnamefont {K.}~\bibnamefont
  {Panchagnula}}, \bibinfo {author} {\bibfnamefont {D.}~\bibnamefont {Graf}},
  \bibinfo {author} {\bibfnamefont {F.~E.~A.}\ \bibnamefont {Albertani}},\ and\
  \bibinfo {author} {\bibfnamefont {A.~J.~W.}\ \bibnamefont {Thom}},\
  }\bibfield  {title} {\enquote {\bibinfo {title} {Translational eigenstates of
  {{He}}@{{C60}} from four-dimensional ab~initio potential energy surfaces
  interpolated using {{Gaussian}} process regression},}\ }\href
  {https://doi.org/10.1063/5.0197903} {\bibfield  {journal} {\bibinfo
  {journal} {The Journal of Chemical Physics}\ }\textbf {\bibinfo {volume}
  {160}},\ \bibinfo {pages} {104303} (\bibinfo {year} {2024})}\BibitemShut
  {NoStop}%
\bibitem [{\citenamefont
  {Behler}(2016)}]{behlerPerspectiveMachineLearning2016}%
  \BibitemOpen
  \bibfield  {author} {\bibinfo {author} {\bibfnamefont {J.}~\bibnamefont
  {Behler}},\ }\bibfield  {title} {\enquote {\bibinfo {title} {Perspective:
  {{Machine}} learning potentials for atomistic simulations},}\ }\href
  {https://doi.org/10.1063/1.4966192} {\bibfield  {journal} {\bibinfo
  {journal} {The Journal of Chemical Physics}\ }\textbf {\bibinfo {volume}
  {145}},\ \bibinfo {pages} {170901} (\bibinfo {year} {2016})}\BibitemShut
  {NoStop}%
\bibitem [{\citenamefont {Musil}\ \emph {et~al.}(2021)\citenamefont {Musil},
  \citenamefont {Grisafi}, \citenamefont {Bart{\'o}k}, \citenamefont {Ortner},
  \citenamefont {Cs{\'a}nyi},\ and\ \citenamefont
  {Ceriotti}}]{musilPhysicsInspiredStructuralRepresentations2021}%
  \BibitemOpen
  \bibfield  {author} {\bibinfo {author} {\bibfnamefont {F.}~\bibnamefont
  {Musil}}, \bibinfo {author} {\bibfnamefont {A.}~\bibnamefont {Grisafi}},
  \bibinfo {author} {\bibfnamefont {A.~P.}\ \bibnamefont {Bart{\'o}k}},
  \bibinfo {author} {\bibfnamefont {C.}~\bibnamefont {Ortner}}, \bibinfo
  {author} {\bibfnamefont {G.}~\bibnamefont {Cs{\'a}nyi}},\ and\ \bibinfo
  {author} {\bibfnamefont {M.}~\bibnamefont {Ceriotti}},\ }\bibfield  {title}
  {\enquote {\bibinfo {title} {Physics-{{Inspired Structural Representations}}
  for {{Molecules}} and {{Materials}}},}\ }\href
  {https://doi.org/10.1021/acs.chemrev.1c00021} {\bibfield  {journal} {\bibinfo
   {journal} {Chemical Reviews}\ }\textbf {\bibinfo {volume} {121}},\ \bibinfo
  {pages} {9759--9815} (\bibinfo {year} {2021})}\BibitemShut {NoStop}%
\bibitem [{\citenamefont {Dral}\ \emph {et~al.}(2017)\citenamefont {Dral},
  \citenamefont {Owens}, \citenamefont {Yurchenko},\ and\ \citenamefont
  {Thiel}}]{dralStructurebasedSamplingSelfcorrecting2017}%
  \BibitemOpen
  \bibfield  {author} {\bibinfo {author} {\bibfnamefont {P.~O.}\ \bibnamefont
  {Dral}}, \bibinfo {author} {\bibfnamefont {A.}~\bibnamefont {Owens}},
  \bibinfo {author} {\bibfnamefont {S.~N.}\ \bibnamefont {Yurchenko}},\ and\
  \bibinfo {author} {\bibfnamefont {W.}~\bibnamefont {Thiel}},\ }\bibfield
  {title} {\enquote {\bibinfo {title} {Structure-based sampling and
  self-correcting machine learning for accurate calculations of potential
  energy surfaces and vibrational levels},}\ }\href
  {https://doi.org/10.1063/1.4989536} {\bibfield  {journal} {\bibinfo
  {journal} {The Journal of Chemical Physics}\ }\textbf {\bibinfo {volume}
  {146}},\ \bibinfo {pages} {244108} (\bibinfo {year} {2017})}\BibitemShut
  {NoStop}%
\bibitem [{\citenamefont {Rasmussen}\ and\ \citenamefont
  {Williams}(2005)}]{rasmussenGaussianProcessesMachine2005}%
  \BibitemOpen
  \bibfield  {author} {\bibinfo {author} {\bibfnamefont {C.~E.}\ \bibnamefont
  {Rasmussen}}\ and\ \bibinfo {author} {\bibfnamefont {C.~K.}\ \bibnamefont
  {Williams}},\ }\href@noop {} {\emph {\bibinfo {title} {Gaussian {{Processes}}
  for {{Machine Learning}}}}}\ (\bibinfo  {publisher} {MIT Press},\ \bibinfo
  {year} {2005})\BibitemShut {NoStop}%
\bibitem [{\citenamefont
  {Cioslowski}(2023)}]{cioslowskiElectronicStructureCalculations2023a}%
  \BibitemOpen
  \bibfield  {author} {\bibinfo {author} {\bibfnamefont {J.}~\bibnamefont
  {Cioslowski}},\ }\bibfield  {title} {\enquote {\bibinfo {title} {Electronic
  {{Structure Calculations}} on {{Endohedral Complexes}} of {{Fullerenes}}:
  {{Reminiscences}} and {{Prospects}}},}\ }\href
  {https://doi.org/10.3390/molecules28031384} {\bibfield  {journal} {\bibinfo
  {journal} {Molecules}\ }\textbf {\bibinfo {volume} {28}},\ \bibinfo {pages}
  {1384} (\bibinfo {year} {2023})}\BibitemShut {NoStop}%
\bibitem [{\citenamefont {Glasbrenner}, \citenamefont {Graf},\ and\
  \citenamefont
  {Ochsenfeld}(2020)}]{glasbrennerEfficientReducedScalingSecondOrder2020}%
  \BibitemOpen
  \bibfield  {author} {\bibinfo {author} {\bibfnamefont {M.}~\bibnamefont
  {Glasbrenner}}, \bibinfo {author} {\bibfnamefont {D.}~\bibnamefont {Graf}},\
  and\ \bibinfo {author} {\bibfnamefont {C.}~\bibnamefont {Ochsenfeld}},\
  }\bibfield  {title} {\enquote {\bibinfo {title} {Efficient {{Reduced-Scaling
  Second-Order M{\o}ller}}--{{Plesset Perturbation Theory}} with
  {{Cholesky-Decomposed Densities}} and an {{Attenuated Coulomb Metric}}},}\
  }\href {https://doi.org/10.1021/acs.jctc.0c00600} {\bibfield  {journal}
  {\bibinfo  {journal} {Journal of Chemical Theory and Computation}\ }\textbf
  {\bibinfo {volume} {16}},\ \bibinfo {pages} {6856--6868} (\bibinfo {year}
  {2020})}\BibitemShut {NoStop}%
\bibitem [{\citenamefont {Jung}\ \emph {et~al.}(2004)\citenamefont {Jung},
  \citenamefont {Lochan}, \citenamefont {Dutoi},\ and\ \citenamefont
  {{Head-Gordon}}}]{jungScaledOppositespinSecond2004}%
  \BibitemOpen
  \bibfield  {author} {\bibinfo {author} {\bibfnamefont {Y.}~\bibnamefont
  {Jung}}, \bibinfo {author} {\bibfnamefont {R.~C.}\ \bibnamefont {Lochan}},
  \bibinfo {author} {\bibfnamefont {A.~D.}\ \bibnamefont {Dutoi}},\ and\
  \bibinfo {author} {\bibfnamefont {M.}~\bibnamefont {{Head-Gordon}}},\
  }\bibfield  {title} {\enquote {\bibinfo {title} {Scaled opposite-spin second
  order {{M{\o}ller}}--{{Plesset}} correlation energy: {{An}} economical
  electronic structure method},}\ }\href {https://doi.org/10.1063/1.1809602}
  {\bibfield  {journal} {\bibinfo  {journal} {The Journal of Chemical Physics}\
  }\textbf {\bibinfo {volume} {121}},\ \bibinfo {pages} {9793--9802} (\bibinfo
  {year} {2004})}\BibitemShut {NoStop}%
\bibitem [{\citenamefont {Grimme}(2003)}]{grimmeImprovedSecondorderMoller2003}%
  \BibitemOpen
  \bibfield  {author} {\bibinfo {author} {\bibfnamefont {S.}~\bibnamefont
  {Grimme}},\ }\bibfield  {title} {\enquote {\bibinfo {title} {Improved
  second-order {{M{\o}ller}}--{{Plesset}} perturbation theory by separate
  scaling of parallel- and antiparallel-spin pair correlation energies},}\
  }\href {https://doi.org/10.1063/1.1569242} {\bibfield  {journal} {\bibinfo
  {journal} {The Journal of Chemical Physics}\ }\textbf {\bibinfo {volume}
  {118}},\ \bibinfo {pages} {9095--9102} (\bibinfo {year} {2003})}\BibitemShut
  {NoStop}%
\bibitem [{\citenamefont {Graf}\ \emph {et~al.}(2018)\citenamefont {Graf},
  \citenamefont {Beuerle}, \citenamefont {Schurkus}, \citenamefont {Luenser},
  \citenamefont {Savasci},\ and\ \citenamefont
  {Ochsenfeld}}]{grafAccurateEfficientParallel2018}%
  \BibitemOpen
  \bibfield  {author} {\bibinfo {author} {\bibfnamefont {D.}~\bibnamefont
  {Graf}}, \bibinfo {author} {\bibfnamefont {M.}~\bibnamefont {Beuerle}},
  \bibinfo {author} {\bibfnamefont {H.~F.}\ \bibnamefont {Schurkus}}, \bibinfo
  {author} {\bibfnamefont {A.}~\bibnamefont {Luenser}}, \bibinfo {author}
  {\bibfnamefont {G.}~\bibnamefont {Savasci}},\ and\ \bibinfo {author}
  {\bibfnamefont {C.}~\bibnamefont {Ochsenfeld}},\ }\bibfield  {title}
  {\enquote {\bibinfo {title} {Accurate and {{Efficient Parallel
  Implementation}} of an {{Effective Linear-Scaling Direct Random Phase
  Approximation Method}}},}\ }\href {https://doi.org/10.1021/acs.jctc.8b00177}
  {\bibfield  {journal} {\bibinfo  {journal} {Journal of Chemical Theory and
  Computation}\ }\textbf {\bibinfo {volume} {14}},\ \bibinfo {pages}
  {2505--2515} (\bibinfo {year} {2018})}\BibitemShut {NoStop}%
\bibitem [{\citenamefont {Graf}\ and\ \citenamefont
  {Thom}(2023)}]{grafCorrectedDensityFunctional2023}%
  \BibitemOpen
  \bibfield  {author} {\bibinfo {author} {\bibfnamefont {D.}~\bibnamefont
  {Graf}}\ and\ \bibinfo {author} {\bibfnamefont {A.~J.~W.}\ \bibnamefont
  {Thom}},\ }\bibfield  {title} {\enquote {\bibinfo {title} {Corrected density
  functional theory and the random phase approximation: {{Improved}} accuracy
  at little extra cost},}\ }\href {https://doi.org/10.1063/5.0168569}
  {\bibfield  {journal} {\bibinfo  {journal} {The Journal of Chemical Physics}\
  }\textbf {\bibinfo {volume} {159}},\ \bibinfo {pages} {174106} (\bibinfo
  {year} {2023})}\BibitemShut {NoStop}%
\bibitem [{\citenamefont {Giner}\ \emph {et~al.}(2018)\citenamefont {Giner},
  \citenamefont {Pradines}, \citenamefont {Fert{\'e}}, \citenamefont {Assaraf},
  \citenamefont {Savin},\ and\ \citenamefont
  {Toulouse}}]{ginerCuringBasissetConvergence2018}%
  \BibitemOpen
  \bibfield  {author} {\bibinfo {author} {\bibfnamefont {E.}~\bibnamefont
  {Giner}}, \bibinfo {author} {\bibfnamefont {B.}~\bibnamefont {Pradines}},
  \bibinfo {author} {\bibfnamefont {A.}~\bibnamefont {Fert{\'e}}}, \bibinfo
  {author} {\bibfnamefont {R.}~\bibnamefont {Assaraf}}, \bibinfo {author}
  {\bibfnamefont {A.}~\bibnamefont {Savin}},\ and\ \bibinfo {author}
  {\bibfnamefont {J.}~\bibnamefont {Toulouse}},\ }\bibfield  {title} {\enquote
  {\bibinfo {title} {Curing basis-set convergence of wave-function theory using
  density-functional theory: {{A}} systematically improvable approach},}\
  }\href {https://doi.org/10.1063/1.5052714} {\bibfield  {journal} {\bibinfo
  {journal} {The Journal of Chemical Physics}\ }\textbf {\bibinfo {volume}
  {149}},\ \bibinfo {pages} {194301} (\bibinfo {year} {2018})}\BibitemShut
  {NoStop}%
\bibitem [{\citenamefont
  {Pulay}(1983)}]{pulayLocalizabilityDynamicElectron1983}%
  \BibitemOpen
  \bibfield  {author} {\bibinfo {author} {\bibfnamefont {P.}~\bibnamefont
  {Pulay}},\ }\bibfield  {title} {\enquote {\bibinfo {title} {Localizability of
  dynamic electron correlation},}\ }\href
  {https://doi.org/10.1016/0009-2614(83)80703-9} {\bibfield  {journal}
  {\bibinfo  {journal} {Chemical Physics Letters}\ }\textbf {\bibinfo {volume}
  {100}},\ \bibinfo {pages} {151--154} (\bibinfo {year} {1983})}\BibitemShut
  {NoStop}%
\bibitem [{\citenamefont
  {H{\"a}ser}(1993)}]{haserMollerPlessetMP2Perturbation1993}%
  \BibitemOpen
  \bibfield  {author} {\bibinfo {author} {\bibfnamefont {M.}~\bibnamefont
  {H{\"a}ser}},\ }\bibfield  {title} {\enquote {\bibinfo {title}
  {M{\o}ller-{{Plesset}} ({{MP2}}) perturbation theory for large molecules},}\
  }\href {https://doi.org/10.1007/BF01113535} {\bibfield  {journal} {\bibinfo
  {journal} {Theoretica chimica acta}\ }\textbf {\bibinfo {volume} {87}},\
  \bibinfo {pages} {147--173} (\bibinfo {year} {1993})}\BibitemShut {NoStop}%
\bibitem [{\citenamefont {Maslen}\ and\ \citenamefont
  {{Head-Gordon}}(1998)}]{maslenNoniterativeLocalSecond1998}%
  \BibitemOpen
  \bibfield  {author} {\bibinfo {author} {\bibfnamefont {P.~E.}\ \bibnamefont
  {Maslen}}\ and\ \bibinfo {author} {\bibfnamefont {M.}~\bibnamefont
  {{Head-Gordon}}},\ }\bibfield  {title} {\enquote {\bibinfo {title}
  {Non-iterative local second order {{M{\o}ller}}--{{Plesset}} theory},}\
  }\href {https://doi.org/10.1016/S0009-2614(97)01333-X} {\bibfield  {journal}
  {\bibinfo  {journal} {Chemical Physics Letters}\ }\textbf {\bibinfo {volume}
  {283}},\ \bibinfo {pages} {102--108} (\bibinfo {year} {1998})}\BibitemShut
  {NoStop}%
\bibitem [{\citenamefont {Ayala}\ and\ \citenamefont
  {Scuseria}(1999)}]{ayalaLinearScalingSecondorder1999}%
  \BibitemOpen
  \bibfield  {author} {\bibinfo {author} {\bibfnamefont {P.~Y.}\ \bibnamefont
  {Ayala}}\ and\ \bibinfo {author} {\bibfnamefont {G.~E.}\ \bibnamefont
  {Scuseria}},\ }\bibfield  {title} {\enquote {\bibinfo {title} {Linear scaling
  second-order {{Moller}}--{{Plesset}} theory in the atomic orbital basis for
  large molecular systems},}\ }\href {https://doi.org/10.1063/1.478256}
  {\bibfield  {journal} {\bibinfo  {journal} {The Journal of Chemical Physics}\
  }\textbf {\bibinfo {volume} {110}},\ \bibinfo {pages} {3660--3671} (\bibinfo
  {year} {1999})}\BibitemShut {NoStop}%
\bibitem [{\citenamefont {Sch{\"u}tz}, \citenamefont {Hetzer},\ and\
  \citenamefont {Werner}(1999)}]{schutzLoworderScalingLocal1999}%
  \BibitemOpen
  \bibfield  {author} {\bibinfo {author} {\bibfnamefont {M.}~\bibnamefont
  {Sch{\"u}tz}}, \bibinfo {author} {\bibfnamefont {G.}~\bibnamefont {Hetzer}},\
  and\ \bibinfo {author} {\bibfnamefont {H.-J.}\ \bibnamefont {Werner}},\
  }\bibfield  {title} {\enquote {\bibinfo {title} {Low-order scaling local
  electron correlation methods. {{I}}. {{Linear}} scaling local {{MP2}}},}\
  }\href {https://doi.org/10.1063/1.479957} {\bibfield  {journal} {\bibinfo
  {journal} {The Journal of Chemical Physics}\ }\textbf {\bibinfo {volume}
  {111}},\ \bibinfo {pages} {5691--5705} (\bibinfo {year} {1999})}\BibitemShut
  {NoStop}%
\bibitem [{\citenamefont {Saeb{\o}}\ and\ \citenamefont
  {Pulay}(2001)}]{saeboLowscalingMethodSecond2001}%
  \BibitemOpen
  \bibfield  {author} {\bibinfo {author} {\bibfnamefont {S.}~\bibnamefont
  {Saeb{\o}}}\ and\ \bibinfo {author} {\bibfnamefont {P.}~\bibnamefont
  {Pulay}},\ }\bibfield  {title} {\enquote {\bibinfo {title} {A low-scaling
  method for second order {{M{\o}ller}}--{{Plesset}} calculations},}\ }\href
  {https://doi.org/10.1063/1.1389291} {\bibfield  {journal} {\bibinfo
  {journal} {The Journal of Chemical Physics}\ }\textbf {\bibinfo {volume}
  {115}},\ \bibinfo {pages} {3975--3983} (\bibinfo {year} {2001})}\BibitemShut
  {NoStop}%
\bibitem [{\citenamefont {Werner}, \citenamefont {Manby},\ and\ \citenamefont
  {Knowles}(2003)}]{wernerFastLinearScaling2003}%
  \BibitemOpen
  \bibfield  {author} {\bibinfo {author} {\bibfnamefont {H.-J.}\ \bibnamefont
  {Werner}}, \bibinfo {author} {\bibfnamefont {F.~R.}\ \bibnamefont {Manby}},\
  and\ \bibinfo {author} {\bibfnamefont {P.~J.}\ \bibnamefont {Knowles}},\
  }\bibfield  {title} {\enquote {\bibinfo {title} {Fast linear scaling
  second-order {{M{\o}ller-Plesset}} perturbation theory ({{MP2}}) using local
  and density fitting approximations},}\ }\href
  {https://doi.org/10.1063/1.1564816} {\bibfield  {journal} {\bibinfo
  {journal} {The Journal of Chemical Physics}\ }\textbf {\bibinfo {volume}
  {118}},\ \bibinfo {pages} {8149--8160} (\bibinfo {year} {2003})}\BibitemShut
  {NoStop}%
\bibitem [{\citenamefont {Jung}\ and\ \citenamefont
  {{Head-Gordon}}(2006)}]{jungFastCorrelatedElectronic2006}%
  \BibitemOpen
  \bibfield  {author} {\bibinfo {author} {\bibfnamefont {Y.}~\bibnamefont
  {Jung}}\ and\ \bibinfo {author} {\bibfnamefont {M.}~\bibnamefont
  {{Head-Gordon}}},\ }\bibfield  {title} {\enquote {\bibinfo {title} {A fast
  correlated electronic structure method for computing interaction energies of
  large van der {{Waals}} complexes applied to the fullerene--porphyrin
  dimer},}\ }\href {https://doi.org/10.1039/B602438F} {\bibfield  {journal}
  {\bibinfo  {journal} {Physical Chemistry Chemical Physics}\ }\textbf
  {\bibinfo {volume} {8}},\ \bibinfo {pages} {2831--2840} (\bibinfo {year}
  {2006})}\BibitemShut {NoStop}%
\bibitem [{\citenamefont {Jung}, \citenamefont {Shao},\ and\ \citenamefont
  {{Head-Gordon}}(2007)}]{jungFastEvaluationScaled2007}%
  \BibitemOpen
  \bibfield  {author} {\bibinfo {author} {\bibfnamefont {Y.}~\bibnamefont
  {Jung}}, \bibinfo {author} {\bibfnamefont {Y.}~\bibnamefont {Shao}},\ and\
  \bibinfo {author} {\bibfnamefont {M.}~\bibnamefont {{Head-Gordon}}},\
  }\bibfield  {title} {\enquote {\bibinfo {title} {Fast evaluation of scaled
  opposite spin second-order {{M{\o}ller}}--{{Plesset}} correlation energies
  using auxiliary basis expansions and exploiting sparsity},}\ }\href
  {https://doi.org/10.1002/jcc.20590} {\bibfield  {journal} {\bibinfo
  {journal} {Journal of Computational Chemistry}\ }\textbf {\bibinfo {volume}
  {28}},\ \bibinfo {pages} {1953--1964} (\bibinfo {year} {2007})}\BibitemShut
  {NoStop}%
\bibitem [{\citenamefont {Doser}, \citenamefont {Lambrecht},\ and\
  \citenamefont {Ochsenfeld}(2008)}]{doserTighterMultipolebasedIntegral2008}%
  \BibitemOpen
  \bibfield  {author} {\bibinfo {author} {\bibfnamefont {B.}~\bibnamefont
  {Doser}}, \bibinfo {author} {\bibfnamefont {D.~S.}\ \bibnamefont
  {Lambrecht}},\ and\ \bibinfo {author} {\bibfnamefont {C.}~\bibnamefont
  {Ochsenfeld}},\ }\bibfield  {title} {\enquote {\bibinfo {title} {Tighter
  multipole-based integral estimates and parallel implementation of
  linear-scaling {{AO}}--{{MP2}} theory},}\ }\href
  {https://doi.org/10.1039/B804110E} {\bibfield  {journal} {\bibinfo  {journal}
  {Physical Chemistry Chemical Physics}\ }\textbf {\bibinfo {volume} {10}},\
  \bibinfo {pages} {3335--3344} (\bibinfo {year} {2008})}\BibitemShut {NoStop}%
\bibitem [{\citenamefont {Doser}\ \emph {et~al.}(2009)\citenamefont {Doser},
  \citenamefont {Lambrecht}, \citenamefont {Kussmann},\ and\ \citenamefont
  {Ochsenfeld}}]{doserLinearscalingAtomicOrbitalbased2009}%
  \BibitemOpen
  \bibfield  {author} {\bibinfo {author} {\bibfnamefont {B.}~\bibnamefont
  {Doser}}, \bibinfo {author} {\bibfnamefont {D.~S.}\ \bibnamefont
  {Lambrecht}}, \bibinfo {author} {\bibfnamefont {J.}~\bibnamefont
  {Kussmann}},\ and\ \bibinfo {author} {\bibfnamefont {C.}~\bibnamefont
  {Ochsenfeld}},\ }\bibfield  {title} {\enquote {\bibinfo {title}
  {Linear-scaling atomic orbital-based second-order {{M{\o}ller}}--{{Plesset}}
  perturbation theory by rigorous integral screening criteria},}\ }\href
  {https://doi.org/10.1063/1.3072903} {\bibfield  {journal} {\bibinfo
  {journal} {The Journal of Chemical Physics}\ }\textbf {\bibinfo {volume}
  {130}},\ \bibinfo {pages} {064107} (\bibinfo {year} {2009})}\BibitemShut
  {NoStop}%
\bibitem [{\citenamefont {Zienau}\ \emph {et~al.}(2009)\citenamefont {Zienau},
  \citenamefont {Clin}, \citenamefont {Doser},\ and\ \citenamefont
  {Ochsenfeld}}]{zienauCholeskydecomposedDensitiesLaplacebased2009}%
  \BibitemOpen
  \bibfield  {author} {\bibinfo {author} {\bibfnamefont {J.}~\bibnamefont
  {Zienau}}, \bibinfo {author} {\bibfnamefont {L.}~\bibnamefont {Clin}},
  \bibinfo {author} {\bibfnamefont {B.}~\bibnamefont {Doser}},\ and\ \bibinfo
  {author} {\bibfnamefont {C.}~\bibnamefont {Ochsenfeld}},\ }\bibfield  {title}
  {\enquote {\bibinfo {title} {Cholesky-decomposed densities in
  {{Laplace-based}} second-order {{M{\o}ller}}--{{Plesset}} perturbation
  theory},}\ }\href {https://doi.org/10.1063/1.3142592} {\bibfield  {journal}
  {\bibinfo  {journal} {The Journal of Chemical Physics}\ }\textbf {\bibinfo
  {volume} {130}},\ \bibinfo {pages} {204112} (\bibinfo {year}
  {2009})}\BibitemShut {NoStop}%
\bibitem [{\citenamefont {Kristensen}\ \emph {et~al.}(2012)\citenamefont
  {Kristensen}, \citenamefont {H{\o}yvik}, \citenamefont {Jansik},
  \citenamefont {J{\o}rgensen}, \citenamefont {Kj{\ae}rgaard}, \citenamefont
  {Reine},\ and\ \citenamefont {Jakowski}}]{kristensenMP2EnergyDensity2012}%
  \BibitemOpen
  \bibfield  {author} {\bibinfo {author} {\bibfnamefont {K.}~\bibnamefont
  {Kristensen}}, \bibinfo {author} {\bibfnamefont {I.-M.}\ \bibnamefont
  {H{\o}yvik}}, \bibinfo {author} {\bibfnamefont {B.}~\bibnamefont {Jansik}},
  \bibinfo {author} {\bibfnamefont {P.}~\bibnamefont {J{\o}rgensen}}, \bibinfo
  {author} {\bibfnamefont {T.}~\bibnamefont {Kj{\ae}rgaard}}, \bibinfo {author}
  {\bibfnamefont {S.}~\bibnamefont {Reine}},\ and\ \bibinfo {author}
  {\bibfnamefont {J.}~\bibnamefont {Jakowski}},\ }\bibfield  {title} {\enquote
  {\bibinfo {title} {{{MP2}} energy and density for large molecular systems
  with internal error control using the {{Divide-Expand-Consolidate}}
  scheme},}\ }\href {https://doi.org/10.1039/C2CP41958K} {\bibfield  {journal}
  {\bibinfo  {journal} {Physical Chemistry Chemical Physics}\ }\textbf
  {\bibinfo {volume} {14}},\ \bibinfo {pages} {15706--15714} (\bibinfo {year}
  {2012})}\BibitemShut {NoStop}%
\bibitem [{\citenamefont {Maurer}, \citenamefont {Clin},\ and\ \citenamefont
  {Ochsenfeld}(2014)}]{maurerCholeskydecomposedDensityMP22014}%
  \BibitemOpen
  \bibfield  {author} {\bibinfo {author} {\bibfnamefont {S.~A.}\ \bibnamefont
  {Maurer}}, \bibinfo {author} {\bibfnamefont {L.}~\bibnamefont {Clin}},\ and\
  \bibinfo {author} {\bibfnamefont {C.}~\bibnamefont {Ochsenfeld}},\ }\bibfield
   {title} {\enquote {\bibinfo {title} {Cholesky-decomposed density {{MP2}}
  with density fitting: {{Accurate MP2}} and double-hybrid {{DFT}} energies for
  large systems},}\ }\href {https://doi.org/10.1063/1.4881144} {\bibfield
  {journal} {\bibinfo  {journal} {The Journal of Chemical Physics}\ }\textbf
  {\bibinfo {volume} {140}},\ \bibinfo {pages} {224112} (\bibinfo {year}
  {2014})}\BibitemShut {NoStop}%
\bibitem [{\citenamefont {Pinski}\ \emph {et~al.}(2015)\citenamefont {Pinski},
  \citenamefont {Riplinger}, \citenamefont {Valeev},\ and\ \citenamefont
  {Neese}}]{pinskiSparseMapsSystematic2015}%
  \BibitemOpen
  \bibfield  {author} {\bibinfo {author} {\bibfnamefont {P.}~\bibnamefont
  {Pinski}}, \bibinfo {author} {\bibfnamefont {C.}~\bibnamefont {Riplinger}},
  \bibinfo {author} {\bibfnamefont {E.~F.}\ \bibnamefont {Valeev}},\ and\
  \bibinfo {author} {\bibfnamefont {F.}~\bibnamefont {Neese}},\ }\bibfield
  {title} {\enquote {\bibinfo {title} {Sparse maps---{{A}} systematic
  infrastructure for reduced-scaling electronic structure methods. {{I}}.
  {{An}} efficient and simple linear scaling local {{MP2}} method that uses an
  intermediate basis of pair natural orbitals},}\ }\href
  {https://doi.org/10.1063/1.4926879} {\bibfield  {journal} {\bibinfo
  {journal} {The Journal of Chemical Physics}\ }\textbf {\bibinfo {volume}
  {143}},\ \bibinfo {pages} {034108} (\bibinfo {year} {2015})}\BibitemShut
  {NoStop}%
\bibitem [{\citenamefont {Nagy}, \citenamefont {Samu},\ and\ \citenamefont
  {K{\'a}llay}(2016)}]{nagyIntegralDirectLinearScalingSecondOrder2016}%
  \BibitemOpen
  \bibfield  {author} {\bibinfo {author} {\bibfnamefont {P.~R.}\ \bibnamefont
  {Nagy}}, \bibinfo {author} {\bibfnamefont {G.}~\bibnamefont {Samu}},\ and\
  \bibinfo {author} {\bibfnamefont {M.}~\bibnamefont {K{\'a}llay}},\ }\bibfield
   {title} {\enquote {\bibinfo {title} {An {{Integral-Direct Linear-Scaling
  Second-Order M{\o}ller}}--{{Plesset Approach}}},}\ }\href
  {https://doi.org/10.1021/acs.jctc.6b00732} {\bibfield  {journal} {\bibinfo
  {journal} {Journal of Chemical Theory and Computation}\ }\textbf {\bibinfo
  {volume} {12}},\ \bibinfo {pages} {4897--4914} (\bibinfo {year}
  {2016})}\BibitemShut {NoStop}%
\bibitem [{\citenamefont {Baudin}\ \emph {et~al.}(2016)\citenamefont {Baudin},
  \citenamefont {Ettenhuber}, \citenamefont {Reine}, \citenamefont
  {Kristensen},\ and\ \citenamefont
  {Kj{\ae}rgaard}}]{baudinEfficientLinearscalingSecondorder2016}%
  \BibitemOpen
  \bibfield  {author} {\bibinfo {author} {\bibfnamefont {P.}~\bibnamefont
  {Baudin}}, \bibinfo {author} {\bibfnamefont {P.}~\bibnamefont {Ettenhuber}},
  \bibinfo {author} {\bibfnamefont {S.}~\bibnamefont {Reine}}, \bibinfo
  {author} {\bibfnamefont {K.}~\bibnamefont {Kristensen}},\ and\ \bibinfo
  {author} {\bibfnamefont {T.}~\bibnamefont {Kj{\ae}rgaard}},\ }\bibfield
  {title} {\enquote {\bibinfo {title} {Efficient linear-scaling second-order
  {{M{\o}ller-Plesset}} perturbation theory: {{The}}
  divide--expand--consolidate {{RI-MP2}} model},}\ }\href
  {https://doi.org/10.1063/1.4940732} {\bibfield  {journal} {\bibinfo
  {journal} {The Journal of Chemical Physics}\ }\textbf {\bibinfo {volume}
  {144}},\ \bibinfo {pages} {054102} (\bibinfo {year} {2016})}\BibitemShut
  {NoStop}%
\bibitem [{\citenamefont {Pham}\ and\ \citenamefont
  {Gordon}(2019)}]{phamHybridDistributedShared2019}%
  \BibitemOpen
  \bibfield  {author} {\bibinfo {author} {\bibfnamefont {B.~Q.}\ \bibnamefont
  {Pham}}\ and\ \bibinfo {author} {\bibfnamefont {M.~S.}\ \bibnamefont
  {Gordon}},\ }\bibfield  {title} {\enquote {\bibinfo {title} {Hybrid
  {{Distributed}}/{{Shared Memory Model}} for the {{RI-MP2 Method}} in the
  {{Fragment Molecular Orbital Framework}}},}\ }\href
  {https://doi.org/10.1021/acs.jctc.9b00409} {\bibfield  {journal} {\bibinfo
  {journal} {Journal of Chemical Theory and Computation}\ }\textbf {\bibinfo
  {volume} {15}},\ \bibinfo {pages} {5252--5258} (\bibinfo {year}
  {2019})}\BibitemShut {NoStop}%
\bibitem [{\citenamefont {Barca}\ \emph {et~al.}(2020)\citenamefont {Barca},
  \citenamefont {McKenzie}, \citenamefont {Bloomfield}, \citenamefont
  {Gilbert},\ and\ \citenamefont {Gill}}]{barcaQMP2OSMollerPlesset2020}%
  \BibitemOpen
  \bibfield  {author} {\bibinfo {author} {\bibfnamefont {G.~M.~J.}\
  \bibnamefont {Barca}}, \bibinfo {author} {\bibfnamefont {S.~C.}\ \bibnamefont
  {McKenzie}}, \bibinfo {author} {\bibfnamefont {N.~J.}\ \bibnamefont
  {Bloomfield}}, \bibinfo {author} {\bibfnamefont {A.~T.~B.}\ \bibnamefont
  {Gilbert}},\ and\ \bibinfo {author} {\bibfnamefont {P.~M.~W.}\ \bibnamefont
  {Gill}},\ }\bibfield  {title} {\enquote {\bibinfo {title} {Q-{{MP2-OS}}:
  {{M{\o}ller}}--{{Plesset Correlation Energy}} by {{Quadrature}}},}\ }\href
  {https://doi.org/10.1021/acs.jctc.9b01142} {\bibfield  {journal} {\bibinfo
  {journal} {Journal of Chemical Theory and Computation}\ }\textbf {\bibinfo
  {volume} {16}},\ \bibinfo {pages} {1568--1577} (\bibinfo {year}
  {2020})}\BibitemShut {NoStop}%
\bibitem [{\citenamefont {F{\"o}rster}\ \emph {et~al.}(2020)\citenamefont
  {F{\"o}rster}, \citenamefont {Franchini}, \citenamefont {{van Lenthe}},\ and\
  \citenamefont {Visscher}}]{forsterQuadraticPairAtomic2020}%
  \BibitemOpen
  \bibfield  {author} {\bibinfo {author} {\bibfnamefont {A.}~\bibnamefont
  {F{\"o}rster}}, \bibinfo {author} {\bibfnamefont {M.}~\bibnamefont
  {Franchini}}, \bibinfo {author} {\bibfnamefont {E.}~\bibnamefont {{van
  Lenthe}}},\ and\ \bibinfo {author} {\bibfnamefont {L.}~\bibnamefont
  {Visscher}},\ }\bibfield  {title} {\enquote {\bibinfo {title} {A {{Quadratic
  Pair Atomic Resolution}} of the {{Identity Based SOS-AO-MP2 Algorithm Using
  Slater Type Orbitals}}},}\ }\href {https://doi.org/10.1021/acs.jctc.9b00854}
  {\bibfield  {journal} {\bibinfo  {journal} {Journal of Chemical Theory and
  Computation}\ }\textbf {\bibinfo {volume} {16}},\ \bibinfo {pages} {875--891}
  (\bibinfo {year} {2020})}\BibitemShut {NoStop}%
\bibitem [{\citenamefont {Nguyen}\ \emph {et~al.}(2020)\citenamefont {Nguyen},
  \citenamefont {Chen}, \citenamefont {Agee}, \citenamefont {Burow},
  \citenamefont {Tang},\ and\ \citenamefont
  {Furche}}]{nguyenDivergenceManyBodyPerturbation2020a}%
  \BibitemOpen
  \bibfield  {author} {\bibinfo {author} {\bibfnamefont {B.~D.}\ \bibnamefont
  {Nguyen}}, \bibinfo {author} {\bibfnamefont {G.~P.}\ \bibnamefont {Chen}},
  \bibinfo {author} {\bibfnamefont {M.~M.}\ \bibnamefont {Agee}}, \bibinfo
  {author} {\bibfnamefont {A.~M.}\ \bibnamefont {Burow}}, \bibinfo {author}
  {\bibfnamefont {M.~P.}\ \bibnamefont {Tang}},\ and\ \bibinfo {author}
  {\bibfnamefont {F.}~\bibnamefont {Furche}},\ }\bibfield  {title} {\enquote
  {\bibinfo {title} {Divergence of {{Many-Body Perturbation Theory}} for
  {{Noncovalent Interactions}} of {{Large Molecules}}},}\ }\href
  {https://doi.org/10.1021/acs.jctc.9b01176} {\bibfield  {journal} {\bibinfo
  {journal} {Journal of Chemical Theory and Computation}\ }\textbf {\bibinfo
  {volume} {16}},\ \bibinfo {pages} {2258--2273} (\bibinfo {year}
  {2020})}\BibitemShut {NoStop}%
\bibitem [{\citenamefont {Perdew}, \citenamefont {Burke},\ and\ \citenamefont
  {Ernzerhof}(1996)}]{perdewGeneralizedGradientApproximation1996}%
  \BibitemOpen
  \bibfield  {author} {\bibinfo {author} {\bibfnamefont {J.~P.}\ \bibnamefont
  {Perdew}}, \bibinfo {author} {\bibfnamefont {K.}~\bibnamefont {Burke}},\ and\
  \bibinfo {author} {\bibfnamefont {M.}~\bibnamefont {Ernzerhof}},\ }\bibfield
  {title} {\enquote {\bibinfo {title} {Generalized {{Gradient Approximation
  Made Simple}}},}\ }\href {https://doi.org/10.1103/PhysRevLett.77.3865}
  {\bibfield  {journal} {\bibinfo  {journal} {Physical Review Letters}\
  }\textbf {\bibinfo {volume} {77}},\ \bibinfo {pages} {3865--3868} (\bibinfo
  {year} {1996})}\BibitemShut {NoStop}%
\bibitem [{\citenamefont {Perdew}, \citenamefont {Burke},\ and\ \citenamefont
  {Ernzerhof}(1997)}]{perdewGeneralizedGradientApproximation1997}%
  \BibitemOpen
  \bibfield  {author} {\bibinfo {author} {\bibfnamefont {J.~P.}\ \bibnamefont
  {Perdew}}, \bibinfo {author} {\bibfnamefont {K.}~\bibnamefont {Burke}},\ and\
  \bibinfo {author} {\bibfnamefont {M.}~\bibnamefont {Ernzerhof}},\ }\bibfield
  {title} {\enquote {\bibinfo {title} {Generalized {{Gradient Approximation
  Made Simple}} [{{Phys}}. {{Rev}}. {{Lett}}. 77, 3865 (1996)]},}\ }\href
  {https://doi.org/10.1103/PhysRevLett.78.1396} {\bibfield  {journal} {\bibinfo
   {journal} {Physical Review Letters}\ }\textbf {\bibinfo {volume} {78}},\
  \bibinfo {pages} {1396--1396} (\bibinfo {year} {1997})}\BibitemShut {NoStop}%
\bibitem [{\citenamefont {Nam}\ \emph {et~al.}(2021)\citenamefont {Nam},
  \citenamefont {Cho}, \citenamefont {Sim},\ and\ \citenamefont
  {Burke}}]{namExplainingFixingDFT2021}%
  \BibitemOpen
  \bibfield  {author} {\bibinfo {author} {\bibfnamefont {S.}~\bibnamefont
  {Nam}}, \bibinfo {author} {\bibfnamefont {E.}~\bibnamefont {Cho}}, \bibinfo
  {author} {\bibfnamefont {E.}~\bibnamefont {Sim}},\ and\ \bibinfo {author}
  {\bibfnamefont {K.}~\bibnamefont {Burke}},\ }\bibfield  {title} {\enquote
  {\bibinfo {title} {Explaining and {{Fixing DFT Failures}} for {{Torsional
  Barriers}}},}\ }\href {https://doi.org/10.1021/acs.jpclett.1c00426}
  {\bibfield  {journal} {\bibinfo  {journal} {The Journal of Physical Chemistry
  Letters}\ }\textbf {\bibinfo {volume} {12}},\ \bibinfo {pages} {2796--2804}
  (\bibinfo {year} {2021})}\BibitemShut {NoStop}%
\bibitem [{\citenamefont {Nam}\ \emph {et~al.}(2020)\citenamefont {Nam},
  \citenamefont {Song}, \citenamefont {Sim},\ and\ \citenamefont
  {Burke}}]{namMeasuringDensityDrivenErrors2020}%
  \BibitemOpen
  \bibfield  {author} {\bibinfo {author} {\bibfnamefont {S.}~\bibnamefont
  {Nam}}, \bibinfo {author} {\bibfnamefont {S.}~\bibnamefont {Song}}, \bibinfo
  {author} {\bibfnamefont {E.}~\bibnamefont {Sim}},\ and\ \bibinfo {author}
  {\bibfnamefont {K.}~\bibnamefont {Burke}},\ }\bibfield  {title} {\enquote
  {\bibinfo {title} {Measuring {{Density-Driven Errors Using Kohn}}--{{Sham
  Inversion}}},}\ }\href {https://doi.org/10.1021/acs.jctc.0c00391} {\bibfield
  {journal} {\bibinfo  {journal} {Journal of Chemical Theory and Computation}\
  }\textbf {\bibinfo {volume} {16}},\ \bibinfo {pages} {5014--5023} (\bibinfo
  {year} {2020})}\BibitemShut {NoStop}%
\bibitem [{\citenamefont {Sim}, \citenamefont {Song},\ and\ \citenamefont
  {Burke}(2018)}]{simQuantifyingDensityErrors2018}%
  \BibitemOpen
  \bibfield  {author} {\bibinfo {author} {\bibfnamefont {E.}~\bibnamefont
  {Sim}}, \bibinfo {author} {\bibfnamefont {S.}~\bibnamefont {Song}},\ and\
  \bibinfo {author} {\bibfnamefont {K.}~\bibnamefont {Burke}},\ }\bibfield
  {title} {\enquote {\bibinfo {title} {Quantifying {{Density Errors}} in
  {{DFT}}},}\ }\href {https://doi.org/10.1021/acs.jpclett.8b02855} {\bibfield
  {journal} {\bibinfo  {journal} {The Journal of Physical Chemistry Letters}\
  }\textbf {\bibinfo {volume} {9}},\ \bibinfo {pages} {6385--6392} (\bibinfo
  {year} {2018})}\BibitemShut {NoStop}%
\bibitem [{\citenamefont {Sim}\ \emph {et~al.}(2022)\citenamefont {Sim},
  \citenamefont {Song}, \citenamefont {Vuckovic},\ and\ \citenamefont
  {Burke}}]{simImprovingResultsImproving2022}%
  \BibitemOpen
  \bibfield  {author} {\bibinfo {author} {\bibfnamefont {E.}~\bibnamefont
  {Sim}}, \bibinfo {author} {\bibfnamefont {S.}~\bibnamefont {Song}}, \bibinfo
  {author} {\bibfnamefont {S.}~\bibnamefont {Vuckovic}},\ and\ \bibinfo
  {author} {\bibfnamefont {K.}~\bibnamefont {Burke}},\ }\bibfield  {title}
  {\enquote {\bibinfo {title} {Improving {{Results}} by {{Improving
  Densities}}: {{Density-Corrected Density Functional Theory}}},}\ }\href
  {https://doi.org/10.1021/jacs.1c11506} {\bibfield  {journal} {\bibinfo
  {journal} {Journal of the American Chemical Society}\ }\textbf {\bibinfo
  {volume} {144}},\ \bibinfo {pages} {6625--6639} (\bibinfo {year}
  {2022})}\BibitemShut {NoStop}%
\bibitem [{\citenamefont {Vuckovic}\ \emph {et~al.}(2019)\citenamefont
  {Vuckovic}, \citenamefont {Song}, \citenamefont {Kozlowski}, \citenamefont
  {Sim},\ and\ \citenamefont {Burke}}]{vuckovicDensityFunctionalAnalysis2019}%
  \BibitemOpen
  \bibfield  {author} {\bibinfo {author} {\bibfnamefont {S.}~\bibnamefont
  {Vuckovic}}, \bibinfo {author} {\bibfnamefont {S.}~\bibnamefont {Song}},
  \bibinfo {author} {\bibfnamefont {J.}~\bibnamefont {Kozlowski}}, \bibinfo
  {author} {\bibfnamefont {E.}~\bibnamefont {Sim}},\ and\ \bibinfo {author}
  {\bibfnamefont {K.}~\bibnamefont {Burke}},\ }\bibfield  {title} {\enquote
  {\bibinfo {title} {Density {{Functional Analysis}}: {{The Theory}} of
  {{Density-Corrected DFT}}},}\ }\href
  {https://doi.org/10.1021/acs.jctc.9b00826} {\bibfield  {journal} {\bibinfo
  {journal} {Journal of Chemical Theory and Computation}\ }\textbf {\bibinfo
  {volume} {15}},\ \bibinfo {pages} {6636--6646} (\bibinfo {year}
  {2019})}\BibitemShut {NoStop}%
\bibitem [{\citenamefont {Song}\ \emph {et~al.}(2021)\citenamefont {Song},
  \citenamefont {Vuckovic}, \citenamefont {Sim},\ and\ \citenamefont
  {Burke}}]{songDensitySensitivityEmpirical2021}%
  \BibitemOpen
  \bibfield  {author} {\bibinfo {author} {\bibfnamefont {S.}~\bibnamefont
  {Song}}, \bibinfo {author} {\bibfnamefont {S.}~\bibnamefont {Vuckovic}},
  \bibinfo {author} {\bibfnamefont {E.}~\bibnamefont {Sim}},\ and\ \bibinfo
  {author} {\bibfnamefont {K.}~\bibnamefont {Burke}},\ }\bibfield  {title}
  {\enquote {\bibinfo {title} {Density {{Sensitivity}} of {{Empirical
  Functionals}}},}\ }\href {https://doi.org/10.1021/acs.jpclett.0c03545}
  {\bibfield  {journal} {\bibinfo  {journal} {The Journal of Physical Chemistry
  Letters}\ }\textbf {\bibinfo {volume} {12}},\ \bibinfo {pages} {800--807}
  (\bibinfo {year} {2021})}\BibitemShut {NoStop}%
\bibitem [{\citenamefont {Kim}, \citenamefont {Sim},\ and\ \citenamefont
  {Burke}(2013)}]{kimUnderstandingReducingErrors2013}%
  \BibitemOpen
  \bibfield  {author} {\bibinfo {author} {\bibfnamefont {M.-C.}\ \bibnamefont
  {Kim}}, \bibinfo {author} {\bibfnamefont {E.}~\bibnamefont {Sim}},\ and\
  \bibinfo {author} {\bibfnamefont {K.}~\bibnamefont {Burke}},\ }\bibfield
  {title} {\enquote {\bibinfo {title} {Understanding and {{Reducing Errors}} in
  {{Density Functional Calculations}}},}\ }\href
  {https://doi.org/10.1103/PhysRevLett.111.073003} {\bibfield  {journal}
  {\bibinfo  {journal} {Physical Review Letters}\ }\textbf {\bibinfo {volume}
  {111}},\ \bibinfo {pages} {073003} (\bibinfo {year} {2013})}\BibitemShut
  {NoStop}%
\bibitem [{\citenamefont {{Mart{\'i}n-Fern{\'a}ndez}}\ and\ \citenamefont
  {Harvey}(2021)}]{martin-fernandezUseNormalizedMetrics2021}%
  \BibitemOpen
  \bibfield  {author} {\bibinfo {author} {\bibfnamefont {C.}~\bibnamefont
  {{Mart{\'i}n-Fern{\'a}ndez}}}\ and\ \bibinfo {author} {\bibfnamefont {J.~N.}\
  \bibnamefont {Harvey}},\ }\bibfield  {title} {\enquote {\bibinfo {title} {On
  the {{Use}} of {{Normalized Metrics}} for {{Density Sensitivity Analysis}} in
  {{DFT}}},}\ }\href {https://doi.org/10.1021/acs.jpca.1c01290} {\bibfield
  {journal} {\bibinfo  {journal} {The Journal of Physical Chemistry A}\
  }\textbf {\bibinfo {volume} {125}},\ \bibinfo {pages} {4639--4652} (\bibinfo
  {year} {2021})}\BibitemShut {NoStop}%
\bibitem [{\citenamefont {Kim}\ \emph {et~al.}(2019)\citenamefont {Kim},
  \citenamefont {Song}, \citenamefont {Sim},\ and\ \citenamefont
  {Burke}}]{kimHalogenChalcogenBinding2019}%
  \BibitemOpen
  \bibfield  {author} {\bibinfo {author} {\bibfnamefont {Y.}~\bibnamefont
  {Kim}}, \bibinfo {author} {\bibfnamefont {S.}~\bibnamefont {Song}}, \bibinfo
  {author} {\bibfnamefont {E.}~\bibnamefont {Sim}},\ and\ \bibinfo {author}
  {\bibfnamefont {K.}~\bibnamefont {Burke}},\ }\bibfield  {title} {\enquote
  {\bibinfo {title} {Halogen and {{Chalcogen Binding Dominated}} by
  {{Density-Driven Errors}}},}\ }\href
  {https://doi.org/10.1021/acs.jpclett.8b03745} {\bibfield  {journal} {\bibinfo
   {journal} {The Journal of Physical Chemistry Letters}\ }\textbf {\bibinfo
  {volume} {10}},\ \bibinfo {pages} {295--301} (\bibinfo {year}
  {2019})}\BibitemShut {NoStop}%
\bibitem [{\citenamefont {Kim}, \citenamefont {Sim},\ and\ \citenamefont
  {Burke}(2014)}]{kimIonsSolutionDensity2014}%
  \BibitemOpen
  \bibfield  {author} {\bibinfo {author} {\bibfnamefont {M.-C.}\ \bibnamefont
  {Kim}}, \bibinfo {author} {\bibfnamefont {E.}~\bibnamefont {Sim}},\ and\
  \bibinfo {author} {\bibfnamefont {K.}~\bibnamefont {Burke}},\ }\bibfield
  {title} {\enquote {\bibinfo {title} {Ions in solution: {{Density}} corrected
  density functional theory ({{DC-DFT}})},}\ }\href
  {https://doi.org/10.1063/1.4869189} {\bibfield  {journal} {\bibinfo
  {journal} {The Journal of Chemical Physics}\ }\textbf {\bibinfo {volume}
  {140}},\ \bibinfo {pages} {18A528} (\bibinfo {year} {2014})}\BibitemShut
  {NoStop}%
\bibitem [{\citenamefont {Wasserman}\ \emph {et~al.}(2017)\citenamefont
  {Wasserman}, \citenamefont {Nafziger}, \citenamefont {Jiang}, \citenamefont
  {Kim}, \citenamefont {Sim},\ and\ \citenamefont
  {Burke}}]{wassermanImportanceBeingInconsistent2017}%
  \BibitemOpen
  \bibfield  {author} {\bibinfo {author} {\bibfnamefont {A.}~\bibnamefont
  {Wasserman}}, \bibinfo {author} {\bibfnamefont {J.}~\bibnamefont {Nafziger}},
  \bibinfo {author} {\bibfnamefont {K.}~\bibnamefont {Jiang}}, \bibinfo
  {author} {\bibfnamefont {M.-C.}\ \bibnamefont {Kim}}, \bibinfo {author}
  {\bibfnamefont {E.}~\bibnamefont {Sim}},\ and\ \bibinfo {author}
  {\bibfnamefont {K.}~\bibnamefont {Burke}},\ }\bibfield  {title} {\enquote
  {\bibinfo {title} {The {{Importance}} of {{Being Inconsistent}}},}\ }\href
  {https://doi.org/10.1146/annurev-physchem-052516-044957} {\bibfield
  {journal} {\bibinfo  {journal} {Annual Review of Physical Chemistry}\
  }\textbf {\bibinfo {volume} {68}},\ \bibinfo {pages} {555--581} (\bibinfo
  {year} {2017})}\BibitemShut {NoStop}%
\bibitem [{\citenamefont {Eshuis}\ and\ \citenamefont
  {Furche}(2012)}]{eshuisBasisSetConvergence2012}%
  \BibitemOpen
  \bibfield  {author} {\bibinfo {author} {\bibfnamefont {H.}~\bibnamefont
  {Eshuis}}\ and\ \bibinfo {author} {\bibfnamefont {F.}~\bibnamefont
  {Furche}},\ }\bibfield  {title} {\enquote {\bibinfo {title} {Basis set
  convergence of molecular correlation energy differences within the random
  phase approximation},}\ }\href {https://doi.org/10.1063/1.3687005} {\bibfield
   {journal} {\bibinfo  {journal} {The Journal of Chemical Physics}\ }\textbf
  {\bibinfo {volume} {136}},\ \bibinfo {pages} {084105} (\bibinfo {year}
  {2012})}\BibitemShut {NoStop}%
\bibitem [{\citenamefont {Feller}(1993)}]{fellerUseSystematicSequences1993}%
  \BibitemOpen
  \bibfield  {author} {\bibinfo {author} {\bibfnamefont {D.}~\bibnamefont
  {Feller}},\ }\bibfield  {title} {\enquote {\bibinfo {title} {The use of
  systematic sequences of wave functions for estimating the complete basis set,
  full configuration interaction limit in water},}\ }\href
  {https://doi.org/10.1063/1.464749} {\bibfield  {journal} {\bibinfo  {journal}
  {The Journal of Chemical Physics}\ }\textbf {\bibinfo {volume} {98}},\
  \bibinfo {pages} {7059--7071} (\bibinfo {year} {1993})}\BibitemShut {NoStop}%
\bibitem [{\citenamefont {Feller}, \citenamefont {Peterson},\ and\
  \citenamefont {Grant~Hill}(2011)}]{fellerEffectivenessCCSDComplete2011}%
  \BibitemOpen
  \bibfield  {author} {\bibinfo {author} {\bibfnamefont {D.}~\bibnamefont
  {Feller}}, \bibinfo {author} {\bibfnamefont {K.~A.}\ \bibnamefont
  {Peterson}},\ and\ \bibinfo {author} {\bibfnamefont {J.}~\bibnamefont
  {Grant~Hill}},\ }\bibfield  {title} {\enquote {\bibinfo {title} {On the
  effectiveness of {{CCSD}}({{T}}) complete basis set extrapolations for
  atomization energies},}\ }\href {https://doi.org/10.1063/1.3613639}
  {\bibfield  {journal} {\bibinfo  {journal} {The Journal of Chemical Physics}\
  }\textbf {\bibinfo {volume} {135}},\ \bibinfo {pages} {044102} (\bibinfo
  {year} {2011})}\BibitemShut {NoStop}%
\bibitem [{\citenamefont {Halkier}\ \emph {et~al.}(1998)\citenamefont
  {Halkier}, \citenamefont {Helgaker}, \citenamefont {J{\o}rgensen},
  \citenamefont {Klopper}, \citenamefont {Koch}, \citenamefont {Olsen},\ and\
  \citenamefont {Wilson}}]{halkierBasissetConvergenceCorrelated1998}%
  \BibitemOpen
  \bibfield  {author} {\bibinfo {author} {\bibfnamefont {A.}~\bibnamefont
  {Halkier}}, \bibinfo {author} {\bibfnamefont {T.}~\bibnamefont {Helgaker}},
  \bibinfo {author} {\bibfnamefont {P.}~\bibnamefont {J{\o}rgensen}}, \bibinfo
  {author} {\bibfnamefont {W.}~\bibnamefont {Klopper}}, \bibinfo {author}
  {\bibfnamefont {H.}~\bibnamefont {Koch}}, \bibinfo {author} {\bibfnamefont
  {J.}~\bibnamefont {Olsen}},\ and\ \bibinfo {author} {\bibfnamefont {A.~K.}\
  \bibnamefont {Wilson}},\ }\bibfield  {title} {\enquote {\bibinfo {title}
  {Basis-set convergence in correlated calculations on {{Ne}}, {{N2}}, and
  {{H2O}}},}\ }\href {https://doi.org/10.1016/S0009-2614(98)00111-0} {\bibfield
   {journal} {\bibinfo  {journal} {Chemical Physics Letters}\ }\textbf
  {\bibinfo {volume} {286}},\ \bibinfo {pages} {243--252} (\bibinfo {year}
  {1998})}\BibitemShut {NoStop}%
\bibitem [{\citenamefont {Helgaker}\ \emph {et~al.}(1997)\citenamefont
  {Helgaker}, \citenamefont {Klopper}, \citenamefont {Koch},\ and\
  \citenamefont {Noga}}]{helgakerBasissetConvergenceCorrelated1997}%
  \BibitemOpen
  \bibfield  {author} {\bibinfo {author} {\bibfnamefont {T.}~\bibnamefont
  {Helgaker}}, \bibinfo {author} {\bibfnamefont {W.}~\bibnamefont {Klopper}},
  \bibinfo {author} {\bibfnamefont {H.}~\bibnamefont {Koch}},\ and\ \bibinfo
  {author} {\bibfnamefont {J.}~\bibnamefont {Noga}},\ }\bibfield  {title}
  {\enquote {\bibinfo {title} {Basis-set convergence of correlated calculations
  on water},}\ }\href {https://doi.org/10.1063/1.473863} {\bibfield  {journal}
  {\bibinfo  {journal} {The Journal of Chemical Physics}\ }\textbf {\bibinfo
  {volume} {106}},\ \bibinfo {pages} {9639--9646} (\bibinfo {year}
  {1997})}\BibitemShut {NoStop}%
\bibitem [{\citenamefont {Balabanov}\ and\ \citenamefont
  {Peterson}(2005)}]{balabanovSystematicallyConvergentBasis2005}%
  \BibitemOpen
  \bibfield  {author} {\bibinfo {author} {\bibfnamefont {N.~B.}\ \bibnamefont
  {Balabanov}}\ and\ \bibinfo {author} {\bibfnamefont {K.~A.}\ \bibnamefont
  {Peterson}},\ }\bibfield  {title} {\enquote {\bibinfo {title} {Systematically
  convergent basis sets for transition metals. {{I}}. {{All-electron}}
  correlation consistent basis sets for the 3d elements {{Sc}}--{{Zn}}},}\
  }\href {https://doi.org/10.1063/1.1998907} {\bibfield  {journal} {\bibinfo
  {journal} {The Journal of Chemical Physics}\ }\textbf {\bibinfo {volume}
  {123}},\ \bibinfo {pages} {064107} (\bibinfo {year} {2005})}\BibitemShut
  {NoStop}%
\bibitem [{\citenamefont {Dunning}(1989)}]{dunningGaussianBasisSets1989}%
  \BibitemOpen
  \bibfield  {author} {\bibinfo {author} {\bibfnamefont {T.~H.}\ \bibnamefont
  {Dunning}, \bibfnamefont {Jr.}},\ }\bibfield  {title} {\enquote {\bibinfo
  {title} {Gaussian basis sets for use in correlated molecular calculations.
  {{I}}. {{The}} atoms boron through neon and hydrogen},}\ }\href
  {https://doi.org/10.1063/1.456153} {\bibfield  {journal} {\bibinfo  {journal}
  {The Journal of Chemical Physics}\ }\textbf {\bibinfo {volume} {90}},\
  \bibinfo {pages} {1007--1023} (\bibinfo {year} {1989})}\BibitemShut {NoStop}%
\bibitem [{\citenamefont {Koput}\ and\ \citenamefont
  {Peterson}(2002)}]{koputInitioPotentialEnergy2002}%
  \BibitemOpen
  \bibfield  {author} {\bibinfo {author} {\bibfnamefont {J.}~\bibnamefont
  {Koput}}\ and\ \bibinfo {author} {\bibfnamefont {K.~A.}\ \bibnamefont
  {Peterson}},\ }\bibfield  {title} {\enquote {\bibinfo {title} {Ab {{Initio
  Potential Energy Surface}} and {{Vibrational}}-{{Rotational Energy Levels}}
  of {{X2$\Sigma$}}+ {{CaOH}}},}\ }\href {https://doi.org/10.1021/jp026283u}
  {\bibfield  {journal} {\bibinfo  {journal} {The Journal of Physical Chemistry
  A}\ }\textbf {\bibinfo {volume} {106}},\ \bibinfo {pages} {9595--9599}
  (\bibinfo {year} {2002})}\BibitemShut {NoStop}%
\bibitem [{\citenamefont {Prascher}\ \emph {et~al.}(2011)\citenamefont
  {Prascher}, \citenamefont {Woon}, \citenamefont {Peterson}, \citenamefont
  {Dunning},\ and\ \citenamefont {Wilson}}]{prascherGaussianBasisSets2011}%
  \BibitemOpen
  \bibfield  {author} {\bibinfo {author} {\bibfnamefont {B.~P.}\ \bibnamefont
  {Prascher}}, \bibinfo {author} {\bibfnamefont {D.~E.}\ \bibnamefont {Woon}},
  \bibinfo {author} {\bibfnamefont {K.~A.}\ \bibnamefont {Peterson}}, \bibinfo
  {author} {\bibfnamefont {T.~H.}\ \bibnamefont {Dunning}},\ and\ \bibinfo
  {author} {\bibfnamefont {A.~K.}\ \bibnamefont {Wilson}},\ }\bibfield  {title}
  {\enquote {\bibinfo {title} {Gaussian basis sets for use in correlated
  molecular calculations. {{VII}}. {{Valence}}, core-valence, and scalar
  relativistic basis sets for {{Li}}, {{Be}}, {{Na}}, and {{Mg}}},}\ }\href
  {https://doi.org/10.1007/s00214-010-0764-0} {\bibfield  {journal} {\bibinfo
  {journal} {Theoretical Chemistry Accounts}\ }\textbf {\bibinfo {volume}
  {128}},\ \bibinfo {pages} {69--82} (\bibinfo {year} {2011})}\BibitemShut
  {NoStop}%
\bibitem [{\citenamefont {Wilson}\ \emph {et~al.}(1999)\citenamefont {Wilson},
  \citenamefont {Woon}, \citenamefont {Peterson},\ and\ \citenamefont
  {Dunning}}]{wilsonGaussianBasisSets1999}%
  \BibitemOpen
  \bibfield  {author} {\bibinfo {author} {\bibfnamefont {A.~K.}\ \bibnamefont
  {Wilson}}, \bibinfo {author} {\bibfnamefont {D.~E.}\ \bibnamefont {Woon}},
  \bibinfo {author} {\bibfnamefont {K.~A.}\ \bibnamefont {Peterson}},\ and\
  \bibinfo {author} {\bibfnamefont {T.~H.}\ \bibnamefont {Dunning},
  \bibfnamefont {Jr.}},\ }\bibfield  {title} {\enquote {\bibinfo {title}
  {Gaussian basis sets for use in correlated molecular calculations. {{IX}}.
  {{The}} atoms gallium through krypton},}\ }\href
  {https://doi.org/10.1063/1.478678} {\bibfield  {journal} {\bibinfo  {journal}
  {The Journal of Chemical Physics}\ }\textbf {\bibinfo {volume} {110}},\
  \bibinfo {pages} {7667--7676} (\bibinfo {year} {1999})}\BibitemShut {NoStop}%
\bibitem [{\citenamefont {Woon}\ and\ \citenamefont
  {Dunning}(1993)}]{woonGaussianBasisSets1993}%
  \BibitemOpen
  \bibfield  {author} {\bibinfo {author} {\bibfnamefont {D.~E.}\ \bibnamefont
  {Woon}}\ and\ \bibinfo {author} {\bibfnamefont {T.~H.}\ \bibnamefont
  {Dunning}, \bibfnamefont {Jr.}},\ }\bibfield  {title} {\enquote {\bibinfo
  {title} {Gaussian basis sets for use in correlated molecular calculations.
  {{III}}. {{The}} atoms aluminum through argon},}\ }\href
  {https://doi.org/10.1063/1.464303} {\bibfield  {journal} {\bibinfo  {journal}
  {The Journal of Chemical Physics}\ }\textbf {\bibinfo {volume} {98}},\
  \bibinfo {pages} {1358--1371} (\bibinfo {year} {1993})}\BibitemShut {NoStop}%
\bibitem [{\citenamefont {Woon}\ and\ \citenamefont
  {Dunning}(1994)}]{woonGaussianBasisSets1994}%
  \BibitemOpen
  \bibfield  {author} {\bibinfo {author} {\bibfnamefont {D.~E.}\ \bibnamefont
  {Woon}}\ and\ \bibinfo {author} {\bibfnamefont {T.~H.}\ \bibnamefont
  {Dunning}, \bibfnamefont {Jr.}},\ }\bibfield  {title} {\enquote {\bibinfo
  {title} {Gaussian basis sets for use in correlated molecular calculations.
  {{IV}}. {{Calculation}} of static electrical response properties},}\ }\href
  {https://doi.org/10.1063/1.466439} {\bibfield  {journal} {\bibinfo  {journal}
  {The Journal of Chemical Physics}\ }\textbf {\bibinfo {volume} {100}},\
  \bibinfo {pages} {2975--2988} (\bibinfo {year} {1994})}\BibitemShut {NoStop}%
\bibitem [{\citenamefont {Price}, \citenamefont {Bryenton},\ and\ \citenamefont
  {Johnson}(2021)}]{priceRequirementsAccurateDispersioncorrected2021}%
  \BibitemOpen
  \bibfield  {author} {\bibinfo {author} {\bibfnamefont {A.~J.~A.}\
  \bibnamefont {Price}}, \bibinfo {author} {\bibfnamefont {K.~R.}\ \bibnamefont
  {Bryenton}},\ and\ \bibinfo {author} {\bibfnamefont {E.~R.}\ \bibnamefont
  {Johnson}},\ }\bibfield  {title} {\enquote {\bibinfo {title} {Requirements
  for an accurate dispersion-corrected density functional},}\ }\href
  {https://doi.org/10.1063/5.0050993} {\bibfield  {journal} {\bibinfo
  {journal} {The Journal of Chemical Physics}\ }\textbf {\bibinfo {volume}
  {154}},\ \bibinfo {pages} {230902} (\bibinfo {year} {2021})}\BibitemShut
  {NoStop}%
\bibitem [{\citenamefont {Blum}\ \emph {et~al.}(2009)\citenamefont {Blum},
  \citenamefont {Gehrke}, \citenamefont {Hanke}, \citenamefont {Havu},
  \citenamefont {Havu}, \citenamefont {Ren}, \citenamefont {Reuter},\ and\
  \citenamefont {Scheffler}}]{blumInitioMolecularSimulations2009}%
  \BibitemOpen
  \bibfield  {author} {\bibinfo {author} {\bibfnamefont {V.}~\bibnamefont
  {Blum}}, \bibinfo {author} {\bibfnamefont {R.}~\bibnamefont {Gehrke}},
  \bibinfo {author} {\bibfnamefont {F.}~\bibnamefont {Hanke}}, \bibinfo
  {author} {\bibfnamefont {P.}~\bibnamefont {Havu}}, \bibinfo {author}
  {\bibfnamefont {V.}~\bibnamefont {Havu}}, \bibinfo {author} {\bibfnamefont
  {X.}~\bibnamefont {Ren}}, \bibinfo {author} {\bibfnamefont {K.}~\bibnamefont
  {Reuter}},\ and\ \bibinfo {author} {\bibfnamefont {M.}~\bibnamefont
  {Scheffler}},\ }\bibfield  {title} {\enquote {\bibinfo {title} {{\emph{Ab
  Initio}} molecular simulations with numeric atom-centered orbitals},}\ }\href
  {https://doi.org/10.1016/j.cpc.2009.06.022} {\bibfield  {journal} {\bibinfo
  {journal} {Computer Physics Communications}\ }\textbf {\bibinfo {volume}
  {180}},\ \bibinfo {pages} {2175--2196} (\bibinfo {year} {2009})}\BibitemShut
  {NoStop}%
\bibitem [{\citenamefont {Ihrig}\ \emph {et~al.}(2015)\citenamefont {Ihrig},
  \citenamefont {Wieferink}, \citenamefont {Zhang}, \citenamefont {Ropo},
  \citenamefont {Ren}, \citenamefont {Rinke}, \citenamefont {Scheffler},\ and\
  \citenamefont {Blum}}]{ihrigAccurateLocalizedResolution2015}%
  \BibitemOpen
  \bibfield  {author} {\bibinfo {author} {\bibfnamefont {A.~C.}\ \bibnamefont
  {Ihrig}}, \bibinfo {author} {\bibfnamefont {J.}~\bibnamefont {Wieferink}},
  \bibinfo {author} {\bibfnamefont {I.~Y.}\ \bibnamefont {Zhang}}, \bibinfo
  {author} {\bibfnamefont {M.}~\bibnamefont {Ropo}}, \bibinfo {author}
  {\bibfnamefont {X.}~\bibnamefont {Ren}}, \bibinfo {author} {\bibfnamefont
  {P.}~\bibnamefont {Rinke}}, \bibinfo {author} {\bibfnamefont
  {M.}~\bibnamefont {Scheffler}},\ and\ \bibinfo {author} {\bibfnamefont
  {V.}~\bibnamefont {Blum}},\ }\bibfield  {title} {\enquote {\bibinfo {title}
  {Accurate localized resolution of identity approach for linear-scaling hybrid
  density functionals and for many-body perturbation theory},}\ }\href
  {https://doi.org/10.1088/1367-2630/17/9/093020} {\bibfield  {journal}
  {\bibinfo  {journal} {New Journal of Physics}\ }\textbf {\bibinfo {volume}
  {17}},\ \bibinfo {pages} {093020} (\bibinfo {year} {2015})}\BibitemShut
  {NoStop}%
\bibitem [{\citenamefont {Price}, \citenamefont {{Otero-de-la-Roza}},\ and\
  \citenamefont {Johnson}(2023)}]{priceXDMcorrectedHybridDFT2023}%
  \BibitemOpen
  \bibfield  {author} {\bibinfo {author} {\bibfnamefont {A.~J.~A.}\
  \bibnamefont {Price}}, \bibinfo {author} {\bibfnamefont {A.}~\bibnamefont
  {{Otero-de-la-Roza}}},\ and\ \bibinfo {author} {\bibfnamefont {E.~R.}\
  \bibnamefont {Johnson}},\ }\bibfield  {title} {\enquote {\bibinfo {title}
  {{{XDM-corrected}} hybrid {{DFT}} with numerical atomic orbitals predicts
  molecular crystal lattice energies with unprecedented accuracy},}\ }\href
  {https://doi.org/10.1039/D2SC05997E} {\bibfield  {journal} {\bibinfo
  {journal} {Chemical Science}\ }\textbf {\bibinfo {volume} {14}},\ \bibinfo
  {pages} {1252--1262} (\bibinfo {year} {2023})}\BibitemShut {NoStop}%
\bibitem [{\citenamefont {Baker}\ \emph {et~al.}(1991)\citenamefont {Baker},
  \citenamefont {Fowler}, \citenamefont {Lazzeretti}, \citenamefont
  {Malagoli},\ and\ \citenamefont {Zanasi}}]{bakerStructurePropertiesC701991}%
  \BibitemOpen
  \bibfield  {author} {\bibinfo {author} {\bibfnamefont {J.}~\bibnamefont
  {Baker}}, \bibinfo {author} {\bibfnamefont {P.~W.}\ \bibnamefont {Fowler}},
  \bibinfo {author} {\bibfnamefont {P.}~\bibnamefont {Lazzeretti}}, \bibinfo
  {author} {\bibfnamefont {M.}~\bibnamefont {Malagoli}},\ and\ \bibinfo
  {author} {\bibfnamefont {R.}~\bibnamefont {Zanasi}},\ }\bibfield  {title}
  {\enquote {\bibinfo {title} {Structure and properties of {{C70}}},}\ }\href
  {https://doi.org/10.1016/0009-2614(91)87184-D} {\bibfield  {journal}
  {\bibinfo  {journal} {Chemical Physics Letters}\ }\textbf {\bibinfo {volume}
  {184}},\ \bibinfo {pages} {182--186} (\bibinfo {year} {1991})}\BibitemShut
  {NoStop}%
\bibitem [{\citenamefont {Huynh}, \citenamefont {{Wibowo-Teale}},\ and\
  \citenamefont {{Wibowo-Teale}}(2024)}]{huynhQSym2QuantumSymbolic2024}%
  \BibitemOpen
  \bibfield  {author} {\bibinfo {author} {\bibfnamefont {B.~C.}\ \bibnamefont
  {Huynh}}, \bibinfo {author} {\bibfnamefont {M.}~\bibnamefont
  {{Wibowo-Teale}}},\ and\ \bibinfo {author} {\bibfnamefont {A.~M.}\
  \bibnamefont {{Wibowo-Teale}}},\ }\bibfield  {title} {\enquote {\bibinfo
  {title} {{{QSym2}}: {{A Quantum Symbolic Symmetry Analysis Program}} for
  {{Electronic Structure}}},}\ }\href
  {https://doi.org/10.1021/acs.jctc.3c01118} {\bibfield  {journal} {\bibinfo
  {journal} {Journal of Chemical Theory and Computation}\ }\textbf {\bibinfo
  {volume} {20}},\ \bibinfo {pages} {114--133} (\bibinfo {year}
  {2024})}\BibitemShut {NoStop}%
\bibitem [{\citenamefont {Light}, \citenamefont {Hamilton},\ and\ \citenamefont
  {Lill}(1985)}]{lightGeneralizedDiscreteVariable1985}%
  \BibitemOpen
  \bibfield  {author} {\bibinfo {author} {\bibfnamefont {J.~C.}\ \bibnamefont
  {Light}}, \bibinfo {author} {\bibfnamefont {I.~P.}\ \bibnamefont
  {Hamilton}},\ and\ \bibinfo {author} {\bibfnamefont {J.~V.}\ \bibnamefont
  {Lill}},\ }\bibfield  {title} {\enquote {\bibinfo {title} {Generalized
  discrete variable approximation in quantum mechanics},}\ }\href
  {https://doi.org/10.1063/1.448462} {\bibfield  {journal} {\bibinfo  {journal}
  {The Journal of Chemical Physics}\ }\textbf {\bibinfo {volume} {82}},\
  \bibinfo {pages} {1400--1409} (\bibinfo {year} {1985})}\BibitemShut {NoStop}%
\bibitem [{\citenamefont {Light}\ and\ \citenamefont
  {Carrington}(2000)}]{lightDiscreteVariableRepresentationsTheir2000}%
  \BibitemOpen
  \bibfield  {author} {\bibinfo {author} {\bibfnamefont {J.~C.}\ \bibnamefont
  {Light}}\ and\ \bibinfo {author} {\bibfnamefont {T.}~\bibnamefont
  {Carrington}},\ }\bibfield  {title} {\enquote {\bibinfo {title}
  {Discrete-{{Variable Representations}} and their {{Utilization}}},}\ }in\
  \href {https://doi.org/10.1002/9780470141731.ch4} {\emph {\bibinfo
  {booktitle} {Advances in {{Chemical Physics}}}}}\ (\bibinfo  {publisher}
  {John Wiley \& Sons, Ltd},\ \bibinfo {year} {2000})\ pp.\ \bibinfo {pages}
  {263--310}\BibitemShut {NoStop}%
\bibitem [{\citenamefont {Echave}\ and\ \citenamefont
  {Clary}(1992)}]{echavePotentialOptimizedDiscrete1992}%
  \BibitemOpen
  \bibfield  {author} {\bibinfo {author} {\bibfnamefont {J.}~\bibnamefont
  {Echave}}\ and\ \bibinfo {author} {\bibfnamefont {D.~C.}\ \bibnamefont
  {Clary}},\ }\bibfield  {title} {\enquote {\bibinfo {title} {Potential
  optimized discrete variable representation},}\ }\href
  {https://doi.org/10.1016/0009-2614(92)85330-D} {\bibfield  {journal}
  {\bibinfo  {journal} {Chemical Physics Letters}\ }\textbf {\bibinfo {volume}
  {190}},\ \bibinfo {pages} {225--230} (\bibinfo {year} {1992})}\BibitemShut
  {NoStop}%
\bibitem [{\citenamefont {Bryenton}\ \emph {et~al.}(2023)\citenamefont
  {Bryenton}, \citenamefont {Adeleke}, \citenamefont {Dale},\ and\
  \citenamefont {Johnson}}]{bryentonDelocalizationErrorGreatest2023}%
  \BibitemOpen
  \bibfield  {author} {\bibinfo {author} {\bibfnamefont {K.~R.}\ \bibnamefont
  {Bryenton}}, \bibinfo {author} {\bibfnamefont {A.~A.}\ \bibnamefont
  {Adeleke}}, \bibinfo {author} {\bibfnamefont {S.~G.}\ \bibnamefont {Dale}},\
  and\ \bibinfo {author} {\bibfnamefont {E.~R.}\ \bibnamefont {Johnson}},\
  }\bibfield  {title} {\enquote {\bibinfo {title} {Delocalization error:
  {{The}} greatest outstanding challenge in density-functional theory},}\
  }\href {https://doi.org/10.1002/wcms.1631} {\bibfield  {journal} {\bibinfo
  {journal} {WIREs Computational Molecular Science}\ }\textbf {\bibinfo
  {volume} {13}},\ \bibinfo {pages} {e1631} (\bibinfo {year}
  {2023})}\BibitemShut {NoStop}%
\bibitem [{\citenamefont {Gould}\ and\ \citenamefont
  {Dale}(2022)}]{gouldPoisoningDensityFunctional2022}%
  \BibitemOpen
  \bibfield  {author} {\bibinfo {author} {\bibfnamefont {T.}~\bibnamefont
  {Gould}}\ and\ \bibinfo {author} {\bibfnamefont {S.~G.}\ \bibnamefont
  {Dale}},\ }\bibfield  {title} {\enquote {\bibinfo {title} {Poisoning density
  functional theory with benchmark sets of difficult systems},}\ }\href
  {https://doi.org/10.1039/D2CP00268J} {\bibfield  {journal} {\bibinfo
  {journal} {Physical Chemistry Chemical Physics}\ }\textbf {\bibinfo {volume}
  {24}},\ \bibinfo {pages} {6398--6403} (\bibinfo {year} {2022})}\BibitemShut
  {NoStop}%
\bibitem [{\citenamefont {Khan}\ \emph {et~al.}(2024)\citenamefont {Khan},
  \citenamefont {Price}, \citenamefont {Ach},\ and\ \citenamefont {{von
  Lilienfeld}}}]{khanAdaptiveHybridDensity2024}%
  \BibitemOpen
  \bibfield  {author} {\bibinfo {author} {\bibfnamefont {D.}~\bibnamefont
  {Khan}}, \bibinfo {author} {\bibfnamefont {A.~J.~A.}\ \bibnamefont {Price}},
  \bibinfo {author} {\bibfnamefont {M.~L.}\ \bibnamefont {Ach}},\ and\ \bibinfo
  {author} {\bibfnamefont {O.~A.}\ \bibnamefont {{von Lilienfeld}}},\ }\href
  {https://doi.org/10.48550/arXiv.2402.14793} {\enquote {\bibinfo {title}
  {Adaptive hybrid density functionals},}\ } (\bibinfo {year} {2024}),\ \Eprint
  {https://arxiv.org/abs/2402.14793} {arxiv:2402.14793 [physics]} \BibitemShut
  {NoStop}%
\bibitem [{\citenamefont {Johnson}\ and\ \citenamefont
  {DiLabio}(2004)}]{johnsonTheoreticalStudyDispersionbound2004}%
  \BibitemOpen
  \bibfield  {author} {\bibinfo {author} {\bibfnamefont {E.~R.}\ \bibnamefont
  {Johnson}}\ and\ \bibinfo {author} {\bibfnamefont {G.~A.}\ \bibnamefont
  {DiLabio}},\ }\bibfield  {title} {\enquote {\bibinfo {title} {A theoretical
  study of the dispersion-bound silane--methane dimer},}\ }\href
  {https://doi.org/10.1016/j.cplett.2004.08.124} {\bibfield  {journal}
  {\bibinfo  {journal} {Chemical Physics Letters}\ }\textbf {\bibinfo {volume}
  {397}},\ \bibinfo {pages} {314--318} (\bibinfo {year} {2004})}\BibitemShut
  {NoStop}%
\bibitem [{\citenamefont {Panchagnula}, \citenamefont {Graf},\ and\
  \citenamefont {Johnson}(2024)}]{KripaPanchagnulaDataset2024}%
  \BibitemOpen
  \bibfield  {author} {\bibinfo {author} {\bibfnamefont {K.}~\bibnamefont
  {Panchagnula}}, \bibinfo {author} {\bibfnamefont {D.}~\bibnamefont {Graf}},\
  and\ \bibinfo {author} {\bibfnamefont {E.}~\bibnamefont {Johnson}},\
  }\href@noop {} {\enquote {\bibinfo {title} {Research data supporting
  "targeting spectroscopic accuracy for dispersion bound systems from
  \textit{ab initio} techniques: translational eigenstates of ne@c$_{70}$"},}\
  }\bibinfo {howpublished} {https://doi.org/10.17863/CAM.109314} (\bibinfo
  {year} {2024})\BibitemShut {NoStop}%
\end{thebibliography}%


\begin{thebibliography}{9}%
\makeatletter
\providecommand \@ifxundefined [1]{%
 \@ifx{#1\undefined}
}%
\providecommand \@ifnum [1]{%
 \ifnum #1\expandafter \@firstoftwo
 \else \expandafter \@secondoftwo
 \fi
}%
\providecommand \@ifx [1]{%
 \ifx #1\expandafter \@firstoftwo
 \else \expandafter \@secondoftwo
 \fi
}%
\providecommand \natexlab [1]{#1}%
\providecommand \enquote  [1]{``#1''}%
\providecommand \bibnamefont  [1]{#1}%
\providecommand \bibfnamefont [1]{#1}%
\providecommand \citenamefont [1]{#1}%
\providecommand \href@noop [0]{\@secondoftwo}%
\providecommand \href [0]{\begingroup \@sanitize@url \@href}%
\providecommand \@href[1]{\@@startlink{#1}\@@href}%
\providecommand \@@href[1]{\endgroup#1\@@endlink}%
\providecommand \@sanitize@url [0]{\catcode `\\12\catcode `\$12\catcode
  `\&12\catcode `\#12\catcode `\^12\catcode `\_12\catcode `\%12\relax}%
\providecommand \@@startlink[1]{}%
\providecommand \@@endlink[0]{}%
\providecommand \url  [0]{\begingroup\@sanitize@url \@url }%
\providecommand \@url [1]{\endgroup\@href {#1}{\urlprefix }}%
\providecommand \urlprefix  [0]{URL }%
\providecommand \Eprint [0]{\href }%
\providecommand \doibase [0]{https://doi.org/}%
\providecommand \selectlanguage [0]{\@gobble}%
\providecommand \bibinfo  [0]{\@secondoftwo}%
\providecommand \bibfield  [0]{\@secondoftwo}%
\providecommand \translation [1]{[#1]}%
\providecommand \BibitemOpen [0]{}%
\providecommand \bibitemStop [0]{}%
\providecommand \bibitemNoStop [0]{.\EOS\space}%
\providecommand \EOS [0]{\spacefactor3000\relax}%
\providecommand \BibitemShut  [1]{\csname bibitem#1\endcsname}%
\let\auto@bib@innerbib\@empty
\bibitem [{\citenamefont {Panchagnula}\ \emph {et~al.}(2024)\citenamefont
  {Panchagnula}, \citenamefont {Graf}, \citenamefont {Albertani},\ and\
  \citenamefont {Thom}}]{panchagnulaTranslationalEigenstatesHe2024a}%
  \BibitemOpen
  \bibfield  {author} {\bibinfo {author} {\bibfnamefont {K.}~\bibnamefont
  {Panchagnula}}, \bibinfo {author} {\bibfnamefont {D.}~\bibnamefont {Graf}},
  \bibinfo {author} {\bibfnamefont {F.~E.~A.}\ \bibnamefont {Albertani}},\ and\
  \bibinfo {author} {\bibfnamefont {A.~J.~W.}\ \bibnamefont {Thom}},\
  }\bibfield  {title} {\enquote {\bibinfo {title} {Translational eigenstates of
  {{He}}@{{C60}} from four-dimensional ab~initio potential energy surfaces
  interpolated using {{Gaussian}} process regression},}\ }\href
  {https://doi.org/10.1063/5.0197903} {\bibfield  {journal} {\bibinfo
  {journal} {The Journal of Chemical Physics}\ }\textbf {\bibinfo {volume}
  {160}},\ \bibinfo {pages} {104303} (\bibinfo {year} {2024})}\BibitemShut
  {NoStop}%
\bibitem [{\citenamefont {Bacanu}\ \emph {et~al.}(2021)\citenamefont {Bacanu},
  \citenamefont {Jafari}, \citenamefont {Aouane}, \citenamefont {Rantaharju},
  \citenamefont {Walkey}, \citenamefont {Hoffman}, \citenamefont {Shugai},
  \citenamefont {Nagel}, \citenamefont {{Jim{\'e}nez-Ruiz}}, \citenamefont
  {Horsewill}, \citenamefont {Rols}, \citenamefont {R{\~o}{\~o}m},
  \citenamefont {Whitby},\ and\ \citenamefont
  {Levitt}}]{bacanuExperimentalDeterminationInteraction2021c}%
  \BibitemOpen
  \bibfield  {author} {\bibinfo {author} {\bibfnamefont {G.~R.}\ \bibnamefont
  {Bacanu}}, \bibinfo {author} {\bibfnamefont {T.}~\bibnamefont {Jafari}},
  \bibinfo {author} {\bibfnamefont {M.}~\bibnamefont {Aouane}}, \bibinfo
  {author} {\bibfnamefont {J.}~\bibnamefont {Rantaharju}}, \bibinfo {author}
  {\bibfnamefont {M.}~\bibnamefont {Walkey}}, \bibinfo {author} {\bibfnamefont
  {G.}~\bibnamefont {Hoffman}}, \bibinfo {author} {\bibfnamefont
  {A.}~\bibnamefont {Shugai}}, \bibinfo {author} {\bibfnamefont
  {U.}~\bibnamefont {Nagel}}, \bibinfo {author} {\bibfnamefont
  {M.}~\bibnamefont {{Jim{\'e}nez-Ruiz}}}, \bibinfo {author} {\bibfnamefont
  {A.~J.}\ \bibnamefont {Horsewill}}, \bibinfo {author} {\bibfnamefont
  {S.}~\bibnamefont {Rols}}, \bibinfo {author} {\bibfnamefont {T.}~\bibnamefont
  {R{\~o}{\~o}m}}, \bibinfo {author} {\bibfnamefont {R.~J.}\ \bibnamefont
  {Whitby}},\ and\ \bibinfo {author} {\bibfnamefont {M.~H.}\ \bibnamefont
  {Levitt}},\ }\bibfield  {title} {\enquote {\bibinfo {title} {Experimental
  determination of the interaction potential between a helium atom and the
  interior surface of a {{C60}} fullerene molecule},}\ }\href
  {https://doi.org/10.1063/5.0066817} {\bibfield  {journal} {\bibinfo
  {journal} {The Journal of Chemical Physics}\ }\textbf {\bibinfo {volume}
  {155}},\ \bibinfo {pages} {144302} (\bibinfo {year} {2021})}\BibitemShut
  {NoStop}%
\bibitem [{\citenamefont {Jafari}\ \emph {et~al.}(2022)\citenamefont {Jafari},
  \citenamefont {Bacanu}, \citenamefont {Shugai}, \citenamefont {Nagel},
  \citenamefont {Walkey}, \citenamefont {Hoffman}, \citenamefont {Levitt},
  \citenamefont {Whitby},\ and\ \citenamefont
  {R{\~o}{\~o}m}}]{jafariTerahertzSpectroscopyHelium2022}%
  \BibitemOpen
  \bibfield  {author} {\bibinfo {author} {\bibfnamefont {T.}~\bibnamefont
  {Jafari}}, \bibinfo {author} {\bibfnamefont {G.~R.}\ \bibnamefont {Bacanu}},
  \bibinfo {author} {\bibfnamefont {A.}~\bibnamefont {Shugai}}, \bibinfo
  {author} {\bibfnamefont {U.}~\bibnamefont {Nagel}}, \bibinfo {author}
  {\bibfnamefont {M.}~\bibnamefont {Walkey}}, \bibinfo {author} {\bibfnamefont
  {G.}~\bibnamefont {Hoffman}}, \bibinfo {author} {\bibfnamefont {M.~H.}\
  \bibnamefont {Levitt}}, \bibinfo {author} {\bibfnamefont {R.~J.}\
  \bibnamefont {Whitby}},\ and\ \bibinfo {author} {\bibfnamefont
  {T.}~\bibnamefont {R{\~o}{\~o}m}},\ }\bibfield  {title} {\enquote {\bibinfo
  {title} {Terahertz spectroscopy of the helium endofullerene
  {{He}}@{{C60}}},}\ }\href {https://doi.org/10.1039/D2CP00515H} {\bibfield
  {journal} {\bibinfo  {journal} {Physical Chemistry Chemical Physics}\
  }\textbf {\bibinfo {volume} {24}},\ \bibinfo {pages} {9943--9952} (\bibinfo
  {year} {2022})}\BibitemShut {NoStop}%
\bibitem [{\citenamefont {Echave}\ and\ \citenamefont
  {Clary}(1992)}]{echavePotentialOptimizedDiscrete1992}%
  \BibitemOpen
  \bibfield  {author} {\bibinfo {author} {\bibfnamefont {J.}~\bibnamefont
  {Echave}}\ and\ \bibinfo {author} {\bibfnamefont {D.~C.}\ \bibnamefont
  {Clary}},\ }\bibfield  {title} {\enquote {\bibinfo {title} {Potential
  optimized discrete variable representation},}\ }\href
  {https://doi.org/10.1016/0009-2614(92)85330-D} {\bibfield  {journal}
  {\bibinfo  {journal} {Chemical Physics Letters}\ }\textbf {\bibinfo {volume}
  {190}},\ \bibinfo {pages} {225--230} (\bibinfo {year} {1992})}\BibitemShut
  {NoStop}%
\bibitem [{\citenamefont {Panchagnula}\ and\ \citenamefont
  {Thom}(2023)}]{panchagnulaExploringParameterSpace2023b}%
  \BibitemOpen
  \bibfield  {author} {\bibinfo {author} {\bibfnamefont {K.}~\bibnamefont
  {Panchagnula}}\ and\ \bibinfo {author} {\bibfnamefont {A.~J.~W.}\
  \bibnamefont {Thom}},\ }\bibfield  {title} {\enquote {\bibinfo {title}
  {Exploring the parameter space of an endohedral atom in a cylindrical
  cavity},}\ }\href {https://doi.org/10.1063/5.0170010} {\bibfield  {journal}
  {\bibinfo  {journal} {The Journal of Chemical Physics}\ }\textbf {\bibinfo
  {volume} {159}},\ \bibinfo {pages} {164308} (\bibinfo {year}
  {2023})}\BibitemShut {NoStop}%
\bibitem [{\citenamefont {Abramowitz}\ and\ \citenamefont
  {Stegun}(1983)}]{abramowitzHandbookMathematicalFunctions1983}%
  \BibitemOpen
  \bibfield  {author} {\bibinfo {author} {\bibfnamefont {M.}~\bibnamefont
  {Abramowitz}}\ and\ \bibinfo {author} {\bibfnamefont {I.}~\bibnamefont
  {Stegun}},\ }\href@noop {} {\emph {\bibinfo {title} {Handbook of
  {{Mathematical Functions}} with {{Formulas}}, {{Graphs}}, and {{Mathematical
  Tables}}}}}\ (\bibinfo  {publisher} {Dover Publications},\ \bibinfo {year}
  {1983})\BibitemShut {NoStop}%
\bibitem [{\citenamefont {Huynh}, \citenamefont {{Wibowo-Teale}},\ and\
  \citenamefont {{Wibowo-Teale}}(2024)}]{huynhQSym2QuantumSymbolic2024}%
  \BibitemOpen
  \bibfield  {author} {\bibinfo {author} {\bibfnamefont {B.~C.}\ \bibnamefont
  {Huynh}}, \bibinfo {author} {\bibfnamefont {M.}~\bibnamefont
  {{Wibowo-Teale}}},\ and\ \bibinfo {author} {\bibfnamefont {A.~M.}\
  \bibnamefont {{Wibowo-Teale}}},\ }\bibfield  {title} {\enquote {\bibinfo
  {title} {{{QSym2}}: {{A Quantum Symbolic Symmetry Analysis Program}} for
  {{Electronic Structure}}},}\ }\href
  {https://doi.org/10.1021/acs.jctc.3c01118} {\bibfield  {journal} {\bibinfo
  {journal} {Journal of Chemical Theory and Computation}\ }\textbf {\bibinfo
  {volume} {20}},\ \bibinfo {pages} {114--133} (\bibinfo {year}
  {2024})}\BibitemShut {NoStop}%
\bibitem [{\citenamefont {Pedregosa}\ \emph {et~al.}(2011)\citenamefont
  {Pedregosa}, \citenamefont {Varoquaux}, \citenamefont {Gramfort},
  \citenamefont {Michel}, \citenamefont {Thirion}, \citenamefont {Grisel},
  \citenamefont {Blondel}, \citenamefont {Prettenhofer}, \citenamefont {Weiss},
  \citenamefont {Dubourg}, \citenamefont {Vanderplas}, \citenamefont {Passos},
  \citenamefont {Cournapeau}, \citenamefont {Brucher}, \citenamefont {Perrot},\
  and\ \citenamefont {Duchesnay}}]{pedregosaScikitlearnMachineLearning2011}%
  \BibitemOpen
  \bibfield  {author} {\bibinfo {author} {\bibfnamefont {F.}~\bibnamefont
  {Pedregosa}}, \bibinfo {author} {\bibfnamefont {G.}~\bibnamefont
  {Varoquaux}}, \bibinfo {author} {\bibfnamefont {A.}~\bibnamefont {Gramfort}},
  \bibinfo {author} {\bibfnamefont {V.}~\bibnamefont {Michel}}, \bibinfo
  {author} {\bibfnamefont {B.}~\bibnamefont {Thirion}}, \bibinfo {author}
  {\bibfnamefont {O.}~\bibnamefont {Grisel}}, \bibinfo {author} {\bibfnamefont
  {M.}~\bibnamefont {Blondel}}, \bibinfo {author} {\bibfnamefont
  {P.}~\bibnamefont {Prettenhofer}}, \bibinfo {author} {\bibfnamefont
  {R.}~\bibnamefont {Weiss}}, \bibinfo {author} {\bibfnamefont
  {V.}~\bibnamefont {Dubourg}}, \bibinfo {author} {\bibfnamefont
  {J.}~\bibnamefont {Vanderplas}}, \bibinfo {author} {\bibfnamefont
  {A.}~\bibnamefont {Passos}}, \bibinfo {author} {\bibfnamefont
  {D.}~\bibnamefont {Cournapeau}}, \bibinfo {author} {\bibfnamefont
  {M.}~\bibnamefont {Brucher}}, \bibinfo {author} {\bibfnamefont
  {M.}~\bibnamefont {Perrot}},\ and\ \bibinfo {author} {\bibfnamefont
  {{\'E}.}~\bibnamefont {Duchesnay}},\ }\bibfield  {title} {\enquote {\bibinfo
  {title} {Scikit-learn: {{Machine Learning}} in {{Python}}},}\ }\href@noop {}
  {\bibfield  {journal} {\bibinfo  {journal} {Journal of Machine Learning
  Research}\ }\textbf {\bibinfo {volume} {12}},\ \bibinfo {pages} {2825--2830}
  (\bibinfo {year} {2011})}\BibitemShut {NoStop}%
\bibitem [{\citenamefont
  {Cioslowski}(2023)}]{cioslowskiElectronicStructureCalculations2023a}%
  \BibitemOpen
  \bibfield  {author} {\bibinfo {author} {\bibfnamefont {J.}~\bibnamefont
  {Cioslowski}},\ }\bibfield  {title} {\enquote {\bibinfo {title} {Electronic
  {{Structure Calculations}} on {{Endohedral Complexes}} of {{Fullerenes}}:
  {{Reminiscences}} and {{Prospects}}},}\ }\href
  {https://doi.org/10.3390/molecules28031384} {\bibfield  {journal} {\bibinfo
  {journal} {Molecules}\ }\textbf {\bibinfo {volume} {28}},\ \bibinfo {pages}
  {1384} (\bibinfo {year} {2023})}\BibitemShut {NoStop}%
\end{thebibliography}%

\end{document}


\title[SI: \Endo{Ne}{C70}]{Supplementary Information: Targeting spectroscopic accuracy for dispersion bound systems from \textit{ab initio} techniques: translational eigenstates of \Endo{Ne}{C70}}
    \author{K. Panchagnula*\orcidlink{0009-0004-3952-073X}}
    \email{ksp31@cam.ac.uk}
        \affiliation{Yusuf Hamied Department of Chemistry, University of Cambridge, Cambridge, United Kingdom}
    \author{D. Graf \orcidlink{0000-0002-7640-4162}}
        \affiliation{Yusuf Hamied Department of Chemistry, University of Cambridge, Cambridge, United Kingdom}
        \affiliation{Department of Chemistry, University of Munich (LMU), Munich, Germany}
    \author{E.R. Johnson \orcidlink{0000-0002-5651-468X}}
        \affiliation{Yusuf Hamied Department of Chemistry, University of Cambridge, Cambridge, United Kingdom}
        \affiliation{Department of Chemistry, Dalhousie University, 6243 Alumni Crescent, Halifax, Nova Scotia, Canada }
    \author{A.J.W. Thom \orcidlink{0000-0002-2417-7869}}
        \affiliation{Yusuf Hamied Department of Chemistry, University of Cambridge, Cambridge, United Kingdom}%
    
    \date{\today}

    \maketitle

    \section{Potential Energy Surface \label{sec:SI PES}}
        \begin{table}
           \centering
           \begin{tabular}{cccccc}
                \hline Method&$\sigma^2$&$l_x$/\AA&$l_y$/\AA&$l_z$\AA&$\nu^2$  \\
                \hline MP2&$11.7^2$&38.7&36.8&41.6&1e-9 \\
                SCS-MP2&$13.2^2$&40.5&38.5&41.5&1e-9\\
                SOS-MP2&$13.8^2$&41.1&39.1&44.1&1e-9\\
                RPA@PBE&$12.6^2$&39.6&37.7&42.4&1e-9\\
                C(HF)-RPA&$12.8^2$&39.7&37.8&42.6&1e-9\\
                B86bPBE-25X-XDM&$13.6^2$&41&38.9&44.1&1e-9\\
                B86bPBE-50X-XDM&$12.1^2$&39.1&37&42.1&1e-9\\
                \hline
           \end{tabular}
           \caption{Hyperparameters found while training the GP on the unsymmetrised data for each ES method.}
           \label{table:SI Hyperparameters}
       \end{table}
        
        \begin{figure*}
            \centering
            \setcounter{subfigure}{0}
            \subfloat[SCS-MP2]{\includegraphics[scale=0.42]{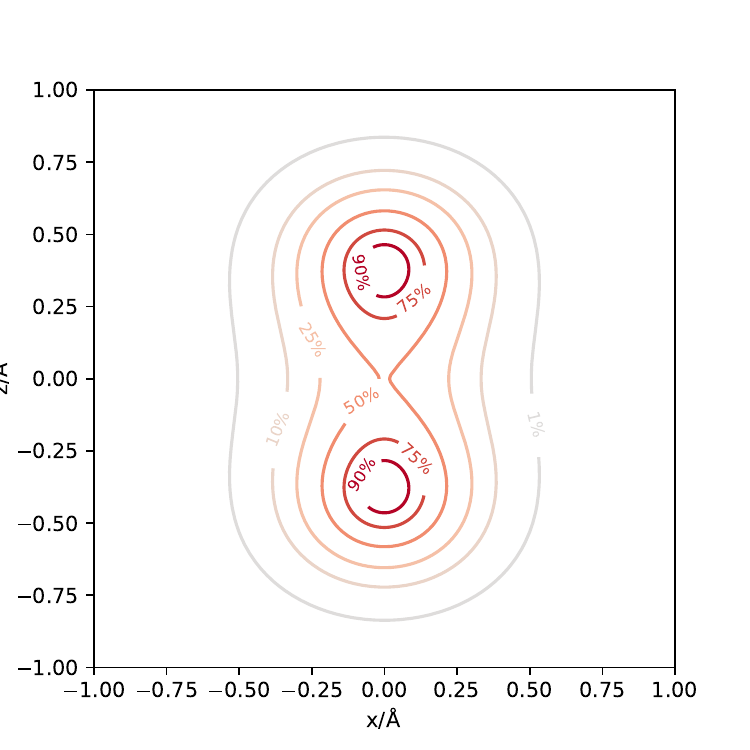}}\,
            \subfloat[SOS-MP2]{\includegraphics[scale=0.42]{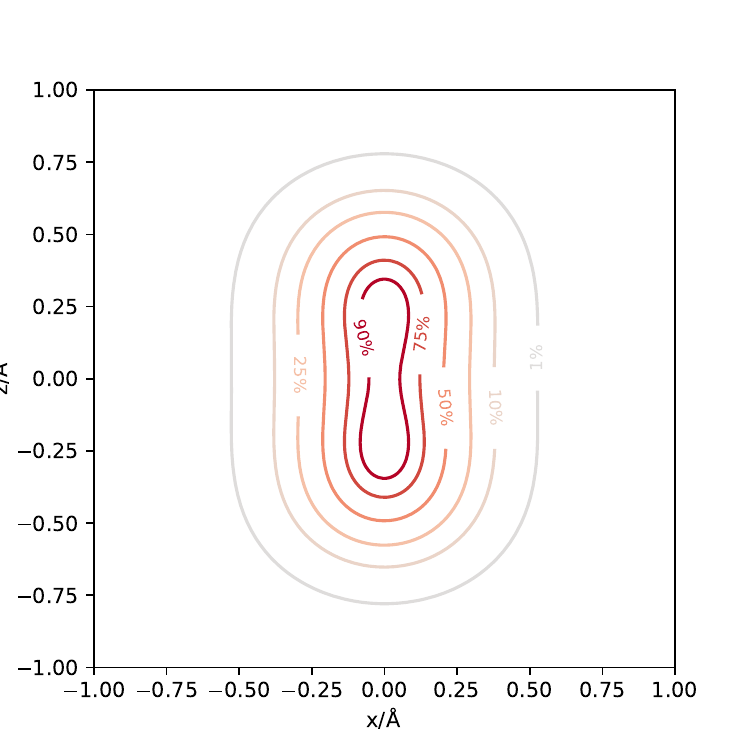}}\\
            \subfloat[C(HF)-RPA]{\includegraphics[scale=0.42]{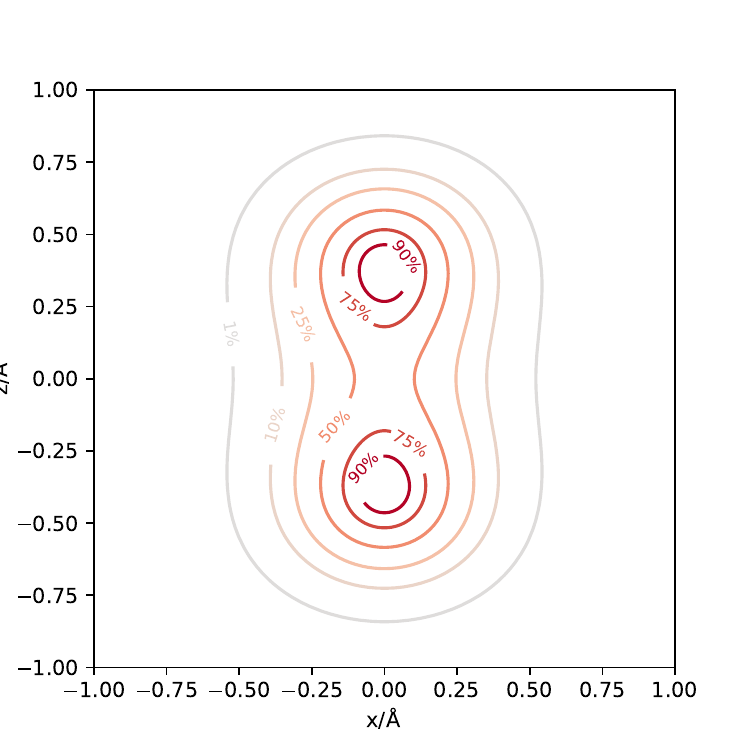}}\,
            \subfloat[B86bPBE-50X-XDM]{\includegraphics[scale=0.42]{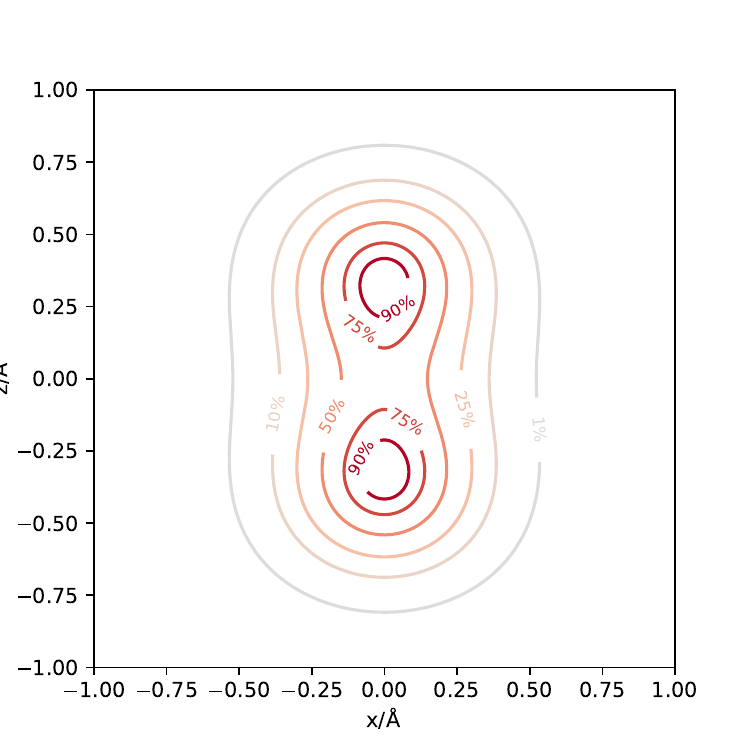}}
            \caption{Two dimensional slices of the ground state wavefunction in the $xz$ plane for (a) SCS-MP2, (b) SOS-MP2, (c) C(HF)-RPA, and (d) B86bPBE-50X-XDM ES methods. Contours are taken at 1\%, 10\%, 25\%, 50\%, 75\%, 90\% and 99\% of the maximum amplitude of the wavefunction in this plane.}
            \label{fig:SI wavefunctions}
        \end{figure*}

        \begin{figure*}
            \centering
            \setcounter{subfigure}{0}
            \subfloat[MP2]{\includegraphics[scale=0.5]{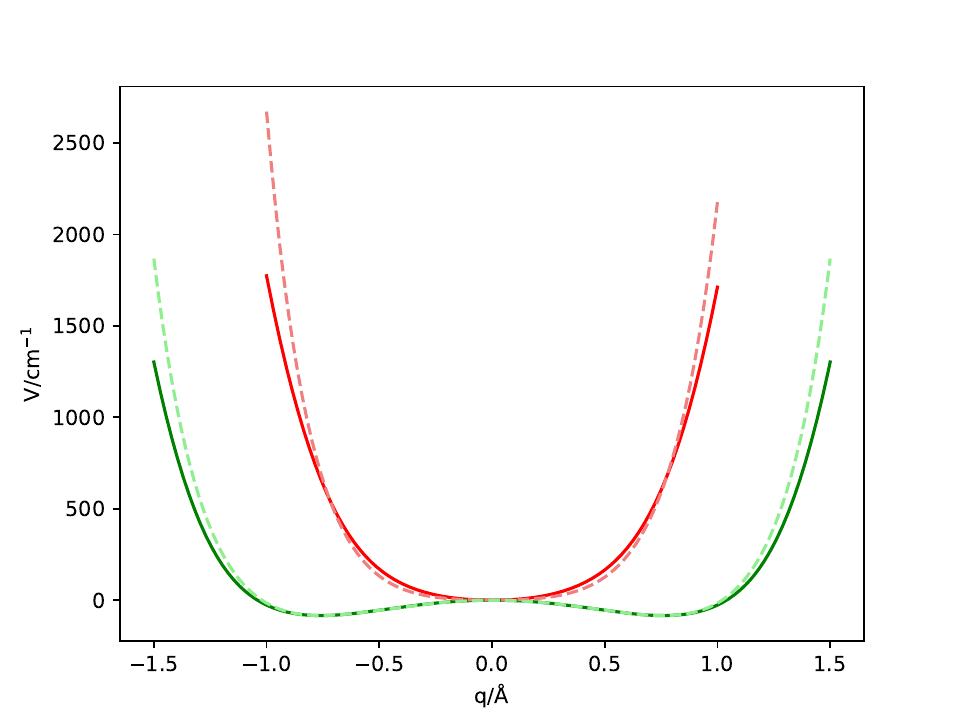}}\,
            \subfloat[RPA@PBE]{\includegraphics[scale=0.5]{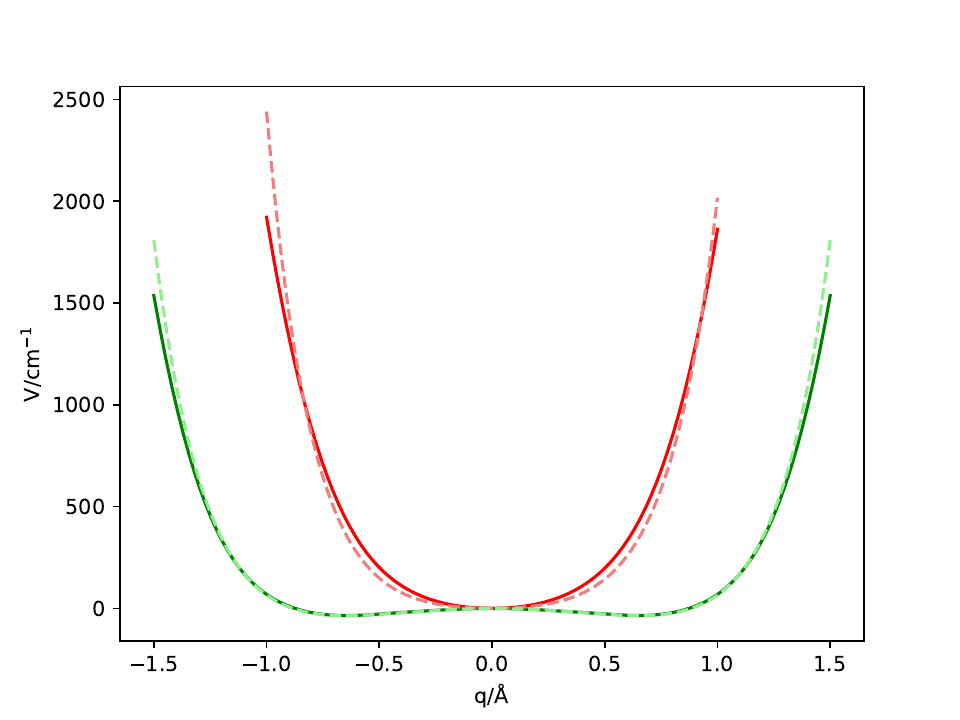}}
            \caption{Comparison of the ES (solid) and LJ (dashed) one-dimensional slices through the $x$ (red) and $z$ (green) directions of the (a) MP2 and (b) RPA@PBE methods and the LJ parameters found by fitting to the minima positions and barrier heights.}
            \label{fig:SI ES LJ}
        \end{figure*}
        
        \begin{figure*}
            \centering
            \includegraphics[scale=1.2]{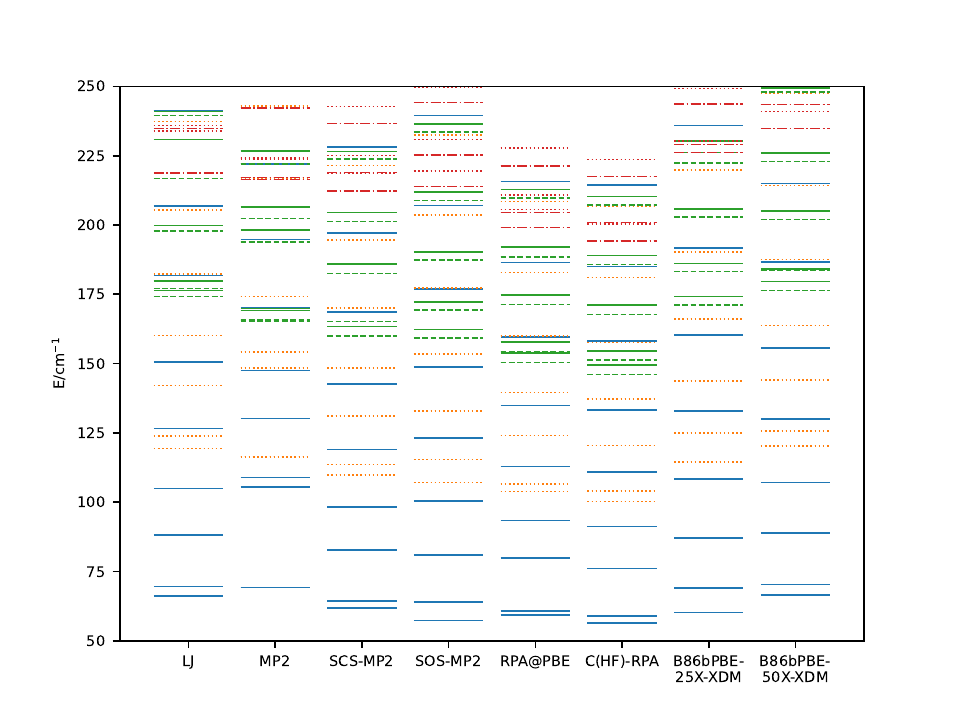}
            \caption{50 lowest translational energies of \Endo{Ne}{C70} for MP2, SCS-MP2, SOS-MP2, RPA@PBE, C(HF)-RPA, B86bPBE-25X-XDM and B86bPBE-50X-XDM PESs using their respective effective LJ parameters, alongside the previously used LJ parameters. The energy zero is set to the minimum of each PES, ensuring this is an absolute energy scale. The colours and linestyle correspond to the principal and angular momentum quantum numbers $(n,|l|)$ respectively. $n\in[0,1,2,3]$ is shown in blue, orange, green and red lines; $|l|\in[0,1,2,3]$ is shown by solid, dotted, dashed, dash-dot lines.}
            \label{fig:SI Energy Levels}
        \end{figure*}

        The set of training data was constructed as described in Eq \eqref{eq:SI Training Data}. The ranges for the uniform distributions were chosen arbitrarily, as with the 1.5\AA\, cutoff. However, this is justified from a discrete variable representation perspective (DVR), as the potential energy becomes quite large and so the endohedral \ce{Ne} atom does not explore the full volume given by these restrictions, as observed in Figs 4 and \ref{fig:SI wavefunctions}. The full training set also included 41 points along the $z$ and 21 points along the $x$ axes respectively, giving a total training set size of 336 points. The points along the axes were to ensure the double well potential which was likely to be present was captured before training the Gaussian Process (GP). This is an important characteristic of the potential, which may have been lost in the interpolation if not enough of it was captured within the training set.

        The training points for this system are based off uniform Cartesian distributions, whereas for \Endo{He}{C60}, they were sampled from adapted Gaussian distributions.\cite{panchagnulaTranslationalEigenstatesHe2024a} This is because for \Endo{He}{C60}, there was some \textit{a priori} knowledge available, of the translational and breathing frequencies due to the existence of spectroscopic data.\cite{bacanuExperimentalDeterminationInteraction2021c, jafariTerahertzSpectroscopyHelium2022} This allowed for a potential optimised discrete variable representation\cite{echavePotentialOptimizedDiscrete1992} (PODVR) type method to pick the training points. The lack thereof for this system suggested using a uniform sampling method instead. This was further compounded by the range of differing barrier heights and minima positions of the double well\cite{panchagnulaExploringParameterSpace2023b} for each electronic structure method, as seen in Table II.

            \begin{equation}
                \text{Training Set} = \{\begin{pmatrix}
                    x\sim U[-0.75,0.75]\\
                    y\sim U[-0.75,0.75]\\
                    z\sim U[-1.5,1.5])
                \end{pmatrix} | \sqrt{x^2+y^2+z^2}<1.5\text{\AA}\} \label{eq:SI Training Data}
            \end{equation}

        The use of randomly distributed training coordinates does not inherently contain information of the symmetry of the system. This implies that this symmetry will not necessarily be enforced by the GP when learning the PES. Ensuring the PES transforms as the totally symmetric irreducible representation (irrep) can be achieved in two different ways, which both involve calculating all the symmetrically equivalent copies of coordinates. For a point group $\mathcal{G}$, it can be constructed from its set of generators $S$. Appropriate combinations of these generators can be applied before training the GP, as described in Eq \eqref{eq:SI Pre symmetrise} where the symmetry element $\hat{g}$ is composed from generators in $S$. This has the disadvantage that this increases the training set size by a factor of $|\mathcal{G}|$. 

        Alternatively, the symmetrisation can occur after training the GP, with the value of the PES and associated covariance calculated by averaging over all symmetric copies as shown in Eqns \eqref{eq:SI Averaged GP mean} and \eqref{eq:SI Averaged GP covariance} respectively. The mean and covariances used in the summations are calculated from the GP as described in Eqns(4) and (5) respectively. This procedure on the other hand increases the prediction set size by a factor of $|\mathcal{G}|$.

            \begin{align}
                \{(\mathbf{X}_t, E_t)\}&=\{(\hat{g}\mathbf{X}, E) | \forall\hat{g}\in \mathcal{G},\, \forall (\mathbf{X},E)\in \text{Training Set}\} \label{eq:SI Pre symmetrise}\\
                E(\mathbf{X})&=\frac{\sum_{\hat{g}\in\mathcal{G}}E(\hat{g}\mathbf{X})}{|\mathcal{G}|},\label{eq:SI Averaged GP mean}\\
                \boldsymbol{\Sigma}(\mathbf{X})&=\frac{\sum_{\hat{g}\in\mathcal{G}}\boldsymbol{\Sigma}(\hat{g}\mathbf{X})}{|\mathcal{G}|},\label{eq:SI Averaged GP covariance}
            \end{align}

        where $(\mathbf{X}_t, E_t)$ is a tuple containing a three-dimensional Cartesian coordinate, with its energy. For the \Endo{Ne}{C70} system, $\mathcal{G}$ is $D_{5h}$, the set of generators $S$ can be given by $\{C_5, C_2', \sigma_h\}$. This is not the simplest set of generators, but given the orientation of the cage, these are very simple to apply. 
        
        Both procedures will ensure the PES transforms as the totally symmetric irrep. However, symmetrising the training set before training the GP is much more computationally intensive, even though the order of $D_{5h}$ is only 20, and may be prone to overfitting. Averaging over symmetric copies of predictions points can result in the GP losing the property that when a prediction point lies within the training set, the GP will return a value within the noise tolerance of the kernel. The averaging of the covariance increases the size of error bars on prediction points due to its positive-definiteness and therefore properties calculated from the PES. However, as seen in Fig 3, the error bars are all under 2\wavenumber, suggesting this is not a substantial effect. 

        As we chose to symmetrise post training, the amount of symmetry breaking within the PES was quantified using the inner product defined in Eq (6). By picking an anisotropic Gaussian as the weight function, this not only ensured convergence of the integral, but also allowed for simple computation using Gauss-Hermite quadrature.\cite{abramowitzHandbookMathematicalFunctions1983} By picking appropriate scale factors $q_i=s_ix_i$, the integral was computed as 
            \begin{align}            
                \braket{\hat{g}V|V}&=\int_{\mathbb{R}^3}V(\hat{g}^{-1}\mathbf{X})V(\mathbf{X})e^{-q_x^2}e^{-q_y^2}e^{-q_z^2}d^3\mathbf{q}\nonumber\\
                &=\int_{-\infty}^{\infty}\int_{-\infty}^{\infty}\int_{-\infty}^{\infty}V(\hat{g}^{-1}\left(\frac{q_{x_i}}{s_x},\frac{q_{y_j}}{s_y},\frac{q_{z_k}}{s_z}\right))V\left(\frac{q_{x_i}}{s_x},\frac{q_{y_j}}{s_y},\frac{q_{z_k}}{s_z}\right)\nonumber\\
                &\times e^{-q_x^2}e^{-q_y^2}e^{-q_z^2}dq_xdq_ydq_z\nonumber\\
                &=\sum_{ijk} V(\hat{g}^{-1}\left(\frac{q_{x_i}}{s_x},\frac{q_{y_j}}{s_y},\frac{q_{z_k}}{s_z}\right))V\left(\frac{q_{x_i}}{s_x},\frac{q_{y_j}}{s_y},\frac{q_{z_k}}{s_z}\right)w_{x_i}w_{y_j}w_{z_k}.\label{eq:SI PES symmetry}
            \end{align}
        These scale factors were chosen as in a discrete variable representation framework (DVR), with $|x|<1$\AA,\, $|y|<1$\AA,\, and $|z|<2$\AA. The exact values for this symmetry breaking given in Table I will depend on the precise definition of inner product.\cite{huynhQSym2QuantumSymbolic2024}

        The GP was trained using the functions given in scikit-learn. The hyperparameters of the optimised kernel defined in Eq (3) are given in Table \ref{table:SI Hyperparameters}.\cite{pedregosaScikitlearnMachineLearning2011}

    \section{\Endo{Ne}{C70} Properties \label{sec:SI Properties}}
        \subsection{Binding Energies \label{sec:SI Binding Energies}}
            \begin{table}
                \centering
                \begin{tabular}{ccccc}
                     \hline Method& $z_0$ & $z_m$ & $z_0$ CPC & $z_m$ CPC  \\
                     \hline MP2&-4.959&-5.200&-4.527&-4.789\\
                     SCS-MP2&-3.642&-3.730&-3.183&-3.278\\
                     SOS-MP2&-2.983&-3.020&-2.511&-2.554\\
                     RPA@PBE&-3.277&-3.382&-1.918&-1.985\\
                     C(HF)-RPA&-2.915&-2.999&-1.755&-1.806\\
                     B86bPBE-25X-XDM&-4.995&-5.024&-&-\\
                     B86bPBE-50X-XDM&-5.248&-5.324&-&-\\
                     \hline
                \end{tabular}
                \caption{Binding energies of \Endo{Ne}{C70} in kcal/mol. The position of the \ce{Ne} atom is at both $z_0$ the origin, and $z_m$ the minimum of the ES method to the nearest 0.05\AA\, from what is given in Table II. Counterpoise corrected (CPC) and uncorrected values are presented.
                }
                \label{table:SI Binding Energies}
            \end{table}

            In order to gauge the magnitude of the basis set superposition error (BSSE), the complexation (or binding) energy of the \Endo{Ne}{C70} system can be calculated as 
                \begin{equation}
                    E_{\text{binding}}(z)=E(\Endo{Ne}{C70}(z)) - E(\ce{Ne}(z)) - E(\ce{C70}(z)) \label{eq:SI binding}
                \end{equation}
            where $z$ is the position of the \ce{Ne} atom along the unique direction. In order to calculate the counterpoise correction (CPC), the $E(\ce{Ne}(z))$ and $E(\ce{C70}(z))$ can be replaced by $E(\Endo{Ne}{gC70}(z))$ and $E(\Endo{gNe}{C70}(z))$ where the preceding $\ce{g}$ refers to ghost atoms/molecules. These binding energies are shown in Table \ref{table:SI Binding Energies} where $z$ is at the origin, and at the minimum of the ES potential to the nearest 0.05\AA\, as these values are in training data. Noticably, the CPC values for the B86bPBE methods are missing which is due to the XDM damping parameters specifically being fit to minimise errors in the binding energies of dimers without the need to account for BSSE. Therefore correcting for it would effectively double count the correction, making the quantities less accurate and so we do not account for it.

            All the methods predict the encapsulation of the \ce{Ne} atom to be favourable, but they are not within the regime of thermochemical agreement of 1kcal/mol (approximately 350\wavenumber). However, these values are comparable with the values predicted for \Endo{Ne}{C60} whose values take a much wider range with a larger set of ES methods.\cite{cioslowskiElectronicStructureCalculations2023a} While it is the case that the magnitude of the BSSE can vary between methods with MP2 type methods being approximately 0.5kcal/mol and the RPA type methods being approximately 1.2kcal/mol, for the purpose of calculating quantities from the PES, it is more important to see how the correction varies across the surface. Although only two points are calculated, we see the BSSE differs by under 10\wavenumber in all the methods between $z_0$ and $z_m$ which span over 0.5\AA\, in space and so we approximate this correction to be flat throughout the surface and absorb it into defining the energy zero of the system.

        \subsection{Translational Eigenstates \label{sec:SI Translational Eigenstates}}

            The error bars in the translational eigenstates in Fig 3 arise due to the $\Delta V$ component in the Hamiltonian given in Eq (7). This can be evaluated as
                \begin{align}
                    \braket{m_xm_ym_z|\Delta V|n_xn_yn_z} &=\mathbf{T}\boldsymbol{\Delta}\mathbf{V}^{DVR}\mathbf{T}^T \label{eq:SI Potential Matrix}\\
                    \boldsymbol{\Delta}\mathbf{V}_{(m_xm_ym_z),(n_xn_yn_z)}&=\sum_{ijk}m_x(x_i)m_y(y_i)m_z(z_i)\Delta V(x_i,y_j,z_k)\times \nonumber\\
                    & n_x(x_i)n_y(y_i)n_z(z_i)w_{x_i}w_{y_j}w_{z_k}. \label{eq:SI Potential Matrix Element}
                \end{align}

            Even though the potential matrix is evaluated in the finite basis representation as defined in Eq \eqref{eq:SI Potential Matrix}, the matrix multiplication is disguising that under the hood the matrix elements are calculated via Gauss-Hermite quadrature\cite{abramowitzHandbookMathematicalFunctions1983} as in Eq \eqref{eq:SI Potential Matrix Element}. The GP can sample the value of the potential at the quadrature points $(x_i,y_j,x_k)$ using the mean and covariance as defined in Eqns (4) and (5) as
                \begin{equation}
                    \Delta \mathbf{V}(\mathbf{X}_q) \sim N (E(\mathbf{X}_q), \boldsymbol{\Sigma}(\mathbf{X}_q)) \label{eq:SI Potential Sample}
                \end{equation}
            where $\mathbf{X}_q$ refers to the full set of quadrature points used in order for the integral in Eq \eqref{eq:SI Potential Matrix Element} to converge. This can be sampled $N$ times, which generates $N$ different copies of the Hamiltonian, which can all be diagonalised. This leads to a set of cardinality $N$ for each eigenvalue whose standard deviation can be calculated as an error bar which is seen in Fig 3.

            Contours of the ground state wavefunction for SCS-MP2, SOS-MP2, C(HF)-RPA, and B86bPBE-50X-XDM are shown in Fig \ref{fig:SI wavefunctions}. The grouping of ES methods by barrier height into four sets is further justified, with the SCS-MP2, C(HF)-RPA and B86bPBE-50X-XDM wavefunctions displaying similar characteristics to each other. These wavefunctions exhibit two maxima, but not as prominently as the MP2 in Fig 4, with more wavefunction density towards the centre of the cage, analogous to the RPA@PBE. The SCS-MP2 wavefunction appears more contracted in the $x$ direction as compared to the C(HF)-RPA, which is backed up by its larger prolateness, $\varsigma$, given in Table III. With the two minima much closer together and smaller barrier height, the SOS-MP2 has the two maxima coalesce into a single peak at the origin with the \ce{Ne} atom more delocalised, comparable to the B86bPBE-25X-XDM.

    \section{Effective Lennard--Jones Parameters \label{sec:SI Eff LJ}}

       \begin{table}
           \centering
           \begin{tabular}{ccc}
                \hline Method&$\varepsilon$/\wavenumber&$\sigma$/\AA  \\
                \hline MP2&67.41&2.897\\
                SCS-MP2&49.90&2.963\\
                SOS-MP2&42.29&3.003\\
                RPA@PBE&46.45&2.947\\
                C(HF)-RPA&42.27&2.957\\
                B86bPBE-25X-XDM&45.47&3.021\\
                B86bPBE-50X-XDM&56.66&2.984\\
                LJ&43.79&3.03\\
                \hline
           \end{tabular}
           \caption{Effective LJ parameters for all ES methods found by matching the barrier height and minima positions, alongside the previously used LJ paramters for this system.}
           \label{table:SI EffectiveLJ}
       \end{table}
       \begin{table}
           \centering
           \begin{tabular}{cc}
                \hline Method&Hellinger Distance  \\
                \hline MP2&0.054\\
                SCS-MP2&0.049\\
                SOS-MP2&0.044\\
                RPA@PBE&0.048\\
                C(HF)-RPA&0.056\\
                B86bPBE-25X-XDM&0.021\\
                B86bPBE-50X-XDM&0.032\\
                \hline
           \end{tabular}
           \caption{Hellinger distance between the ground state of the ES method to its effective LJ counterpart.}
           \label{table:SI LJ Hellinger}
       \end{table}

       The effective LJ parameters for the ES methods found by matching the barrier height and minima positions are given in Table \ref{table:SI EffectiveLJ}. They reveal a wide spread in both LJ parameters with $\varepsilon$ ranging between 42.27\wavenumber and 67.41\wavenumber and $\sigma$ in the interval [2.897, 3.003]\AA. Two examples of how well these effective LJ PESes match their counterparts are shown in Fig \ref{fig:SI ES LJ} with the MP2 and RPA@PBE. Along the $z$ direction in green, the dashed LJ potential matches well in both cases between the minima, as expected as this is how it was constructed. Further out than this, they deviate initially by only on the order of 10\wavenumber but this increases to over 100\wavenumber at $\pm1.5$\AA. The $x$ direction in red however highlights the main deficiencies of this effective PES. The MP2 matches well to about $\pm0.6$\AA\, before the LJ grows much more quickly and the difference at -1\AA\, is close to 1000\wavenumber compared to around 500\wavenumber at +1\AA. This highlights the lack of pure radial symmetry in the LJ PES, as there is a stronger repulsive interaction when the central \ce{Ne} is moving towards a cage \ce{C} atom than towards a bond. For the RPA@PBE, it is a less good match as at even small deviations along the $x$ direction the potential differs substantially more than the MP2 equivalent. Eventually the LJ catches up and grows quicker, but the difference at -1\AA\, and +1\AA\, is much lower for the RPA@PBE than the MP2.

        As discussed in Sec IIIA, picking LJ parameters by matching the double well features is not the only option. Another property of the PES which could be chosen to ensure a good match of the fundamental frequency, could be the curvature in the $x$ and $z$ directions of the PES at the origin. When the zero point energy lies above the barrier height, this will ensure the energies of the lowest few eigenstates match well. However, this has its own drawbacks of incorrectly predicting the barrier height and minima positions, which could be measured experimentally, using X-Ray or neutron diffraction.\cite{panchagnulaExploringParameterSpace2023b} 
        
        The lowest 50 energy levels of each ES method up to 250\wavenumber using their effective LJ parameters is given in Fig \ref{fig:SI Energy Levels}, which can be compared to their true energy levels in Fig 3. The eigenspectra suggest that the grouping of the ES methods by using the barrier height as a discriminator is not as effective. While this still works well for the fundamental frequencies, the zero-point energies (ZPEs) do not match as well as the true ES methods. The quality of match is dependent on the interplay between the attractive $R^{-6}$ term and repulsive $R^{-12}$ term, and how well these LJ terms coincide with the electronic structure description of these effects. MP2 has the best agreement between the ES and effective LJ eigenspectra, with the lowest lying eigenstates differing by approximately 3\wavenumber, and the highest within the figure by 10\wavenumber. The MP2 derivatives SCS-MP2 and SOS-MP2 perform similarly, with the differences ranging between 6\wavenumber and 15\wavenumber. The RPA@PBE and C(HF)-RPA follow slightly behind, with differences lying in the intervals [7,20]\wavenumber, and [8,30]\wavenumber respectively. The B86bPBE methods have difference in the intervals [1,3]\wavenumber and [0.4,9]\wavenumber for the 25\% and 50\% Hartree--Fock exchange respectively. This is indicative of the LJ most accurately describing the B86bPBE methods, followed by the MP2 with the worst match with the RPA type methods. This is compounded with the Hellinger distance of the ground state ES wavefunction to its effective LJ counterpart given in Table \ref{table:SI LJ Hellinger}. The B86bPBE are clearly closer than the other methods, but the MP2 type methods are generally closer than the RPA type methods. An interesting outlier is that the RPA@PBE is closer than the MP2, which is unexpected given the energy match is the other way round. Another feature to notice in the effective LJ eigenspectra is the ordering of states by quantum numbers differs to what is seen for the ES. This is to be expected given the curvature of the LJ type PES in the radial direction as seen in Fig \ref{fig:SI ES LJ}, suggesting the ordering of $(n,l)$ quantum numbers to be different.

        The large spread of possible LJ parameters, which are very sensitive to how they are calculated and the lack of transferability indicates that they should not be used. This is exacerbated in the difference between these LJ type PESes and their ES equivalents shown in Fig \ref{fig:SI contours}, and the difference in the translational eigenstates in Figs 3 and \ref{fig:SI Energy Levels}.

    \bibliography{NeC70}